\documentclass[aps, prc, twocolumn, tightenlines, letterpaper, amsmath, amssymb, preprintnumbers, superscriptaddress, floatfix, longbibliography, nofootinbib]{revtex4-2}
\usepackage[T1]{fontenc}
\usepackage{bm}
\usepackage{xspace}
\usepackage[dvipsnames]{xcolor}
\usepackage{graphicx}
\usepackage{tabularx}
\usepackage{enumitem}
\usepackage{qcircuit}
\usepackage{braket}
\usepackage{dsfont}
\usepackage[normalem]{ulem}
\usepackage{longtable}

\usepackage[hypertexnames=false]{hyperref}
\hypersetup{
    colorlinks=true,       
    linkcolor=blue,          
    citecolor=blue,        
    filecolor=blue,      
    urlcolor=blue           
}

\newcolumntype{Y}{>{\centering\arraybackslash}X}
\newcolumntype{L}{>{\raggedright\arraybackslash}X}

\usepackage{mathtools}
\usepackage{orcidlink}

\begin{document}

\begin{figure}
\vskip -1.cm
\leftline{\includegraphics[width=0.15\textwidth]{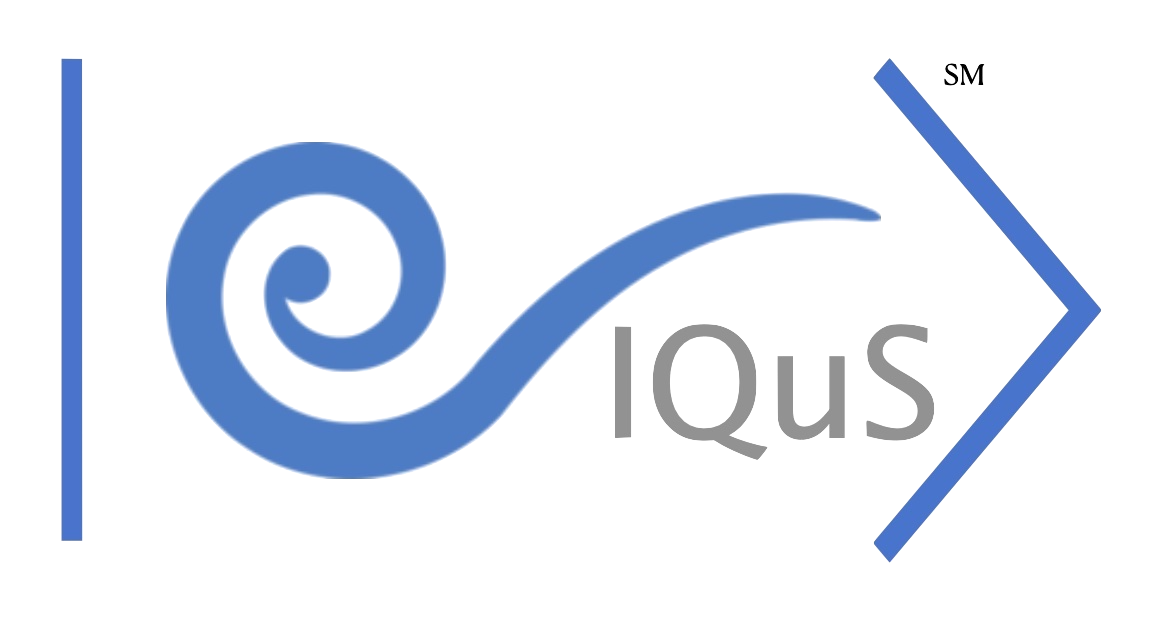}}
\vskip -2.cm
\end{figure}

\title{Quantum Simulations in Effective Model Spaces (I): Hamiltonian-Learning VQE using Digital Quantum Computers and Application to the Lipkin-Meshkov-Glick Model}


\author{Caroline E. P. Robin\,\orcidlink{0000-0001-5487-270X}}
\email{crobin@physik.uni-bielefeld.de}
\affiliation{Fakult\"at f\"ur Physik, Universit\"at Bielefeld, D-33615, Bielefeld, Germany}
\affiliation{GSI Helmholtzzentrum f\"ur Schwerionenforschung, Planckstra{\ss}e 1, 64291 Darmstadt, Germany}

\author{Martin J.~Savage\,\orcidlink{0000-0001-6502-7106}}
\email{mjs5@uw.edu}
\affiliation{InQubator for Quantum Simulation (IQuS), Department of Physics, University of Washington, Seattle, WA 98195}
\thanks{On leave from the Institute for Nuclear Theory.}

\preprint{IQuS@UW-21-041}
\date{\today}

\begin{abstract}
\begin{description}
\item[Background]
Quantum simulations offer the potential to predict the structure and dynamics of nuclear many-body systems that are beyond the capabilities of classical computing.
Generally, preparing the ground state of strongly-interacting many-body systems relevant to nuclear physics is however inefficient, even using ideal quantum computers.
In addition, currently available NISQ-era quantum devices possess modest numbers of qubits, limiting the size of quantum many-body systems that can be simulated.
In this context, a reformulation of the quantum many-body problems using truncated model spaces and Hamiltonians is desirable to make them more amenable to near-term quantum computers.
The importance of symmetries in low-energy theories, including effective field theories (EFTs), lattice quantum chromodynamics (QCD), and effective model spaces for nuclear systems, 
in particular their interplay with the reduction of active Hilbert spaces, is well known.
Lesser known is the fact that the non-commutivity of some symmetries and truncations of the model space can be profitably combined with variational calculations to rearrange the entanglement into localized structures and enable more efficient simulations.
\item[Purpose] 
The goal of the present study is to explore and utilize
the non-commutivity of symmetries and model-space truncations of quantum many-body systems important to nuclear physics, particularly in combination with variational algorithms for quantum simulations and effective Hamiltonian learning.
\item[Method]
We introduce an iterative hybrid classical-quantum algorithm,
Hamiltonian learning variational quantum eigensolver (HL-VQE),
that simultaneously optimizes an effective Hamiltonian,
thereby rearranging entanglement into the effective model space,
and the associated ground-state wavefunction.
Quantum simulations, using classical computers and IBM's superconducting-qubit quantum computers, are performed to demonstrate the HL-VQE algorithm, 
in the context of the Lipkin-Meshkov-Glick (LMG) model of interacting fermions,
where the Hamiltonian transformation corresponds to an orbital rotation.
We use a mapping where the number of qubits scales with the 
$\log$ of the size of the effective model space, rather than the particle number.
\item[Results]
HL-VQE is found to provide an exponential improvement in LMG-model calculations of the ground-state energy and wavefunction,
compared to naive truncations without Hamiltonian learning, throughout a significant fraction of the Hilbert space.
In the context of EFT, this corresponds to counterterms scaling exponentially with the cut-off as opposed to power law.
Implementations on IBM's QExperience quantum computers and simulators for 1- and 2-qubit effective model spaces
are shown to provide accurate and precise results, reproducing classical predictions. 
\item[Conclusions] 
For a range of parameters defining the LMG model,
the HL-VQE algorithm is found to have better scaling of quantum resources requirements than previously explored algorithms.
In particular, the HL-VQE scales efficiently over a large fraction of the model space, in contrast to VQE alone.
This work constitutes a step in the development of entanglement-driven quantum algorithms for descriptions of nuclear many-body systems.
This, in part, leverages the potential of noisy intermediate-scale quantum (NISQ) devices.
 The exponential scaling of counterterms observed in this study suggests the possibility of more general applicability to other non-perturbative EFTs.
\end{description}
\end{abstract}

\maketitle

\section{Introduction}
\label{sec:intro}
\noindent
Future quantum computers will allow us to predict the properties and dynamics of physically relevant quantum many-body systems that are inaccessible to classical computing~\cite{5392446,5391327,Benioff1980,Manin1980,Feynman1982,Fredkin1982,Feynman1986,doi:10.1063/1.881299,williamsNASAconference,Lloyd1073}, 
including nuclei and dense matter.
It is the capability to control and maintain coherence, entanglement and quantum correlations 
that allows some classically inefficient simulations to be performed efficiently 
with (ideal) quantum computers.
One of the major features of such devices is the linear scaling of the memory requirement (number of qubits) needed to map a full $N$-particle Hilbert space.
However, even with this capability, the required time for solving 
some important nuclear many-body problems on quantum computers in an exact way 
generally
remains super-polynomial in the number of particles, 
residing in the {\bf QMA} complexity class~\cite{10.1007/978-3-540-30538-5_31}, rather than in {\bf BQP}.
This scaling can be mitigated by applying 
truncations of the full Hilbert space 
and Hamiltonian
to render the problem solvable in polynomial time.
Such truncations, which can be made in space-time, of the continuous fields, in the dimensionality of local Hilbert spaces, and so forth, are usually justified because 
only a subset of the Hilbert space is expected to be relevant to 
the description of ground or low-lying excited states.
Furthermore, currently available quantum devices, in addition, suffer
from important limitations such as noise, limited number of qubits and connectivity, as well as gate and measurement errors. 
In this context, truncations of the Hilbert space allow for a reduction of the circuit sizes, 
in terms of both qubit and gate numbers, decreasing the noise and errors, thus making such circuits more suitable for implementation on current Noisy Intermediate Scale Quantum (NISQ)~\cite{Preskill2018quantumcomputingin} devices.
Extracting meaningful predictions from the truncated spaces
then requires a complete quantification of uncertainties in the simulations, 
and therefore theoretical and algorithmic understandings of these truncations is essential 
(for a recent review, see, for example,  Ref.~\cite{Klco:2021lap}).
\\ \\
\indent
In addition to truncations of the model space, simulations of systems with truncated "nearby" Hamiltonians, with systematically removable deviations from a target Hamiltonian, can also provide reliable results~\cite{Klco:2021lap}, if the associated uncertainty is within the 
target uncertainty
of the computation.
In this context, the use of effective field theory (EFT) naturally lends itself, as, in principle, the order of truncation of the EFT can be matched to the size of the simulation error through an iterative tuning procedure~\cite{Klco:2021lap}.
EFTs, effective interactions and effective model spaces have been extensively employed in modern nuclear physics for the simulations of few- and many-baryon systems, and have provided new pathways for classical simulations to address quantities of interest, well before reliable predictions from quantum chromodynamics (QCD) become possible~\cite{Beane:2008dv,Beane:2010em,Davoudi:2020ngi,Drischler:2019xuo}.
It is also the means by which lattice QCD calculations furnish results for low-lying processes. The simulated Hamiltonian(s) derived from, for example, EFT and variants thereof, can be systematically improved in support of more precise numerical computations of the systems of interest as they become practical.
In the context of quantum simulations, an expansion of the EFT to higher orders can be made as experimental progress is achieved in building quantum devices.
\\ \\
\indent
For an EFT to be predictive,
it must capture the low-energy degrees of freedom with effective interactions that systematically recover S-matrix elements order-by-order in an expansion (that may or may not involve non-perturbative resummations), with a power-counting defined in the context of a regularization and renormalization scheme.
The challenge in applying EFTs to nuclear systems is the size of the expansion parameters, 
which can be uncomfortably large,
and establishing a consistent power-counting scheme.
Classic examples 
of EFTs that are useful for nuclear physics are 
chiral perturbation theory ($\chi$PT)~\cite{Weinberg:1978kz,Gasser:1983yg}, 
heavy-baryon chiral perturbation theory (HB$\chi$PT)~\cite{Jenkins:1990jv} 
and nuclear EFT~\cite{Weinberg:1990rz,vanKolck:1994yi,Kaplan:1998tg,Epelbaum:1998ka}, 
which are designed to implement the approximate global chiral symmetries of QCD.
These techniques continue to be merged with more 
traditional nuclear shell-model and effective model-space frameworks 
(see {\it e.g.} Ref.~\cite{10.3389/fphy.2020.00379} for a recent review),
so that predictions for nuclei and multi-nucleon systems are consistent with QCD.
Ultimately, the objective of the EFT program is to match  effective nuclear forces to QCD, which can then be used in nuclear many-body calculations performed in effective model spaces, for the description of both structure properties and dynamical processes, 
{\it e.g.} as discussed in Refs.~\cite{Beane:2008dv,Savage:2010hp}. 
In the area of quantum simulations of dynamics, it has recently been shown that product formula can be used to evolve states in a low-energy space forward in time with only exponentially small leakage into the high-energy space~\cite{Sahinoglu:2021},
consistent with the rigorous use of effective model spaces and EFTs.
\\ \\
\indent 
The present work has emerged from considerations of entanglement as an organizing principle for nuclear structure calculations~\cite{PhysRevC.67.051303,PhysRevC.69.024312,Papenbrock_2005,PhysRevC.92.051303,Johnson:2022mzk,Robin:2020aeh,Kruppa_2021,Kruppa:2021yqs,PhysRevA.103.032426,PhysRevA.104.032428,PhysRevA.105.062449,Pazy:2022mmg,Tichai:2022bxr,Bulgac:2022cjg,Bulgac:2022ygo}.
In a previous study~\cite{Robin:2020aeh}, we analysed the entanglement structures of single-particle orbitals emerging from {\it ab-initio} no-core configuration-interaction calculations of light nuclei.
In particular, we investigated how transformations of the single-particle basis 
into a natural basis
could minimize and rearrange entanglement into a localized part of the Hilbert space. 
Including two-nucleon correlations in light nuclei
to define variational natural (VNAT)
single-particle orbitals localized the one-orbital entanglement entropy around the Fermi level, and
eliminated two-orbital negativity in the nuclear 
wavefunction. 
The variational character of the VNAT basis was found to restrict the two-orbital mutual information to the truncated space and decoupled the inactive space.
In that sense, the 
VNAT basis minimizes the number of many-body basis states necessary to describe the nuclear wavefunction, 
and reduces the loss of information originating from the model-space truncation.
More recently, the valence-space density matrix renormalization group (VS-DMRG) has been used to compute orbital entanglement entropy and mutual information in {\it sd}-shell nuclei, to motivate a re-organization of orbitals~\cite{Tichai:2022bxr}.

Following our earlier work, we investigate whether the localization of entanglement structures 
can be utilized in designing efficient quantum simulations of nuclear many-body systems amenable to quantum computers, including NISQ devices. 
Our present study takes a step in that direction,
and demonstrates that 
an orbital transformation, equivalent to transformation of the Hamiltonian, provides an efficient mechanism for including quantum correlations in low-lying states of a model many-body system in quantum simulations, 
and is also potentially useful for classical simulations. 
\\ \\
\indent
As widely appreciated, the Lipkin-Meshkov-Glick (LMG)  model is a useful "sandbox" for testing 
new ideas that may have applicability (to varying degrees) to realistic nuclear systems and forces.
The separation of scales in realistic systems means that the ideas developed within the LMG model have also been helpful in those systems. 
Its rich phenomenology from a simple Hamiltonian provides sufficient complexity that has led to 
a number of previous studies that explore 
quantum correlations and entanglement~\cite{PhysRevA.69.022107,PhysRevA.69.054101,PhysRevA.70.062304,PhysRevA.71.064101,PhysRevA.77.052105,PhysRevA.100.062104,PhysRevA.100.062104,PhysRevA.103.032426,PhysRevA.104.032428,PhysRevA.105.062449}, 
quantum algorithms~\cite{PhysRevC.104.024305,Chikaoka:2022kff,PhysRevC.105.064317,PhysRevC.106.024319},  
and more, to develop understanding and techniques that can be applied to quantum simulations of nuclei and multi-nucleon systems.   
Previous quantum simulations of the LMG model~\cite{PhysRevC.104.024305,Chikaoka:2022kff,PhysRevC.105.064317,PhysRevC.106.024319}
have directly mapped the elementary SU(2)-spaces associated with each fermion to qubits in the quantum computer (or classical simulator).
In this way VQE has been used to determine the ground state of few-nucleon systems in the model~\cite{PhysRevC.104.024305}  
using IBM's quantum computers~\cite{IBMQ},
and ADAPT-VQE~\cite{Grimsley_2019,Anastasiou:2022swg} has been used 
to examine systems of up to $N=12$ nucleons using a classical simulator~\cite{PhysRevC.105.064317}.
That extensive study explored the behavior of ADAPT-VQE building upon the trivial 0p-0h state and also the Hartee-Fock ground state.  
The authors also applied their techniques to valence model spaces of nuclei in the {\it sd}- and {\it pf}-shells, with interesting results, concluding that there could be substantial benefits to applying such quantum algorithms to nuclear structure calculations.
One of these studies~\cite{PhysRevC.106.024319} also considered a number of other mappings, including 
the J-space mapping (that we utilize in this work) and 
binary encoding of the occupation number with consideration of the efficient Gray code,
with good results obtained for the ground states, and also excited states, for small systems.
\\ \\
\indent
In this work, we examine the utility of effective model spaces 
for hybrid classical-quantum simulations of nuclear many-body systems.  
Specifically, we explore potential advantages of simultaneously learning the effective-model-space Hamiltonian and finding the ground-state wavefunction in the LMG model,
introducing the Hamiltonian-Learning-VQE algorithm, which is similar to the orbital-optimized VQE (oo-VQE) algorithm used by quantum chemists~ \cite{doi:10.1063/1.5141835,PhysRevResearch.2.033421}, but differing in the truncation used in 
defining the effective model space(s).  
The truncation we employ more closely resembles that employed in the EFTs mentioned above, aligning to a greater extent with the Wilsonian renormalization group organization, than with the coupled-clusters truncation(s) used in quantum chemistry.
The optimized ground state in the one-dimensional effective model space corresponds to the Hartee-Fock state, while the ground state in larger spaces systematically and self-consistently includes higher-body correlations.
The exponential improvement in convergence of the estimated ground-state energy that we find from the 
optimization, means that the VQE algorithm, which naively scales poorly on quantum devices, 
scales favorably when the Hamiltonian in the effective model space is also learned.
\\ \\
\indent 
The main points emerging from the present work are:
\begin{itemize}
\item 
    The use of EFTs and effective model spaces for quantum simulations of nuclear structure calculations is explored in the context of the Lipkin-Meshkov-Glick model.
    Simultaneously optimizing the effective Hamiltonian and ground-state wavefunction in 
    (truncated) effective model spaces, 
    using the ground-state energy as the cost-function, 
    leads to exponential improvements in the ground-state energies and wavefunctions 
    throughout a significant fraction of the Hilbert space 
    over naive truncations of the Hamiltonian.
    We anticipate that the ideas and results are  broadly applicable.
\item 
    A Hamiltonian learning variation of the VQE algorithm, HL-VQE, is introduced that simultaneously optimizes the Hamiltonian and variational ground-state wavefunction in a truncated model space,
    that is amenable to quantum simulations and scales efficiently (because of the above obervation(s)).
    The algorithm is hybrid-classical-quantum, where expectation values of the Hamiltonian and its gradients are evaluated using a quantum computer, while the variations of the Hamiltonian and wavefunction parameters are performed classically.
\item
    Quantum simulations in effective model spaces using 1- and 2-qubits
    are performed using IBM's quantum computer 
    {\tt ibm\_lagos}, which is one of the IBM Quantum Falcon Processors~\cite{IBMQ},   
    and simulator {\tt AER} using the {\tt qiskit}~\cite{matthew_treinish_2022_6924865} API.
    These simulations demonstrate techniques and workflows
    that could be implemented on more capable quantum computers,  
    providing results that recover the exponential convergence of 
    ground-state energies and wavefunction fidelities.
\end{itemize}
\indent
The outline of the paper is as follows: 
Section~\ref{sec:LGM_ex} introduces the LMG model in detail, particularly its representation in terms of 
SU(2) angular momentum states and the corresponding Hamiltonian in that basis.
These results naturally lend themselves to considering effective model spaces 
and effective Hamiltonians with a variational parameter associated with global rotations. This is detailed in
Section~\ref{sec:LGM_eff}, which presents results obtained with classical calculations for the convergence of the ground-state energy and wavefunction fidelity for selected model parameters and system sizes,
along with an overview of the theoretical fabric defining the logic of the effective model space calculations.
The algorithms, techniques, quantum circuits, gradient-descent, and more, 
associated with the HL-VQE algorithm that we use for quantum simulations performed using IBM's quantum computers are detailed in Sec.~\ref{sec:QuSimFormal}.
Results obtained from 1- and 2-qubit quantum simulations using {\tt ibm\_lagos} and classical simulator {\tt AER} are presented and discussed in Sec.~\ref{sec:QuSimProduction}, where estimates of quantum resources required for simulations using more capable quantum computers are also provided.
In Sec.~\ref{sec:conclusions}, we conclude and summarize the results of this work.

\section{The Lipkin-Meshkov-Glick Model}
\label{sec:LGM_ex}
\begin{figure}[h]
\centering{\includegraphics[width=\columnwidth] {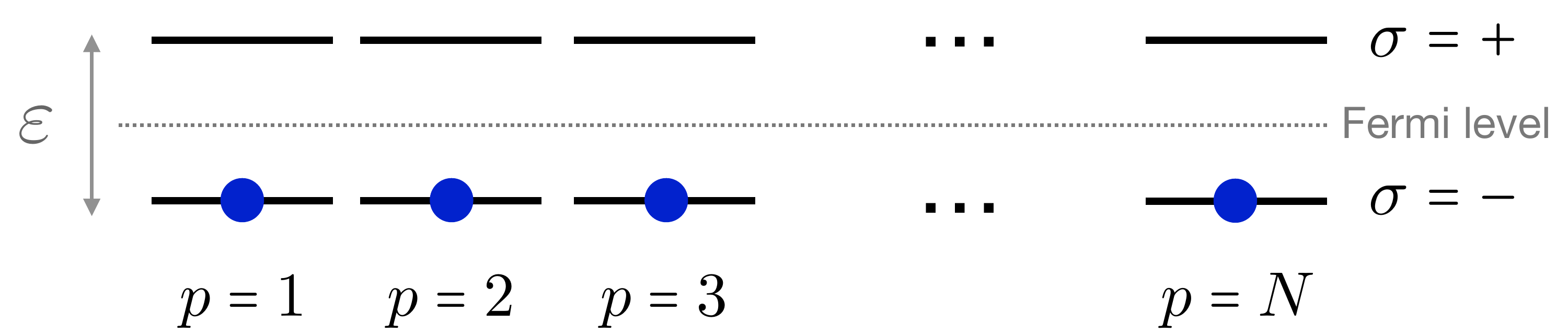}} 
\caption{Lowest-energy non-interacting (0p-0h) configuration of a system of $N$ particles in the LMG model.}
\label{f:Lipkin}
\end{figure}

\noindent
The Lipkin-Meshkov-Glick (LMG) model~\cite{LIPKIN1965188}
describes a system of $N$ interacting fermions, each distributed between two levels separated by an energy $\varepsilon$, labeled 
by $\sigma=\pm$.
Each level is $N$-fold degenerate, with (non-interacting) single-particle states labeled by $p=1,2,...N$.~\footnote{
This is equivalent to a system of $N$ $s={1\over 2}$ particles immersed in a magnetic field,
where each spin interacts with all of the others~\cite{PhysRevA.69.022107,PhysRevA.69.054101}.}
This is illustrated in Fig.~\ref{f:Lipkin}, 
which shows the lowest-energy, or zero-particle-zero-hole (0p-0h), 
non-interacting configuration of the system.  
In the LMG model, the particles are assumed to interact via a monopole-monopole interaction that scatters pairs of particles between states of the upper and lower levels having the same value of $p$. 
The Hamiltonian governing the system has the form
\begin{eqnarray}
\hat H &=& \frac{\varepsilon}{2} \sum_{\sigma p} \sigma c^\dagger_{p\sigma} c_{p\sigma} - \frac{V}{2} \sum_{p q \sigma} c^\dagger_{p\sigma} c^\dagger_{q\sigma} c_{q-\sigma} c_{p-\sigma} \; ,
\nonumber \\
    &=& \varepsilon \hat{J}_z - \frac{V}{2} \left( \hat{J}_+^2 + \hat{J}_-^2 \right) \; ,
\label{eq:H_unrot}
\end{eqnarray}
where the operators $c^\dagger_{p\sigma}$ and $c_{p\sigma}$ create and destroy a fermion in 
level $\sigma$ of state $p$,
respectively, and $\hat{J}_z$, $\hat{J}_+$, $\hat{J}_-$ are collective quasi-spin operators defined as
\begin{eqnarray}
\hat{J}_z &=& \frac{1}{2} \sum_{p\sigma} \sigma c^\dagger_{p\sigma}c_{p\sigma} \; ,
\nonumber \\
\hat{J}_+ &=&  \sum_{p}  c^\dagger_{p+} c_{p-} \; ,
\nonumber \\
\hat{J}_- &=& (J_+)^\dagger =  \sum_{p} c^\dagger_{p-} c_{p+} \; , 
\label{eq:Js}
\end{eqnarray}
which
generate
an $su(2)$ algebra. 
\\ \\
\indent
The Hamiltonian in Eq.~(\ref{eq:H_unrot}) preserves a number of symmetries. In particular, it conserves rotational invariance so that (exact) eigenstates $\ket{\Psi^{(J)}_{ex}}$ of $\hat H$ can be expanded in the basis $\{ \ket{J,M} \}$ formed by eigenstates of $\hat{J}^2$ and $\hat{J}_z$ as
\begin{equation}
\ket{\Psi^{(J)}_{ex}} = \sum_{M=-J}^J A_{J,M} \ket{J,M} \; .
\label{eq:exact_state}
\end{equation} 
Exact solutions can then 
be obtained via diagonalization of the Hamiltonian in the $\{ \ket{J,M} \}$ basis,
which is equivalent to minimizing the energy of the system with respect to the coefficients $\{ A_{J,M}\}$.
In the present work, 
we are interested in determining the ground state, which corresponds to diagonalization in the block characterized by $J=N/2$.
The quantum number $M$, eigenvalue of $\hat{J}_z$, denotes the different non-interacting configurations of the system. 
For instance, the configuration where all particles are in the lower level (0p-0h configuration) corresponds to $\ket{J=N/2,M=-N/2}$ (see Fig.~\ref{f:Lipkin}), the configurations with one particle in the upper level (1p-1h configurations) are contained in $\ket{J=N/2,M=1-N/2}$, and so on, 
up to the configuration where all $N$ particles are in the upper level ($N$p-$N$h), which corresponds to $\ket{J=N/2,M=N/2}$.
In the rest of the manuscript, we will thus adopt the following short-hand notation to denote the many-body basis states,
\begin{equation}
\ket{n} \equiv \ket{J=\frac{N}{2},M = n-J = n - \frac{N}{2}} 
\; ,
\end{equation}
where $n$ denotes the excitation order ($n$p-$n$h) of the state.
\\ \\
\indent
The specific form of the interaction in Eq.~(\ref{eq:H_unrot}) also preserves the number of particles in a given state $p$ (because $V$ scatters particles between states with same value of $p$), which is associated with the operator
\begin{equation}
\hat{N}_p = \sum_\sigma c^\dagger_{p\sigma} c_{p\sigma} 
\; ,
\label{eq:N_p}
\end{equation}
with $\braket{\Psi^{(J)}_{ex} | \hat{N}_p | \Psi^{(J)}_{ex}} = 1$,
as well as "parity" symmetry 
(sometimes referred to as number-parity symmetry~\cite{PhysRevC.105.064317}) associated with
\begin{equation}
\hat{\Pi} = e^{i \pi \hat{N}_+} \; ,
\label{eq:parity_op}
\end{equation}
where 
\begin{equation}
\hat{N}_+ = \sum_p c^\dagger_{p+} c_{p+} = \hat{J}_z + \frac{N}{2} \; ,
\end{equation}
counts the number of particles in the 
$\sigma=+$
upper level. 
The configurations with an 
even (odd, respectively)
number of particles in the upper level are eigenstates of $\hat \Pi$ with eigenvalue $+1$ 
({-1}, respectively):
\begin{equation}
\hat\Pi \ket{n}  = (-1)^n \ket{n} \; .
\end{equation}
This parity symmetry is due to the fact that $V$ only scatters pairs of particles.
\\ \\
\indent
The matrix elements of the Hamiltonian in Eq.~(\ref{eq:H_unrot}) are
\begin{eqnarray}
\braket{n' | \hat{H} | n} &=& \varepsilon \, C^z_{n} \, \delta_{n',n}  - \frac{V}{2} \left( C^{+}_{n} \, \delta_{n',n+2} + C^{-}_{n} \, \delta_{n',n-2} \right) 
\; , \nonumber \\
\end{eqnarray}
%
 where
\begin{eqnarray}
C^{z}_{n}  &=& M = n-J  \; ,
\nonumber\\
C^{+}_{n}  &=& \sqrt{J(J+1) - M (M+1)} \nonumber \\
			&& \times \sqrt{J(J+1) - (M+1)(M+2)} \; ,\nonumber \\			
                      &=& \sqrt{J(J+1) - (n-J) (n-J+1)} \nonumber \\
                         && \times \sqrt{J(J+1) - (n-J+1)(n-J+2)} \; , 
                         \nonumber\\
C^{-}_{n}  &=& \sqrt{J(J+1) - M (M-1)} \nonumber \\
			  && \times  \sqrt{J(J+1) - (M-1)(M-2)} \nonumber \\			  
                      &=& \sqrt{J(J+1) - (n-J) (n-J-1)} \nonumber \\
                           && \times \sqrt{J(J+1) - (n-J-1)(n-J-2)} \; .              
\end{eqnarray}
Thus the Hamiltonian only connects configurations with excitation orders differing by zero or two units, as imposed by the parity symmetry.
Consequently, the interacting ground state will be restricted to a superposition of even-$n$ components 
(0p-0h, 2p-2h, $\cdots$), 
while states with odd-$n$ configurations 
(1p-1h, 3p-3h, $\cdots$) will correspond to excited states.
\\ \\
\indent
It is worth briefly considering the behavior of the system in the $N\rightarrow\infty$ limit.
With appropriate re-scaling, the discrete states map toward a continuum in $n$-space, and the Hamiltonian matrix 
can be identified with the finite-difference form of a second-order differential equation.
For a range of Hamiltonian parameters, 
the lowest-lying solutions of the equation have wavefunctions that fall exponentially for large-$n$, 
and consequently, the energy difference between the lowest-lying even- and odd-parity 
states falls exponentially with $N$, while the gap to the first excited states 
(with same parity as the ground state) tends 
to a nearly $N$-independent value.
Thus, the ground state of the LMG-Hamiltonian is doubly degenerate in the large-$N$ limit, 
with states of opposite parity that are gapped to nearest excitations~\footnote{
The ground-state sector of the LMG model with particular Hamiltonian parameters
possesses properties for a variant of a parity-encoded logical qubit~\cite{https://doi.org/10.48550/arxiv.quant-ph/0006088,Ralph_2005}.}.

\section{Effective Model Spaces for the Lipkin-Meshkov-Glick Model}
\label{sec:LGM_eff}
\noindent
When all configurations are included in Eq.~(\ref{eq:exact_state}), the solution does not depend on the single-particle basis that is used to build the configurations $\ket{n}$.
In realistic many-body calculations, however, one is typically forced to truncate such 
an expansion in order to make the diagonalization problem tractable.
Once a truncation is made, the solution will depend on the nature of the single-particle states,
which can also be optimized via a variational principle.
Minimizing the energy with respect to both expansion coefficients and orbitals corresponds to the approach known as Multi-Configuration Self-Consistent Field (MCSCF) method in Quantum Chemistry (see, {\it e.g.}, Ref.~\cite{doi:10.1021/cr200137a}). 
This approach has also been applied to classical calculations of nuclear systems~\cite{Robin:2015lba,Robin:2016wsx},
and showed good entanglement properties in {\it ab initio} calculations of light nuclei \cite{Robin:2020aeh}. In particular,
it was found that the optimized orbitals led to an increased localization of quantum correlations within the basis.
\\ \\
\indent
In the LMG model, in order to preserve the rotational invariance and the symmetry associated with the operator $\hat{N}_p$ in Eq.~(\ref{eq:N_p}), the orbital transformation between the original and optimized single-particle bases is restricted to only mix states of the lower and upper levels with same value of $p$. This can be written as a unitary transformation 
corresponding to a rotation of the individual quasi-spins around the $y$ axis by an angle $\beta$, as
\begin{eqnarray}
\begin{pmatrix}
c_{p+} (\beta) \\
c_{p-} (\beta)
\end{pmatrix}
= 
\begin{pmatrix}
\cos(\beta/2) & -\sin(\beta/2) \\
\sin(\beta/2) & \cos(\beta/2) \\
\end{pmatrix}
\begin{pmatrix}
c_{p+} \\
c_{p-}
\end{pmatrix}
\; .
\label{eq:transfo}
\end{eqnarray}
where $c_{p\sigma} \equiv c_{p\sigma} (\beta=0)$ is the operator annihilating a particle in the original single-particle state $(p,\sigma)$.
After this rotation,
a truncation can be made by imposing a cut-off $\Lambda$ in the summation in Eq.~(\ref{eq:exact_state}) to include $n$p-$n$h configurations with $n \le \Lambda -1$.
The effective (truncated) many-body state becomes
\begin{equation}
\ket{\Psi}^{(\Lambda)}  = \sum_{n =0}^{ \Lambda -1} A^{(\beta)}_n \ket{n, \beta} \; ,
\label{eq:eff_wf}
\end{equation}
where the basis states $\ket{n,\beta}$ are $n$p-$n$h configurations built on the rotated single-particle basis.
In Eq.~(\ref{eq:eff_wf}), $\ket{\Psi}^{(\Lambda)} \equiv \ket{\Psi}^{(\Lambda)}(\beta)$ with the explicit $\beta$ dependence omitted in what follows.
\\ \\
\indent
The effective Hamiltonian can then be written in terms of the rotated collective quasi-spin operators,
$\hat{\vec{J}}(\beta) = \hat{U}^\dagger(\beta) \, \hat{\vec{J}} \, \hat{U}(\beta)$, where $ \hat{U}(\beta) = e^{-i \hat{J}_y \beta } $, as
\begin{eqnarray}
\hat H(\beta) &&\equiv U^\dagger(\beta) \hat{H} U(\beta) \nonumber \\
&& = \varepsilon \Bigl[ \cos\beta \hat{J}_z(\beta) + \frac{1}{2} \sin\beta \bigl( \hat{J}_+(\beta) + \hat{J}_-(\beta) \bigr) \Bigr] \nonumber \\
&& - \frac{V}{4} \Bigl[ \sin^2\beta \bigl( 4 \hat{J}_z(\beta)^2 - \{ \hat{J}_+(\beta), \hat{J}_-(\beta) \} \bigr) \nonumber \\
&& \hspace{1cm} + (1 + \cos^2\beta) \bigl(\hat{J}_+(\beta)^2 + \hat{J}_-(\beta)^2 \bigr) \nonumber \\
&& \hspace{1cm} - 2 \sin\beta \cos\beta \bigl( \{ \hat{J}_z(\beta),\hat{J}_+(\beta)\} \nonumber \\
&& \hspace{1cm} + \{ \hat{J}_z(\beta), \hat{J}_-(\beta) \} \bigr) \Bigr] \; ,
\label{eq:eff_H}
\end{eqnarray}
where the $\hat{J}_{z,\pm}(\beta)$ operators can be expressed in terms of $c_{p\sigma}(\beta)$ and $c^\dagger_{p\sigma}(\beta)$ in analogy with Eq.~(\ref{eq:Js}) (see Appendix~\ref{app:Lipkin_formalism} for details), 
and $\{ \hat{A}, \hat{B}\} = \hat{A}\hat{B} + \hat{B}\hat{A}$ denotes an anti-commutator.
This unitary transformation of the Hamiltonian preserves the energy eigenvalues in the absence of truncation, but modifies the eigenvalues for arbitrary $\Lambda$.
$\hat H(\beta)$ thus represents an effective Hamiltonian acting in the truncated many-body space.
We see from Eq.~(\ref{eq:eff_H}) that this effective Hamiltonian connects states $\ket{n,\beta}$ and $\ket{n \pm 1, \beta}$ (when $\beta \ne 0$), thus the summation in Eq.~(\ref{eq:eff_wf}) now runs over both even and odd values of $n$.
The corresponding matrix elements of $\hat H(\beta)$ are given in Appendix~\ref{app:Lipkin_formalism}.
The angle $\beta$ and set of coefficients $\{ A_n^{(\beta)}\}$ are determined by applying a variational principle to the energy of the system $E(\Lambda) = \braket{ \Psi^{(\Lambda)} | \hat H (\beta)| \Psi^{(\Lambda)} } - \eta \braket{ \Psi^{(\Lambda)} |  \Psi^{(\Lambda)} } $, where $\eta$ is a Lagrange parameter ensuring normalization of the many-body state $\ket{\Psi}^{(\Lambda)}$.
\\ \\
\indent
The transformed and truncated effective Hamiltonian in Eq.~(\ref{eq:eff_H}) is not expected to be complete in the 
sense of exactly reproducing energy eigenvalues for the $\Lambda$ levels, 
but is expected to capture the 1-body
mean-field contributions and systematically include the effect of correlations beyond mean-field with increasing $\Lambda$.
To put this construction in the framework of low-energy EFT, 
beyond the leading-order (LO) Hamiltonian in Eq.~(\ref{eq:eff_H}) there are additional operators with coefficients (counterterms)
that are required to be determined by matching to the full theory.
In the  analysis that follows, 
these coefficients are set equal to zero, 
and we focus on optimizing 
results from the LO Hamiltonian.  
Interestingly, comparisons with the exact results show that the omitted contributions from the 
counterterms are exponentially suppressed with increasing $\Lambda$ for $\Lambda \ll N$, 
above which 
an unexpected plateau region is found,
see Sec.~\ref{sec:convergence}. 
In the context of mapping to the register of a quantum computer, 
this exponential behavior corresponds to double-exponential 
behavior with respect to the number of qubits.
\\ \\
\indent
To recover the approximation to the exact state in Eq.~(\ref{eq:exact_state}) from the effective state $\ket{\Psi}^{(\Lambda)}$ in Eq.~(\ref{eq:eff_wf}),
it is useful to re-express $\ket{\Psi}^{(\Lambda)}$ in the original unrotated basis ($\beta=0$) as
\begin{eqnarray}
\ket{\Psi}^{(\Lambda)} = \sum_{m =0}^N  A^{(\beta=0)}_m \ket{m, \beta=0} 
\; ,
\label{eq:eff_wf_full}
\end{eqnarray}
where 
\begin{eqnarray}
A^{(\beta=0)}_m  = \sum_{n =0}^{\Lambda -1} A^{(\beta)}_n \braket{m,\beta=0 | n, \beta} 
\; .
\end{eqnarray}
Note that the summation in Eq.~(\ref{eq:eff_wf_full}) is not limited to a certain cut-off, 
and thus includes all possible configurations (up to $N$p-$N$h configurations).
The overlap between many-body basis states built on the original and rotated single-particle states are simply given by the matrix representing the rotation of angle $\beta$ around axis $y$ as
\begin{eqnarray}
\braket{m,\beta=0 | n, \beta} 
                                               &=& \braket{J,M'=m-J | e^{i J_y \beta} | J,M=n-J } \nonumber \\
                                               &\equiv&  d^{J}_{M'=m-J,M=n-J} (\beta) \; , 
                                               \nonumber \\
\label{eq:innerD}
\end{eqnarray}
with $J=N/2$ and (see {\it e.g.}, Ref.~\cite{edmonds2016angular})
\begin{eqnarray} 
 d^J_{M',M} (\beta) &=& \left( \frac{(J+M')! (J-M')!}{(J+M)! (J-M)!} \right)^{1/2} \sum_s \binom{J+M}{J-M'-s} 
 \nonumber \\ 
&& \times \binom{J-M}{s} (-1)^{J-M'-s}  \left(\cos \beta/2 \right)^{2s + M' +M} \nonumber \\ 
&& \times  \left( \sin \beta/2 \right)^{2J-2s-M'-M} \; ,
\label{eq:rotation_mat}
\end{eqnarray}
where the sum over $s$ is restricted by $1/(n!) =0$ for $n<0$.
\\ \\
\indent
Given the nature of the wavefunction obtained in an effective model space,
$\ket{\Psi}^{(\Lambda)}$, 
a projection onto states of good parity,
$\ket{\Psi}^{(\Lambda)}_\pm$, should be subsequently performed,
which can be obtained from
\begin{eqnarray}
\ket{\Psi}^{(\Lambda)}_\pm &=& \frac{1}{2} \left( \mathds{1} \pm \Pi \right) \ket{\Psi}^{(\Lambda)} \; . 
\label{eq:projected_state}
\end{eqnarray}
It is straightforward to see from the matrix elements of $H(\beta)$ (see Appendix \ref{app:Lipkin_formalism}) that the states $\ket{\Psi}^{(\Lambda)}$ and $\hat{\Pi} \ket{\Psi}^{(\Lambda)} = \sum_{n \le \Lambda -1} A^{(\beta)}_n (-1)^n  \ket{n, -\beta}$ are degenerate. 
In the unrotated basis, the projection procedure amounts to canceling the coefficients $A^{(\beta=0)}_m$ for odd values of $m$ for the case of $ \ket{\Psi}^{(\Lambda)}_+$, and even values of $m$ for $\ket{\Psi}^{(\Lambda)}_-$, followed by a rescaling of the coefficients to normalize $\ket{\Psi}^{(\Lambda)}_\pm$ to 1.
Note that in this work, we perform the projection after the minimization procedure. 
In nuclear physics this is usually referred to as "projection after variation"~\cite{ring2004nuclear}. 
Techniques for projecting quantum states in quantum simulations of nuclear many-body systems have been developed in Refs.~\cite{PhysRevLett.125.230502,PhysRevC.105.024324,Lacroix:2022vmg,PhysRevC.107.034310}.
\\ \\
\indent
To quantify the rate of convergence of the approximate 
wavefunction with respect to the size of the model space, we will compute the Bures distance~\cite{Bures}
$D_B(\Lambda)$ as a measure of the distance between the exact state $\ket{ \Psi_{ex} }$ 
given in Eq.~(\ref{eq:exact_state}) 
and the effective projected state $\ket{\Psi}^{(\Lambda)}_+$ given in Eq.~(\ref{eq:projected_state}):
\begin{equation}
D_B(\Lambda) = \sqrt{2 (1-|\braket{\Psi_{ex}|\Psi}^{(\Lambda)}_+|)} 
\; ,
\label{eq:fidelityDB}
\end{equation}
which we find to be a more useful measure of fidelity for our purposes, 
as opposed to the usual overlap of states.
\\ \\
\indent
\begin{figure}[t]
\centering{\includegraphics[width=\columnwidth] {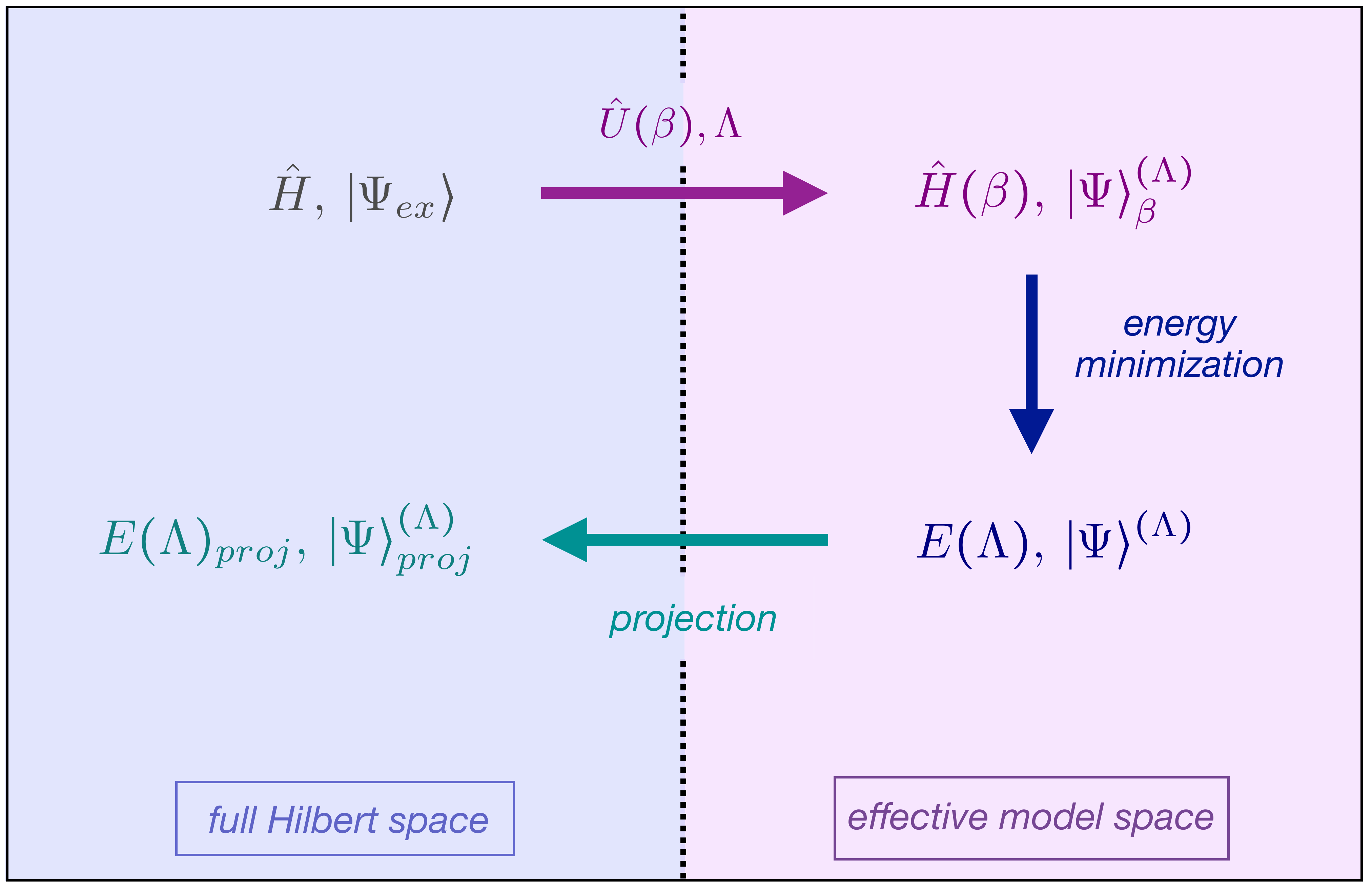}} 
\caption{Flow chart of the calculation.
Starting from the full-space Hamiltonian $\hat{H}$ given in Eq.~(\ref{eq:H_unrot}), with exact eigenvectors $\ket{\Psi_{ex}}$, an effective model space is defined, characterized by a given cut-off $\Lambda$, as well as a unitary transformation $\hat{U}(\beta)$ corresponding to a rotation of the quasi-spin operators by angle $\beta$. 
The transformed Hamiltonian $\hat{H}(\beta) = \hat{U}^\dagger(\beta) \hat{H} \hat{U}(\beta)$, defined in 
Eq.~(\ref{eq:eff_H}), possesses the same eigenvalues as $\hat{H}$. 
However, the goal is to determine the optimal transformation $\hat{U}(\beta)$ so that the truncated transformed Hamiltonian, diagonalized in the effective model space $\Lambda$, reproduces the ground-state energy of the full Hamiltonian with only a small error.
Such an optimal value of the angle $\beta$ is determined via a variational principle applied to the energy of the system. 
}
\label{fig:flow_chart}
\end{figure}
Figure~\ref{fig:flow_chart} summarizes the successive steps of the calculation.
The minimization of the ground-state energy  in an effective model space
can be performed using classical computers (this section), 
or using quantum or hybrid classical-quantum computations (see sections \ref{sec:QuSimFormal} and \ref{sec:QuSimProduction}).
As mentioned above, in the case of the LMG model, the Hamiltonian $\hat{H}$ preserves the parity symmetry associated with the operator $\hat{\Pi}$ in Eq.~(\ref{eq:parity_op}). This symmetry is broken by the transformation $\hat{U}(\beta)$ and thus needs to be restored. We perform this restoration subsequently to the calculation according to Eq.~(\ref{eq:projected_state}). Note that the projection introduces components outside of the truncated model space $\Lambda$. 
We emphasize, however, that the need for projection is specific to the LMG model studied here. 
The forces governing realistic nuclear systems do not possess the symmetry associated with the operator $\hat{\Pi}$, and the effective state in the MCSCF approach typically preserves the symmetries of the Hamiltonian~\cite{Robin:2015lba}.

\subsection{Example System: $N=30$}
\label{sec:EffectiveSpacesLGM}
\begin{figure}[!ht]
\centering{\includegraphics[width=\columnwidth] {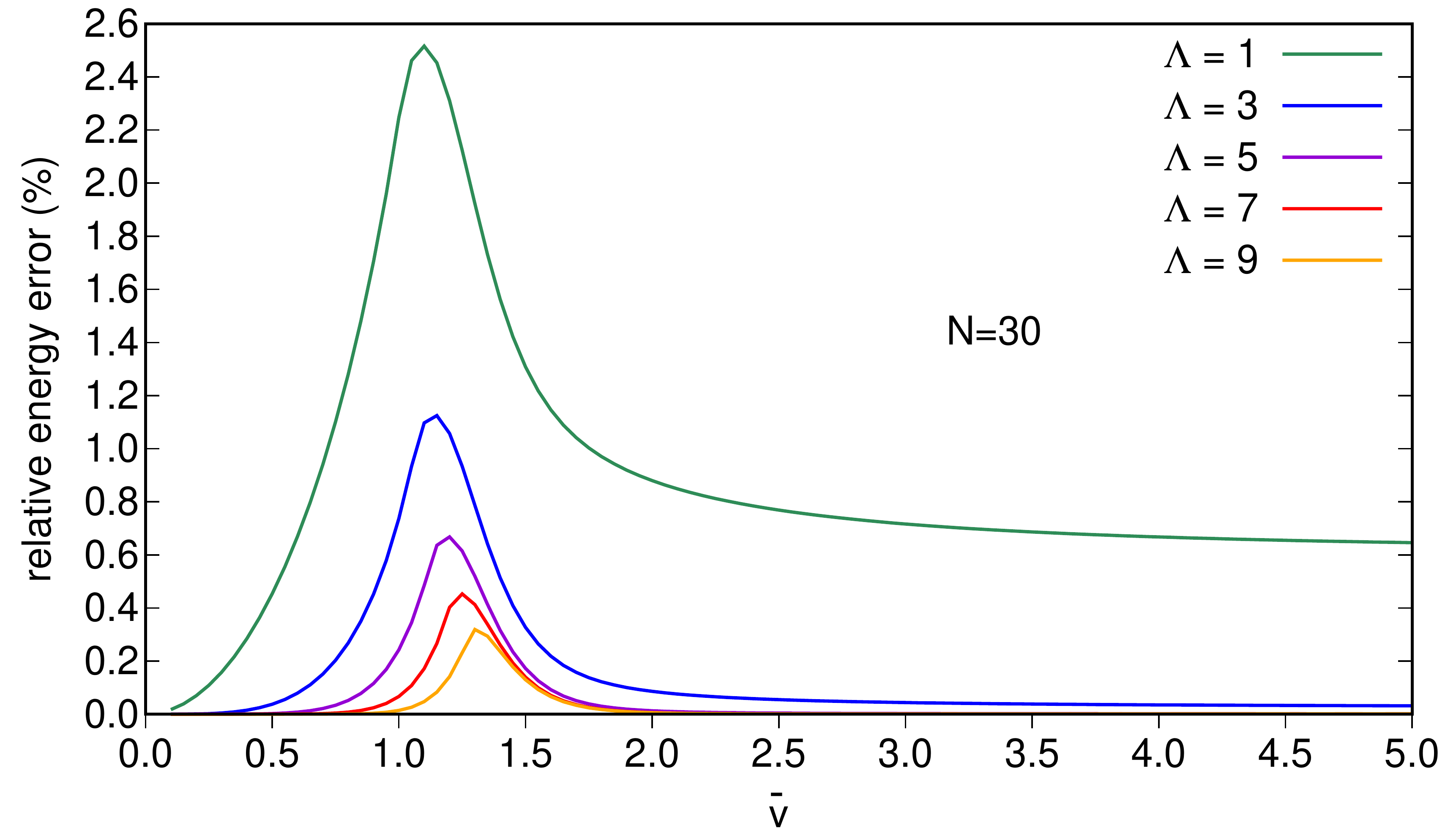}} 
\caption{Relative error $\left| \frac{E_{ex} - E(\Lambda)}{E_{ex}} \right|$ (in $\%$) in the ground state energy as a function of $\bar{v}$, obtained for different values of the cut-off $\Lambda$, for a system of $N=30$ particles. 
}
\label{f:energy_N30}
\end{figure}

\begin{figure}[t]
\centering{\includegraphics[width=\columnwidth] {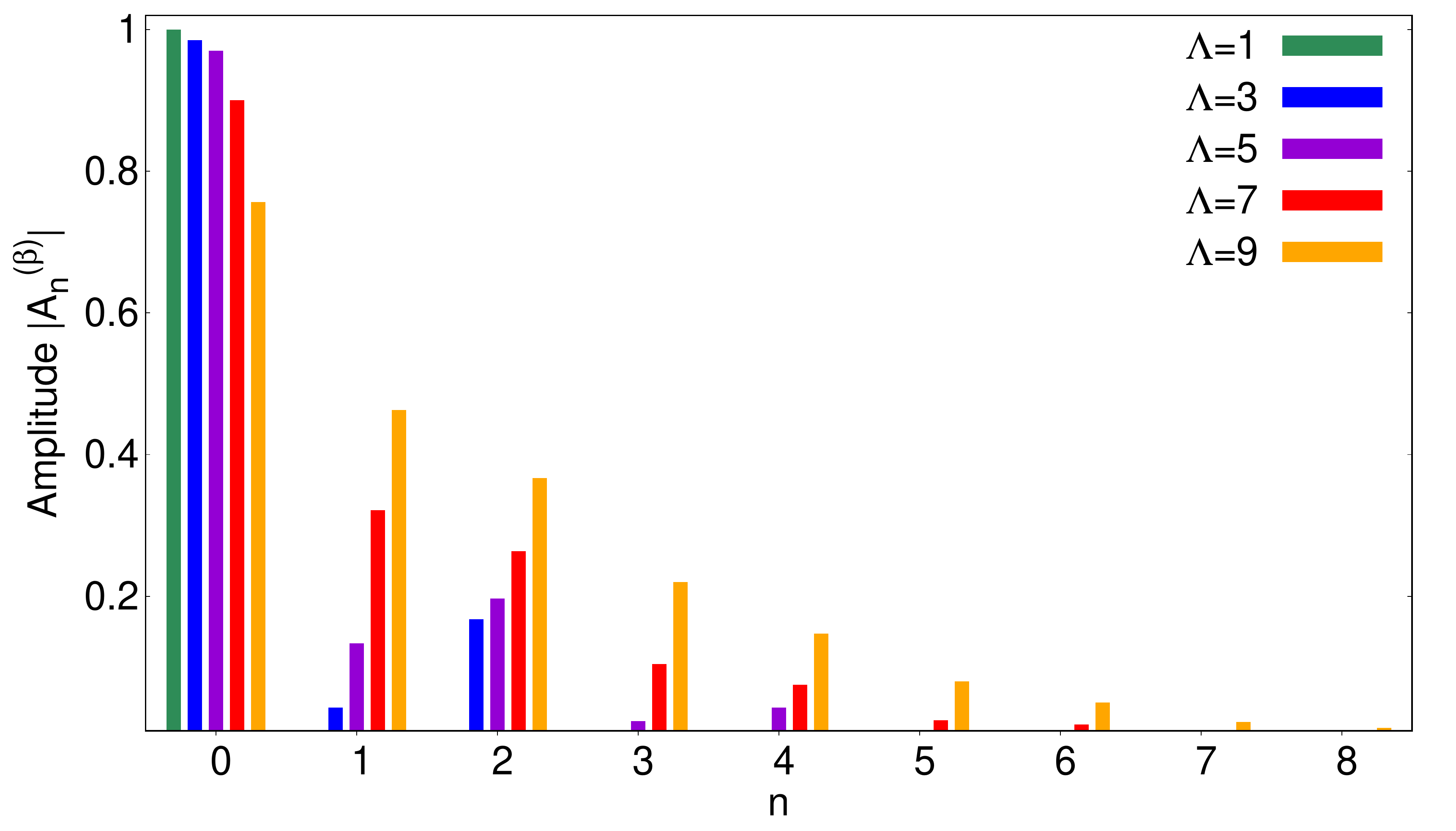}} 
\caption{Absolute value of the amplitudes $A_n^{(\beta)}$ in the effective wavefunction, obtained for $N=30$, $\bar{v}=2.0$ and different cut-off values. }
\label{f:wf_new_v2p0}
\end{figure}

\begin{figure}[t]
\centering{\includegraphics[width=\columnwidth] {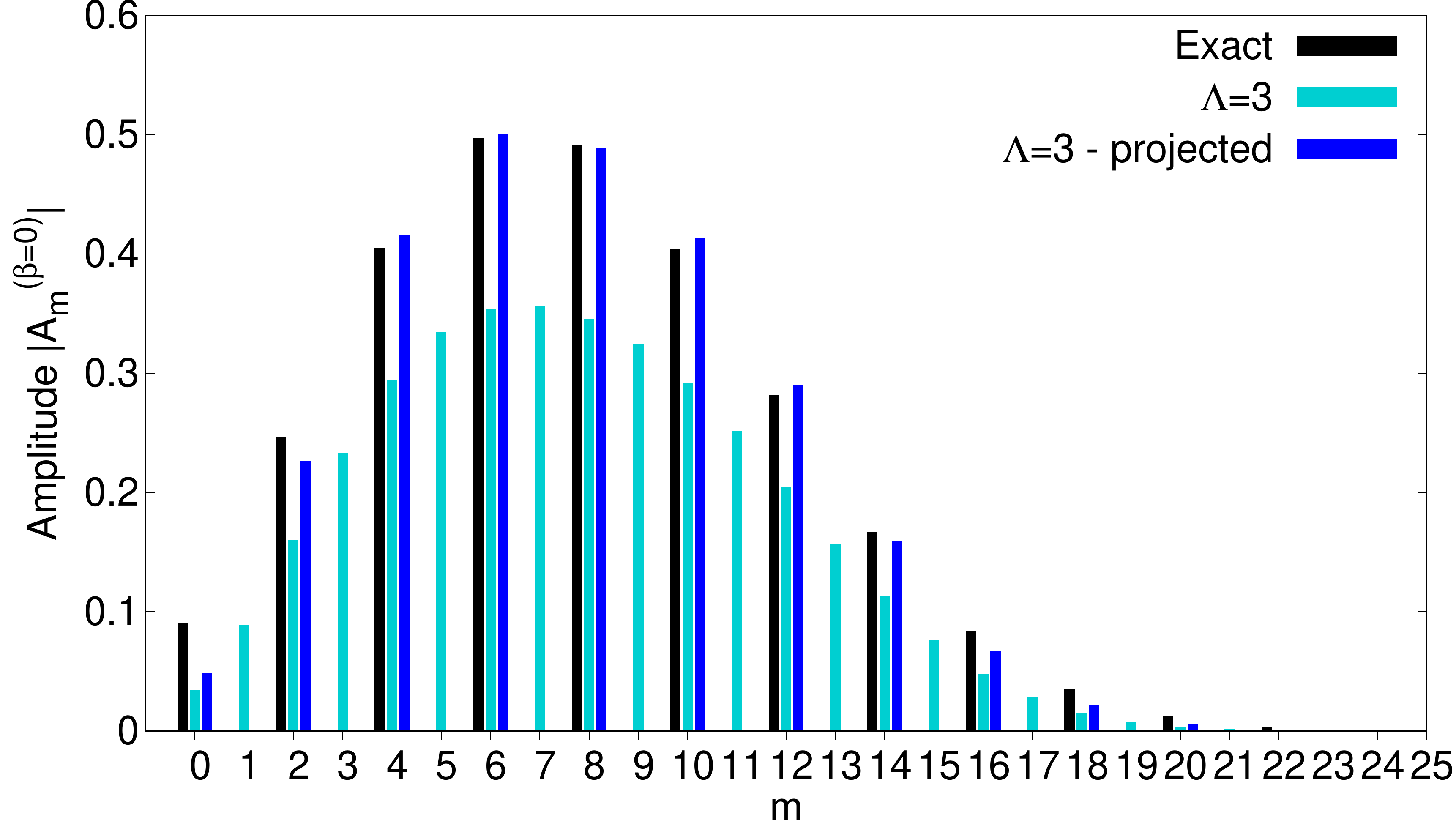}} 
\caption{Absolute value of the amplitudes $A_m^{(\beta=0)}$ in the effective wavefunction (see Eq.~(\ref{eq:eff_wf_full})), obtained for $N=30$, $\bar{v}=2.0$ and $\Lambda =3$. The light are dark blue histograms show the results before and after projection onto a good-parity state, respectively. For comparison the exact wavefunction is shown in black.}
\label{f:wf_proj_L3}
\end{figure}

\begin{figure}[h]
\centering{\includegraphics[width=\columnwidth] {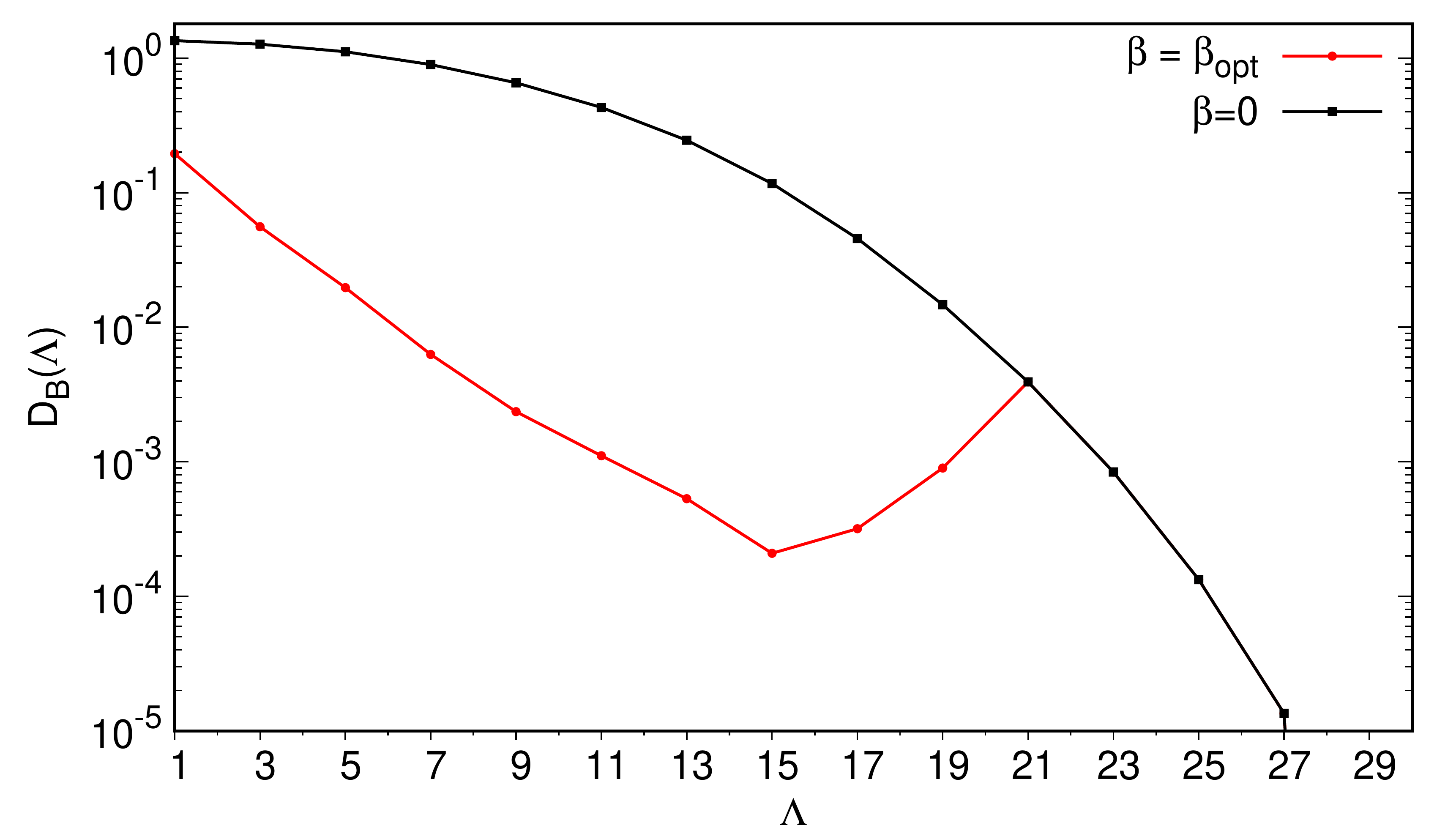}}
\caption{Bures distance $D_B(\Lambda)$ (Eq.~(\ref{eq:fidelityDB})) as a function of the cut-off 
$\Lambda$ for $N=30$ and 
$\bar{v}=2.0$. The red points show the results obtained when optimizing the angle $\beta$ via the variational principle, while the black points are the results obtained with $\beta=0$.}
\label{f:Bures_N30}
\end{figure}

\begin{figure}[h]
\centering{\includegraphics[width=\columnwidth] {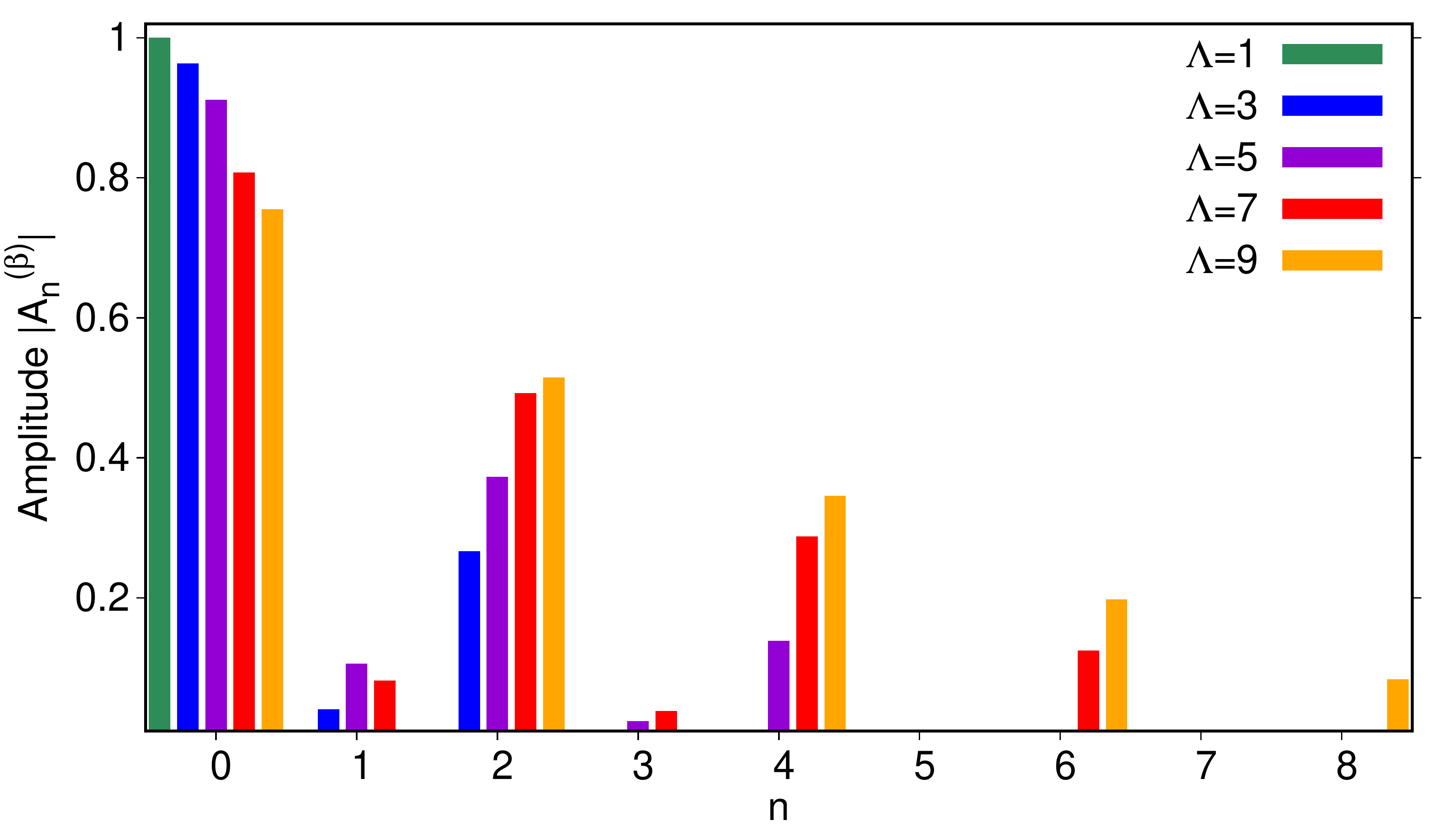}}
\caption{Absolute value of the amplitudes $A_n^{(\beta)}$ in the effective wavefunction, obtained for $N=30$, $\bar{v}=1.2$ and different cut-off values.}
\label{f:wf_new_v1p2}
\end{figure}

\begin{figure}[t]
\centering{\includegraphics[width=\columnwidth] {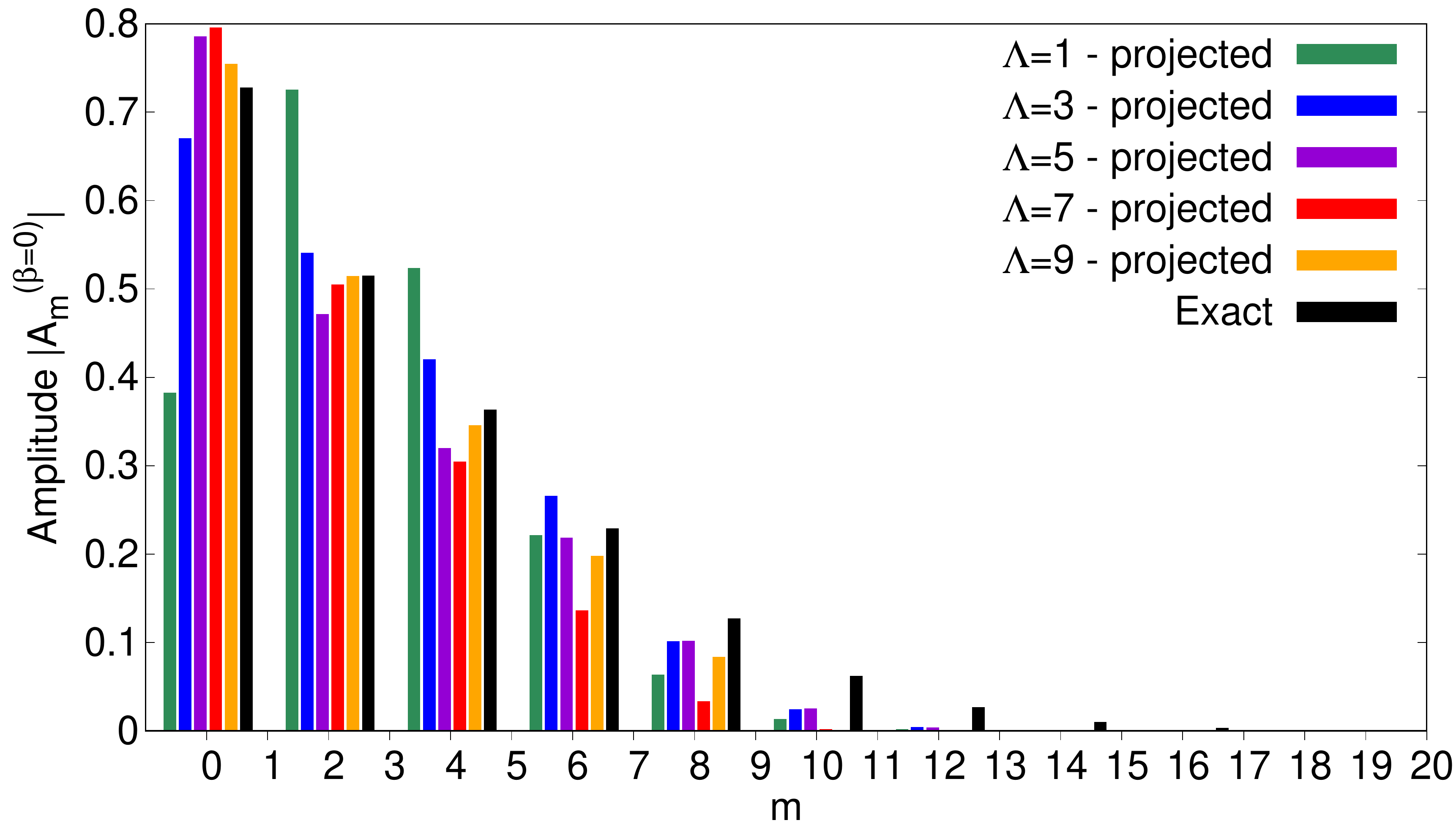}}
\caption{Absolute value of the amplitudes $A_m^{(\beta=0)}$ in the effective wavefunction (see Eq.~(\ref{eq:eff_wf_full})), obtained for $N=30$, $\bar{v}=1.2$ and different cut-off values.}
\label{f:wf_v1p2}
\end{figure}

\noindent
In order to gain physical insight into the model, and to illustrate the capability of the method, we first investigate both the
energy and wavefunction of the ground state, obtained from classical calculations, using different values of the cut-off $\Lambda$ for a system of $N=30$ particles.
\\ \\
\indent
The truncation $\Lambda =1$ corresponds to the Hartree-Fock (HF) limit, {\it i.e.}, 
when the many-body state is restricted to the configuration $\ket{n=0,\beta}$,
where all particles occupy the lower level of the rotated basis. 
This approximation has been extensively studied (see, {\it e.g.} Ref.~\cite{AGASSI1966321}), 
and it is well known that in this case $\beta$ becomes non-zero above a critical value of the ratio of the single-particle and interaction term in the Hamiltonian 
$\bar{v} =(N-1)V/\varepsilon$. 
In particular, the transition to a parity-broken ("deformed") phase occurs at $\bar{v}=1.0$. 
Figure~\ref{f:energy_N30} shows
the relative difference between the exact and HF energy (green curve), as a function of $\bar{v}$. As is well known, the HF approximation works best for large values of $\bar{v}$, away from the phase transition, where the error drops below $1\%$. Close to the phase transition, correlations beyond HF are required for a correct description of the system. This is illustrated by the other curves of Fig.~\ref{f:energy_N30}, which have been obtained with cut-off values $\Lambda=3,5,7,9$. 
\\ \\
\indent
In the following, we examine the behaviour of the ground-state wavefunction away from
and around the phase transition, obtained with two different values of the interaction ratio $\bar{v}=2.0$ and $\bar{v}=1.2$, respectively.
Figure~\ref{f:wf_new_v2p0} shows
the composition of the many-body wavefunction in terms of the components $\ket{n,\beta}$ built on the rotated basis, for the case $\bar{v}=2.0$. The corresponding values of $\beta$ obtained variationally for each cut-off $\Lambda$ are shown in Table~\ref{t:beta_v2p0}. Note that when 
$\Lambda$ takes an even value, the configuration with maximal excitation order $n=\Lambda-1$ is always found to have zero amplitude in the wavefunction (see App.~\ref{app:GSforms} for discussion). 
This is why the results in Table~\ref{t:beta_v2p0} do not change 
when $\Lambda\rightarrow\Lambda+1$ for odd $\Lambda$. 
\begin{table}[t]
\centering
\begin{tabular}{ll||ll} 

$\Lambda$ & $\beta$ & $\Lambda$ & $\beta$ \\
\hline 
\hline 
1,2 & 1.047 & 17,18 & 0.289\\
3,4 & 1.016 & 19,20 & 0.150\\
5,6 & 0.977 & 21,22 & 0.0\\
7,8 & 0.906 & 23,24 & 0.0\\
9,10 & 0.791 & 25,26 & 0.0\\
11,12 & 0.664 & 27,28 & 0.0\\
13,14 & 0.538 & 29,30 $\;\;\;$& 0.0 \\
15,16 $\;\;\;$& 0.415 & 31  & 0.0\\
\hline 
\end{tabular}
\caption{Values of the angle $\beta$ obtained for different values of the cut-off $\Lambda$, for a system of $N=30$ particles, and $\bar{v}=2.0$. 
The solution obtained for even values of $\Lambda$ is the same as for the odd cut-off $\Lambda'= \Lambda-1$ (see discussion in appendix~\ref{app:GSforms}).}
\label{t:beta_v2p0}
\end{table}
We see that as $\Lambda$
(and thus the model space) increases, the angle $\beta$ decreases, and the wavefunction becomes fragmented into several components. For low values of $\Lambda$, the wavefunction remains well localized around the $n=0$ configuration in the effective model space, and the contribution from higher $n$ states decreases rapidly. This rapid fall-off 
is expected to increase the efficiency of the quantum simulations 
(performed in Sec.~\ref{sec:QuSimProduction}) .
As $\Lambda \rightarrow N$ we observe a transition towards a parity-unbroken phase ($\beta=0$) at $\Lambda=21$, and we find that the wavefunction evolves towards the exact (full-Hilbert-space) solution.
\\ \\
\indent
The transformation in Eq.~(\ref{eq:rotation_mat}) is implemented
in order to compare the exact and effective (truncated) wavefunctions. The results are shown in Fig.~\ref{f:wf_proj_L3} for the case $\Lambda=3$. 
The black histograms show the exact wavefunction, which is distributed over $\sim 10$ states, from 0p-0h to 20p-20h, and is peaked around 6p-6h and 8p-8h. 
\begin{figure}[t]
\centering{\includegraphics[width=\columnwidth] {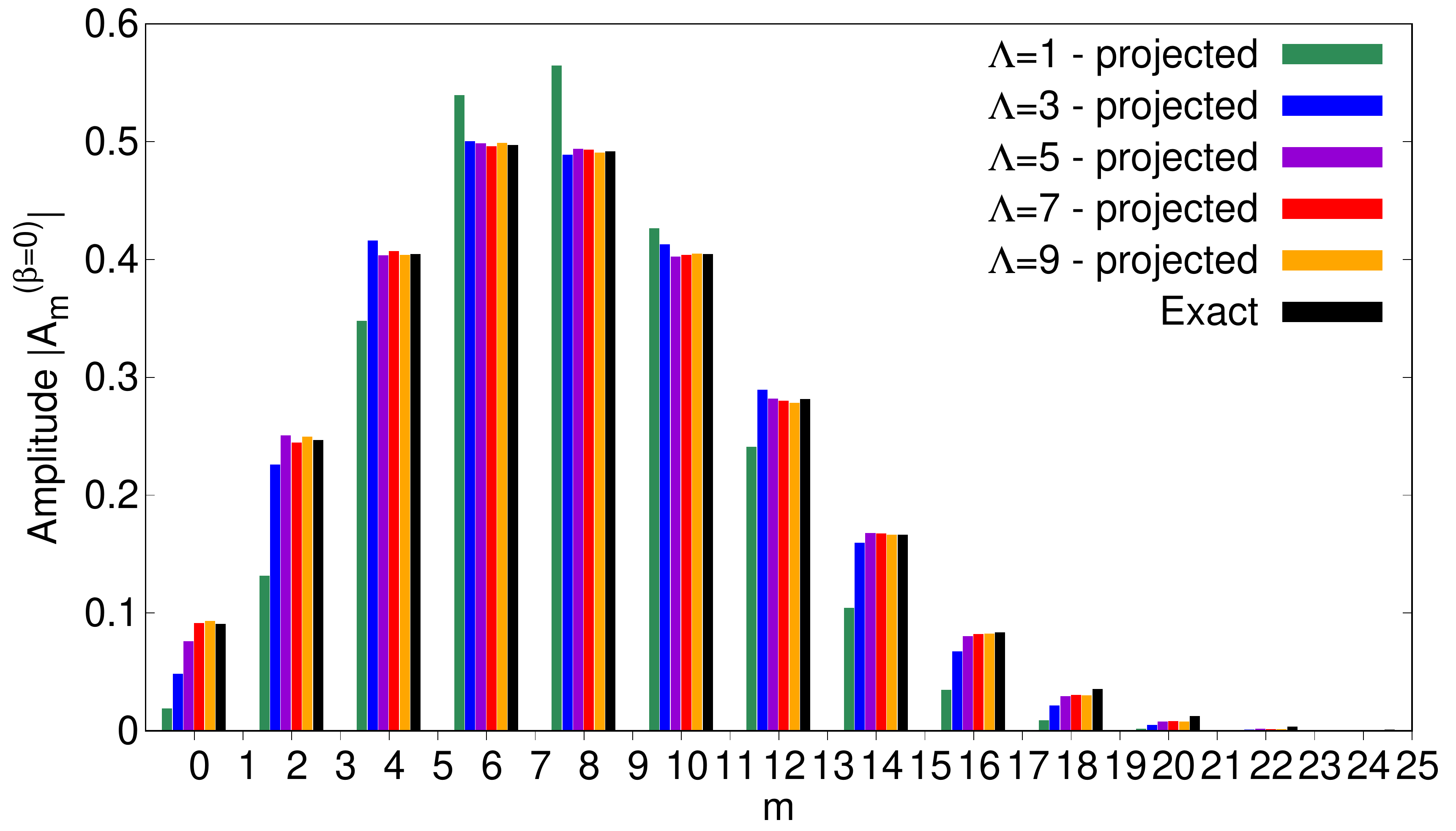}} 
\caption{Absolute value of the amplitudes $A_m^{(\beta=0)}$ obtained after projection of the effective wavefunction, for $N=30$, $\bar{v}=2.0$ and different cut-off values.}
\label{f:wf_cv}
\end{figure}
The light blue histograms correspond to the effective wavefunction expressed in the original $\beta=0$ basis, as in Eq~(\ref{eq:eff_wf_full}). 
The trend of the exact distribution is rather well reproduced, however, we observe non-zero contributions from states characterized by odd values of $m$. These odd-$m$ components are due to the fact that the transformation Eq.~(\ref{eq:transfo}) breaks the parity symmetry associated with the operator $\hat \Pi$ in Eq.~(\ref{eq:parity_op}) since
\begin{equation}
    \hat\Pi \ket{n, \beta} = (-1)^n \ket{n, -\beta} \; ,
\end{equation}
and thus the state $\ket{\Psi}^{(\Lambda)}$ in Eq.~(\ref{eq:eff_wf}) is no longer eigenstate of $\hat\Pi$. 
One should therefore project $\ket{\Psi}^{(\Lambda)}$ onto a parity-preserving state $\ket{\Psi}^{(\Lambda)}_\pm$, 
as discussed in the previous section around Eq.~(\ref{eq:projected_state}).
The dark blue histograms in Fig.~\ref{f:wf_proj_L3} correspond to the effective wavefunction expressed in the original $\beta=0$ basis, after projection onto a good parity state via Eq.~(\ref{eq:projected_state}). 
Even though there remain discrepancies in the tails of the distribution,
the $\Lambda=3$ states in the effective space well recover the exact wavefunction.
These discrepancies are rapidly corrected as $\Lambda$ increases.
This can be seen from 
Fig.~\ref{f:wf_cv} which displays the convergence of the effective wavefunction in the original $\beta=0$ basis. 
The Bures distance is displayed in Fig.~\ref{f:Bures_N30} as a function of the cut-off $\Lambda$ (red points). For comparison we also show the Bures distance between the exact state and the truncated state with fixed value of $\beta=0$ 
(black points). Evidently, applying a truncation to the model space without rotating the single-particle basis only yields a polynomial convergence of the wavefunction. On the contrary, when $\beta$ is determined variationally, the convergence is accelerated and matches an exponential behaviour, up to $\Lambda=15$. For $\Lambda > 15$ the Bures distance increases from its local minimum value
in order to recover the $\beta=0$ (symmetry-unbroken) solution at $\Lambda=21$ as shown in Table~\ref{t:beta_v2p0}.
\\
\\
\indent
It is also interesting to study the behaviour of the system around the phase transition. For example we consider here the case $\bar{v}=1.2$.
Figs.~\ref{f:wf_new_v1p2} and \ref{f:wf_v1p2} show the amplitude of the configurations built on the rotated and original ($\beta=0$) basis, respectively.
The corresponding optimal values of the angle $\beta$ are shown in Table~\ref{t:beta_v1p2}.
\begin{table}[h]
\centering
\begin{tabular}{cc} 
\hline  
$\Lambda$ & $\beta$ \\
\hline 
\hline 
1 & 0.586\\
3 & 0.496\\
5 & 0.371\\
7 & 0.113\\
9 & 0.000 \\
\hline 
\end{tabular}
\caption{Values of the angle $\beta$ obtained for different values of the cut-off $\Lambda$, for a system of $N=30$ particles, and $\bar{v}=1.2$. }
\label{t:beta_v1p2}
\end{table}
Figure~\ref{f:wf_v1p2} shows that the HF ($\Lambda=1$) approximation leads to a poor description of the wavefunction in the region near the phase transition.
This was already pointed out in the original paper~\cite{AGASSI1966321}. In particular, the weight of the $m=0$ basis state (0p-0h) is largely underestimated, while the contributions of the $m=2$ and $m=4$ states (2p-2h and 4p4h) are significantly overestimated, compared to the exact solution. 
Including correlations by increasing 
$\Lambda$ rapidly corrects for this behaviour. However, because the exact wavefunction is already localized around the 0p-0h configuration, and is contained within a few number of basis states, the effective wavefunction quickly converges towards the result characterized by $\beta=0$ (see Table \ref{t:beta_v1p2}). 
Thus, it appears that, near the phase transition, the full-space and truncated-space wavefunctions have comparable support, {\it i.e.}
the number of basis states spanning these two wavefunctions is of the same order, and in that sense, there is no separation of scales to utilize.

\subsection{Convergence of Effective-Model-Space Calculations with $\Lambda$ and $N$} 
\label{sec:convergence}
\noindent
\begin{figure}[h]
\centering{\includegraphics[width=\columnwidth] {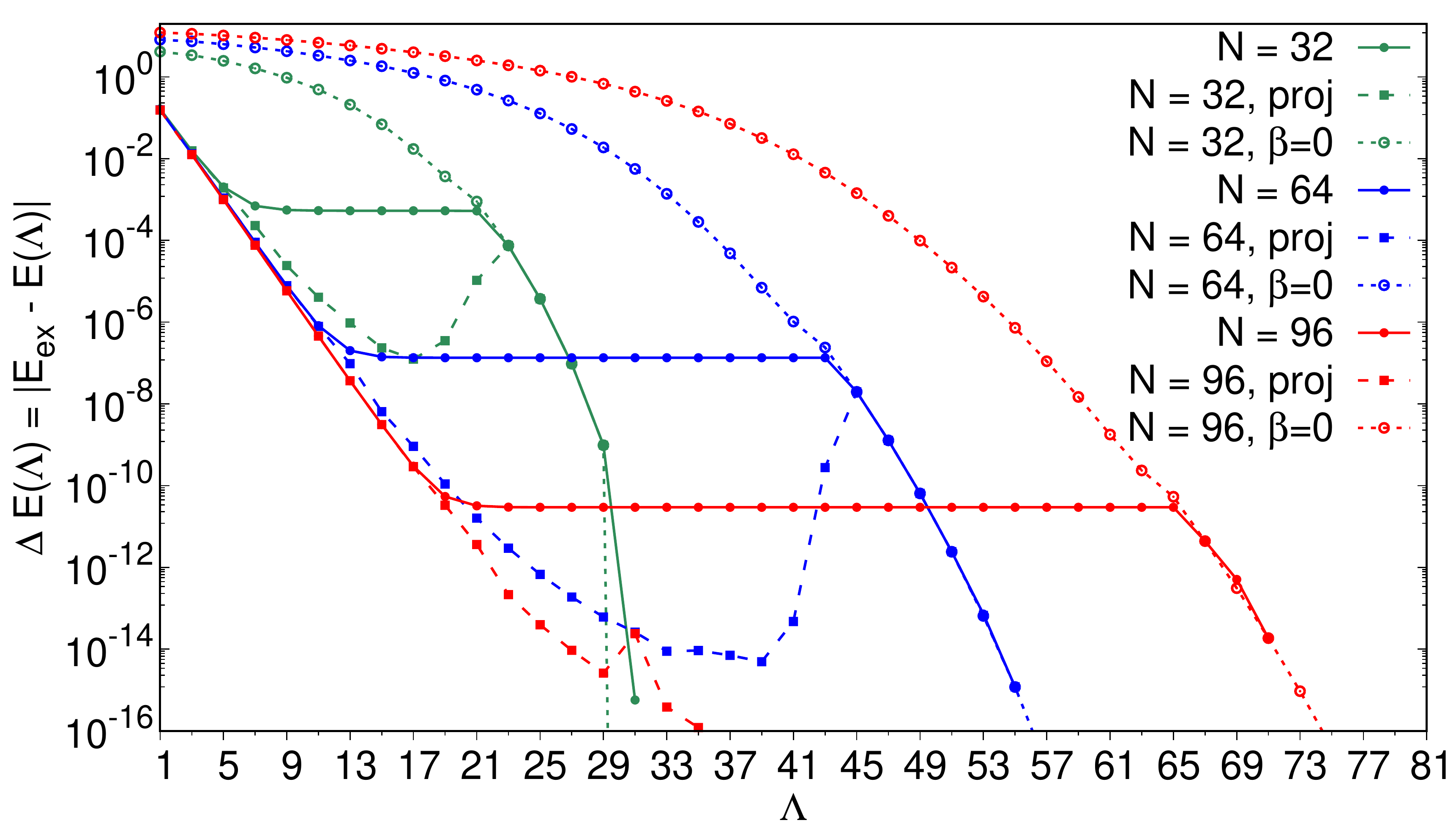}} 
\caption{
The difference between the exact and effective model space ground-state energies $\Delta E(\Lambda)$
as a function of $\Lambda$ for $\bar{v}=2.0$ and for particle numbers $N=32, 64$ and $96$. 
The plain circles and squares are the results obtained without and with projection onto a good-parity state, respectively. 
The empty circles show the results obtained when fixing $\beta=0$ and increasing the model space.
Numerical values for the shown results can be found in Tables~\ref{tab:N32Ediff}, \ref{tab:N64Ediff} and \ref{tab:N96Ediff}
}
\label{fig:delELam}
\end{figure}
Figure~\ref{fig:delELam} shows the difference between the energy of the exact ground-state and the state in the effective model space (plain circles),
\begin{eqnarray}
\Delta E (\Lambda) & = & |E_{ex} - E(\Lambda)|
\ ,
\label{eq:EEdiff}
\end{eqnarray}
obtained for $\bar{v}=2.0$, and for particle numbers $N=32$ (green), $N=64$ (blue) and $N=96$ (red). 
For comparison, $\Delta E (\Lambda)_{\beta=0}$ obtained for $\beta=0$ (empty circles) is also shown.
The latter exhibits a polynomial convergence of the energy with respect to the size of the model space.
This behaviour is consistent with the convergence of the Bures distance of the wavefunction in Fig.~\ref{f:Bures_N30}. 
Similarly, optimizing $\beta$ also accelerates the convergence of the energy is a way that is
consistent with exponential convergence, up to a certain value of $\Lambda$, after which $\Delta E (\Lambda)$ reaches a plateau, before merging with the $\beta=0$ solution.
This behavior can 
again
be easily understood, since for large $\Lambda$ the solution has to converge to the exact one 
with $\beta=0$. 
Because of the faster convergence of $\Delta E (\Lambda) $ for small cut-offs, the plateau region is thus necessary to link the symmetry-broken $\beta\ne 0$ and symmetry-unbroken $\beta=0$ phases. 
The full squares in Fig.~\ref{fig:delELam} display the energy difference $\Delta E (\Lambda)_{proj} = |E_{ex} - E(\Lambda)_{proj}|$ obtained after projecting onto a state of good parity. The projection sustains the exponential convergence to larger values of $\Lambda$, 
and allows for an improvement in precision by several orders of magnitude. 
Around $\Lambda \sim N/2$, the solution starts converging toward the $\beta=0$ solution.~\footnote{For sufficiently large $N$ and $\Lambda$, maintaining precision in the results of matrix diagonalizations 
for the energy difference shown in Fig.~\ref{fig:delELam}
becomes increasingly challenging.}

\section{Quantum Simulations: General Techniques for HL-VQE}
\label{sec:QuSimFormal}
\noindent
As discussed in earlier sections, the Hamiltonian in truncated 
effective model spaces  generally contains 
parameters that are absent in the full Hilbert space, ${\bm \beta}$, 
which relate full-space and effective-space ladder operators.
There are stationary values of ${\bm \beta}$ that, when combined with 
an optimal wavefunction, minimize the total energy of the system.  
We therefore define the expectation value of the energy in the 
effective model space as a cost-function to be minimized with respect 
to both the Hamiltonian parameters, ${\bm \beta}$, 
and the parameters 
defining the wavefunction,
which we will generally denote here as ${\bm\theta}$,
\begin{eqnarray}
E({\bm \beta}, {\bm\theta})  
& = & 
\langle\psi ({\bm\theta}) | \hat H({\bm \beta}) | \psi ({\bm\theta}) \rangle
\ \ \ .
\label{eq:expE}
\end{eqnarray}
Assuming uniquely isolated ground states, this minimization
simultaneously learns the Hamiltonian in the effective model space 
and identifies the associated variational ground-state wavefunction.
We define this to be 
Hamiltonian Learning Variational Quantum Eigensolver (HL-VQE).
A straightforward way to converge to the optimal parameter set 
(for non-pathological systems)
is to use gradient-descent, employing the linear variation of the $E({\bm \beta}, {\bm\theta})$.
Combining the sets of parameters into ${\bf w}=({\bm \beta},{\bm\theta})$, the iteration between adjacent parameter estimates is
\begin{eqnarray}
{\bf w}^{[k+1]} & = & {\bf w}^{[k]} - \eta\ {\bm\nabla}_{\hspace*{-3pt}{\bf w}} \ E( {\bf w}^{[k]})  
\ \ \ ,
\label{eq:witer}
\end{eqnarray}
where $\eta$ is the learning rate.
It is well known how to extend this beyond linear order to quadratic order, 
but we have not implemented this.~\footnote{For the system sizes considered in this work, 
only the simplest classical implementation of gradient-descent methods is used.
For simulations of larger spaces, more sophisticated algorithms will be necessary, such as {\tt Adam}~\cite{https://doi.org/10.48550/arxiv.1412.6980} or {\tt Adagrad}~\cite{JMLR:v12:duchi11a}.
}
\\ \\
\indent
Generally, the Hamiltonian in the effective model spaces can be expanded in terms of
coefficients, that are function of ${\bm \beta}$, times products of Pauli operators, as
\begin{eqnarray}
\hat H({\bm \beta}) & = & 
\sum_{i_1,.., i_M = 1}^4\ h_{i_i,.., i_M}({\bm \beta}) 
\ \overline{\sigma}_{i_1}\otimes\ ... \otimes \overline{\sigma}_{i_M}
\ \ \ ,
\label{eq:HPauli}
\end{eqnarray}
where
$\overline{\sigma}=\{ \hat X , \hat Y , \hat Z , \hat I \}$.
The cost function in Eq.~(\ref{eq:expE}) becomes,
\begin{eqnarray}
E({\bm \beta}, {\bm\theta})  
& = & 
\sum_{i_1,.., i_M=1}^4\
h_{i_i,.., i_M}({\bm \beta}) \ 
\nonumber\\
&& \hspace{0.5cm} \times 
\langle\Psi ({\bm\theta}) | 
\overline{\sigma}_{i_1}\otimes\ ... \otimes \overline{\sigma}_{i_M} 
| \Psi ({\bm\theta}) \rangle
\ \ \ .
\label{eq:expE2}
\end{eqnarray}
The HL-VQE requires the computation of the cost function $E({\bm \beta}, {\bm\theta})$, as well as its derivatives with respect to ${\bm \beta}$ and ${\bm\theta}$ for the gradient descent. This involves computing the coefficients $h_{i_i,.., i_M}({\bm \beta})$ and their derivatives with respect to ${\bm \beta}$, as well as the expectation values of the Pauli operators $\langle\Psi ({\bm\theta}) | \overline{\sigma}_{i_1}\otimes\ ... \otimes \overline{\sigma}_{i_M} | \Psi ({\bm\theta}) \rangle$ and their derivatives with respect to ${\bm\theta}$.
Generally, and as we will assume for the analysis of the LMG model, the coefficients $h_{i_i,.., i_M}({\bm \beta})$ and their derivatives can be evaluated classically. These classical computations become increasingly demanding with increasing dimensionality of the effective space(s).
With the factorization in Eq.~(\ref{eq:expE2}),
it is therefore the expectation values of the strings of 
Pauli operators evaluated in the quantum-many-body wavefunction, as well as their derivatives with respect to ${\bm \theta}$, which will require quantum computation for a sufficiently large effective model space. 
The corresponding procedure for the LMG model\footnote{In future generalizations of the HL-VQE algorithm to more realistic systems than the present LMG model, the use of recently developed techniques, such as classical shadows~\cite{Huang_2020} and importance-sampled classical shadows~\cite{Huang_2022} could potentially provide robust estimates of the expectation values with a reduced number of measurements.}
is detailed in Sec.~\ref{sec:QuSimProduction}.
\\ \\
\indent
\begin{figure}[!ht]
\centering{
\includegraphics[width=\columnwidth]{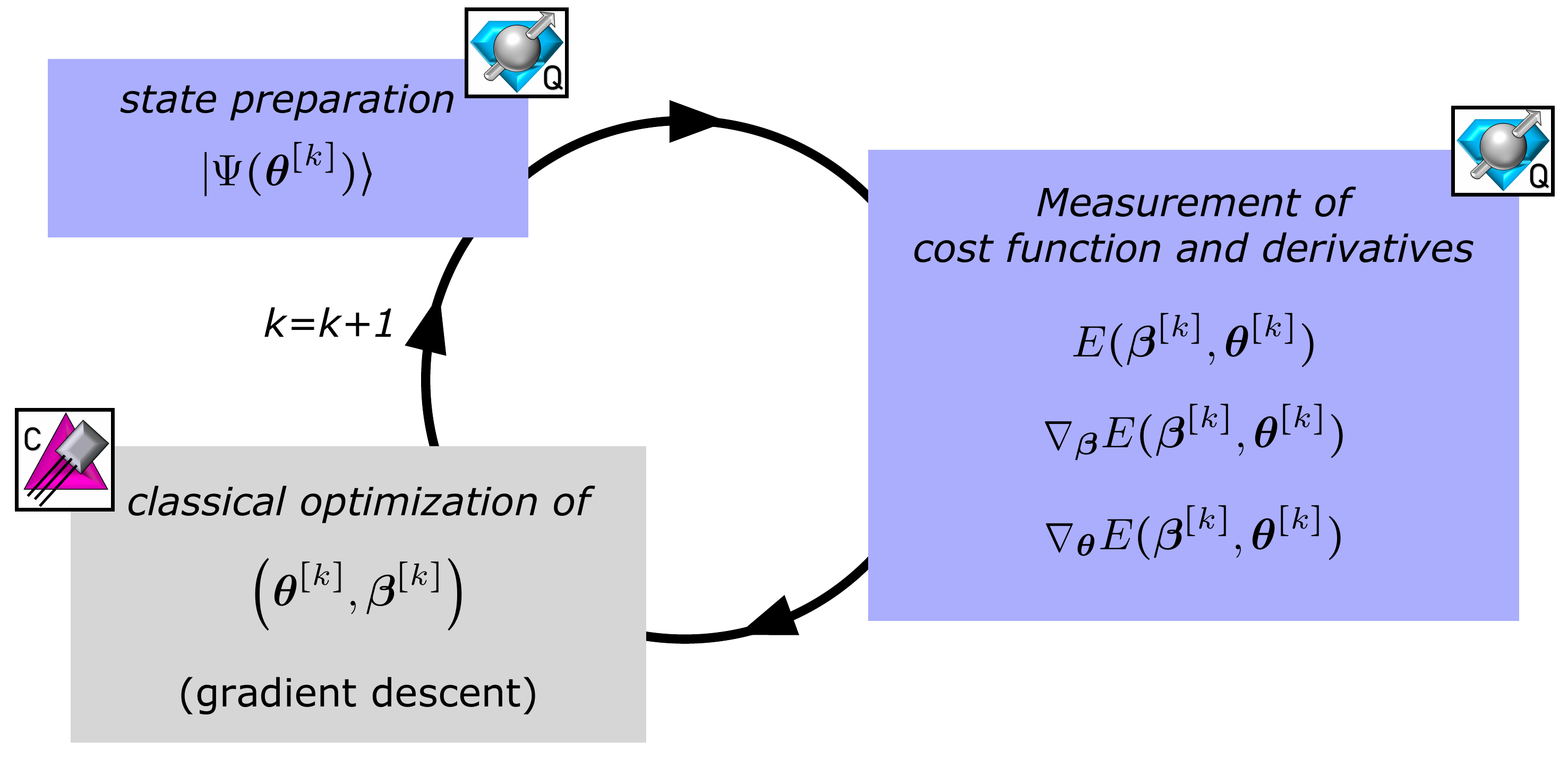}} 
\caption{A schematic of the workflow implementing the HL-VQE algorithm.
The icons denoting classical or quantum computation are defined in Ref.~\cite{Klco:2019xro}.
}
\label{fig:HL_VQE}
\end{figure}
Figure~\ref{fig:HL_VQE} summarizes the general workflow for HL-VQE. 
\begin{itemize}
\item An initial set of parameters $({\bm\beta}^{[0]}, {\bm\theta}^{[0]})$ is chosen. 
\item The algorithm then starts with preparation of the state $\ket{\Psi ({\bm\theta}^{[0]}) }$ which can usually be implemented in many ways with quantum circuits.  
In Sec.~\ref{sec:QuSimProduction}, we will choose quantum circuit structures such that each angle parameterizing the wavefunction appears only in one gate of the 
circuit. Thus, partial derivatives with respect to that angle are localized to that one gate and can be evaluated using a finite-difference relation. See App.~\ref{app:HLVQE} for more details.
\item As is the case for VQE, matrix elements of strings of Pauli operators 
are appropriately transformed into the computational basis, 
where the polarization of each qubit in the register along the z-basis 
is projected for each member of the ensemble.
From these measurements, the expectation values of each operator can be evaluated.
\item Classical computers are used to evaluate $\hat H({\bm\beta}^{[0]})$ and ${\bm\nabla}_{\bm\beta}\hat H({\bm\beta}^{[0]})$, 
so that the cost function $E({\bm \beta}^{[0]}, {\bm\theta}^{[0]})$ and its derivatives, ${\bm\nabla}_{\hspace*{-3pt}{\bf \theta}} E({\bm \beta}^{[0]}, {\bm\theta}^{[0]})$ and ${\bm\nabla}_{\hspace*{-3pt}{\bf \beta}} E({\bm \beta}^{[0]}, {\bm\theta}^{[0]})$, can be reconstructed classically, 
from the outputs of the quantum device.
\item Subsequently, a new set of parameters $({\bm\beta}^{[1]}, {\bm\theta}^{[1]})$ is determined via classical optimization (here gradient descent, as described above). 
\item This procedure is repeated until the effective model space Hamiltonian, wavefunction and cost function converge.
\end{itemize}

We note that similar ideas have been developed and applied in the fields of quantum chemistry and condensed matter.
For instance, Ref.~\cite{PhysRevB.105.115108} introduced an algorithm for determining (non variational) natural orbitals, which was applied to the two-dimensional Hubbard model,
and Refs.~\cite{doi:10.1063/1.5141835,PhysRevResearch.2.033421} developed orbital-optimized VQE (oo-VQE) techniques for molecular calculations, which optimize orbitals via a variational principle, closer to the spirit of our present work, but using a Unitary Coupled Cluster (UCC) ansatz of the wave function.
Contrarily to the present HL-VQE, where both the orbital parameters ${\bm \beta}$ (or, equivalently, the effective Hamiltonian) and the wave function parameters ${\bm \theta}$ are determined simultaneously at each iteration, the orbital optimizations of Refs.~\cite{PhysRevB.105.115108,PhysRevResearch.2.033421} were however performed via two-step procedures, in which the orbitals are determined classically using the outputs of the VQE. 
While the algorithm developed of Ref.~\cite{doi:10.1063/1.5141835} involved a one-step process, it differs from ours in a number of ways.
In particular, it employed a more costly fermion-to-qubit mapping, together with a UCC ansatz for the wavefunction which was further truncated in order to keep the number of wavefunction parameters polynomial. 
\\
\\
\indent
While in this work we focus on the determination of the variational ground state, the HL-VQE algorithm described here can also be applied to the description of excited states, with few modifications. 
This is outlined
in Appendix~\ref{app:excited States}.
We also note that, although we consider
Hamiltonian parameters corresponding to a one-body orbital transformation, the HL-VQE algorithm can also be applied to more general parameter learning, such as unitary transformations of the Hamiltonian generated by two- (or three-) body operators, with obvious connections to Similarity Renormalization Group (SRG) and In-Medium SRG (IM-SRG) methods.

\section{Quantum Simulations: Implementing HL-VQE and Results from IBM's Quantum Computers}
\label{sec:QuSimProduction}
\noindent
In this section, HL-VQE is applied to the LMG model, 
for which the space of Hamiltonian parameters ${\bm \beta}= \beta$ is one-dimensional. We employ a form of the wavefunction $\ket{\Psi}^{(\Lambda)}$ given in Eq.~(\ref{eq:eff_wf}), and present algorithms designed to be executed on IBM's QExperience
quantum computers~\cite{IBMQ}. 
Results obtained for small quantum circuits using the simulator {\tt AER} and quantum computer {\tt ibm\_lagos} are presented.
\\ \\
\subsection{Mapping and state preparation} 
\label{sec:QuSimProduction_stateprep}
\begin{figure}[!ht]
\centering{
\includegraphics[width=0.45\columnwidth]{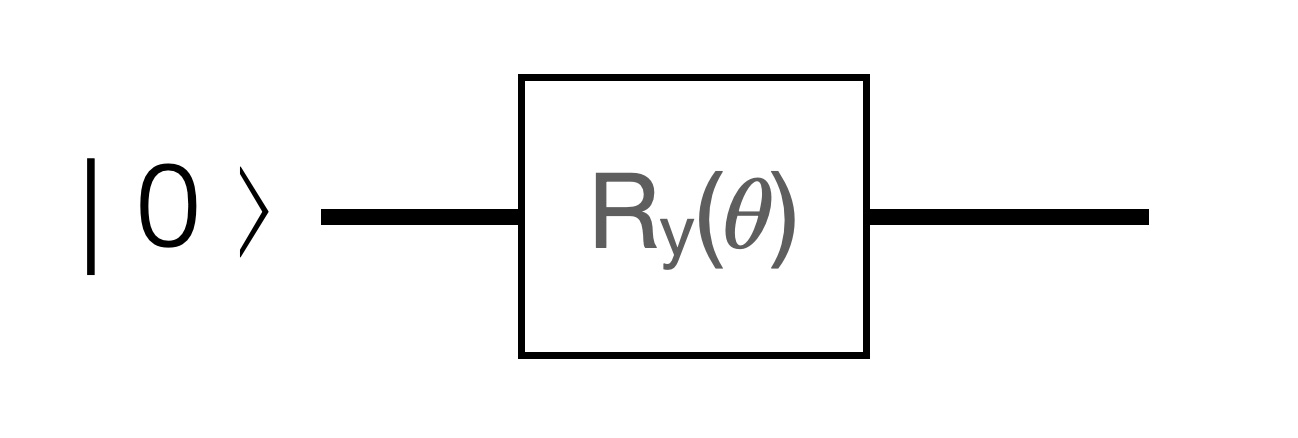}} 
\includegraphics[width=\columnwidth]{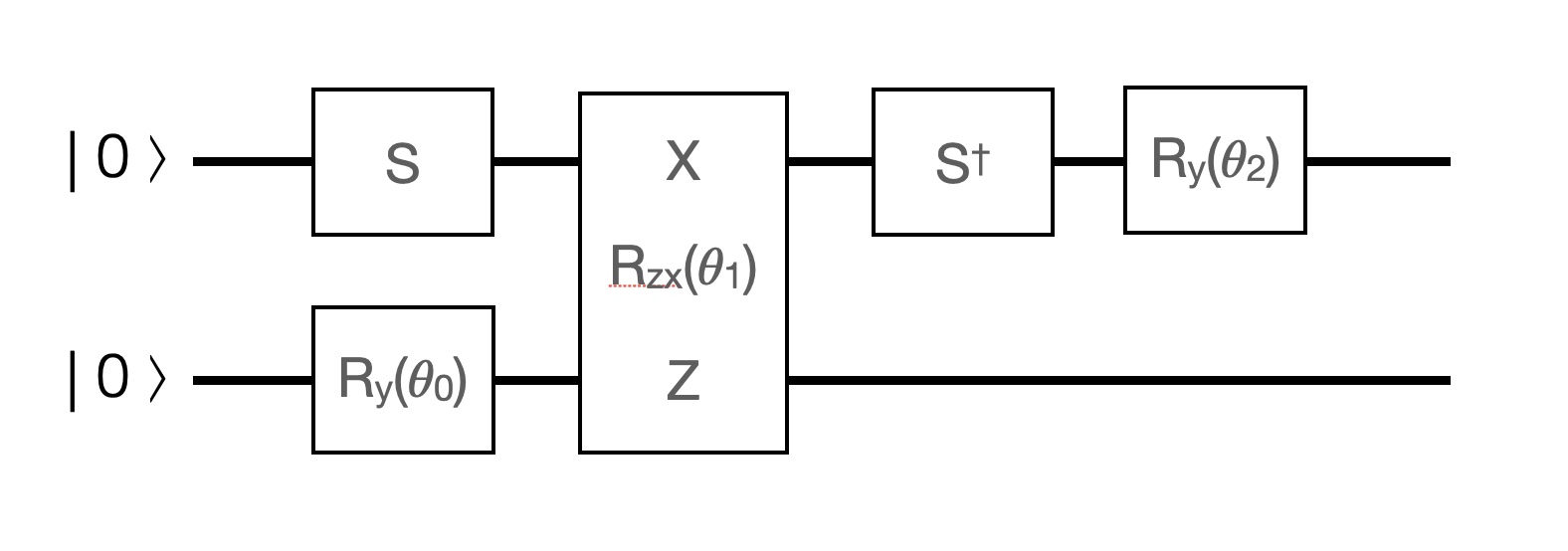} 
\caption{
Quantum circuits that can be used to prepare real wavefunctions on 
one (upper) and two (lower) qubits.
The 2-qubit circuit utilizes IBM's native $R_{ZX}(\theta)$ gate, and
for clarity  the orientation of the $\hat X\otimes \hat Z$ operation is shown explicitly.
}
\label{fig:QCprep12}
\end{figure}

\noindent
In order to implement the HL-VQE method on a digital quantum computer, 
an ansatz for the effective state $\ket{\Psi}^{(\Lambda)}$ that can be obtained by action of a unitary operator is required. 
In general, this is not obviously achievable when the state is written as a linear expansion of Slater determinants.  In the case of the LMG model, this is however possible by adopting a mapping similar to the one used in quantum simulations of quantum field theories~\cite{Jordan1130,DBLP:journals/qic/JordanLP14,Jordan_2018,PhysRevA.99.052335}, where the many-body basis states (as opposed to the fermionic modes) are mapped onto the qubits. In that case the number of required qubits $n_q$ is fixed by the desired cut-off as $\Lambda = 2^{n_q}$. 
\\ \\ 
\indent
For a CP-conserving Hamiltonian, such as the bare or effective Hamiltonians $\hat{H}(\beta)$ of the LMG model, 
the wavefunction in the Hilbert space $\ket{\Psi}^{(\Lambda)}$ in Eq.~(\ref{eq:eff_wf}) can be made relatively real, and, as such, 
there are $\Lambda-1$ angles $\boldsymbol{\theta} = \{ \theta_i \}_{i=1,.., \Lambda-1}$ required to define the $\Lambda$ amplitudes $\{ A_n^{(\beta)} \}$ of the states (unitarity fixes the remaining amplitude). 
The parameterized state $\ket{\Psi}^{(\Lambda)}$ can then be obtained by action of a unitary operator 
$\hat{\mathcal{W}}({\bm \theta})$, 
as
$\ket{\Psi}^{(\Lambda)} \equiv\ket{\Psi(\boldsymbol{\theta})}^{(\Lambda)} = \hat{\mathcal{W}}(\boldsymbol{\theta}) \ket{n=0, \beta}$.
In practice, $\ket{\Psi(\boldsymbol{\theta})}^{(\Lambda)}$ can be prepared by initializing the quantum register in the computational-basis state 
$|\Psi\rangle_{ini} = |0\rangle^{\otimes N}$,
corresponding to the un-entangled 
0p-0h configuration $\ket{n=0, \beta}$, and acting on the register with a quantum circuit involving $\Lambda-1$ angles implementing 
$\hat{\mathcal{W}}({\bm \theta})$.
The are many ways to achieve this. As mentioned above, we designed circuits such that each angle $\theta_i$ appears only once in the gates comprising the circuit. 
In that way, partial derivatives of the cost function with respect to $\theta_i$ are localized to one operator.  
This will allow for any derivative with respect to the angles ${\bm \theta}$ to be evaluated 
in the same way using a finite-difference relation, 
that does not suffer from unnecessarily large statistical uncertainties.  This is discussed in more details in App.~\ref{app:HLVQE}. 
The quantum circuits can be constructed in terms of the available quantum gate set associated with a particular quantum computer.
We chose to work with IBM's QExperience superconducting quantum computers~\cite{IBMQ}.
\\
\\
\indent
To provide elementary examples of the technique, we execute quantum simulations of effective model spaces with $\Lambda=2,4$, corresponding to one and two qubits, respectively, and as such do not consider generation of quantum circuits to furnish real wavefunctions in arbitrary-sized Hilbert spaces. 
Figure~\ref{fig:QCprep12} displays the corresponding quantum circuits used to prepare arbitrary real wavefunctions. The 2-qubit circuit uses IBM's native $R_{ZX}(\theta)$-gate. 
The use of this gate allows us to reduce the number of entangling operations. 
The wavefunctions that are prepared, when the circuits in 
Fig.~\ref{fig:QCprep12} act on the states $|0\rangle$ and $|00\rangle$ respectively are
{\small
\begin{eqnarray}
|\Psi(\theta)\rangle^{(2)}
& = &  
\cos {\theta\over 2} |0\rangle 
\ +\ 
\sin {\theta\over 2} |1\rangle 
\ \ \ ,
\nonumber\\
|\Psi({\bm \theta})\rangle^{(4)}  
& = &  
\cos {\theta_0\over 2} \cos {\theta_2-\theta_1\over 2} |00\rangle 
\ +\ 
\sin {\theta_0\over 2} \cos {\theta_2+\theta_1\over 2} |10\rangle 
\nonumber\\
&+& 
\cos {\theta_0\over 2} \sin {\theta_2-\theta_1\over 2} |01\rangle 
\ +\ 
\sin {\theta_0\over 2} \sin {\theta_2+\theta_1\over 2} |11\rangle 
\ \ \ .
\nonumber \\
\label{eq:psi12}
\end{eqnarray}
}
These correspond to the ansatz in Eq.~(\ref{eq:eff_wf}) 
for the wavefunction, with the mapping
\begin{eqnarray}
|0\rangle &\equiv& \ket{n=0,\beta} ,\nonumber \\
|1\rangle &\equiv& \ket{n=1,\beta} ,
\end{eqnarray}
for the 1-qubit case, and
\begin{eqnarray}
|00\rangle &\equiv& \ket{n=0,\beta} ,\nonumber \\
|01\rangle &\equiv& \ket{n=1,\beta} ,\nonumber \\
|10\rangle &\equiv& \ket{n=2,\beta} ,\nonumber \\
|11\rangle &\equiv& \ket{n=3,\beta} ,
\end{eqnarray}
for the 2-qubit case.

\subsection{Quantum Computation of the Expectation Values and Derivatives} 
\label{sec:QuSimProduction_computation}
\noindent
Once the state  in the effective model state $\ket{\Psi(\bm \theta)}^{(\Lambda)}$
has been prepared, the cost function $E({\bm \beta},{\bm \theta})$ and its derivatives with respect to ${\bm \beta}$ and ${\bm \theta}$ can be computed. 
\\
\\
\indent The two-dimensional effective model space that can be mapped onto 
one qubit
has a Hamiltonian expectation value and $\beta$-derivative of the form, when given in terms of Pauli operators,
\begin{eqnarray}
\langle \hat H^{(2)}(\beta) \rangle & = & 
- {N-1\over 4} \left(  (N-3) V \sin^2\beta +2 \varepsilon \cos\beta \right) \langle \hat I \rangle 
\nonumber\\
& - & 
{1\over 4} \left(  3 (N-1) V \sin^2\beta +2 \varepsilon \cos\beta \right) \langle \hat Z \rangle 
\nonumber\\
& + & 
{\sqrt{N}\over 2} \left(  \varepsilon -(N-1) V \cos\beta\right)\sin\beta\  \langle\hat X \rangle 
\ \ ,
\nonumber\\
\nabla_\beta \langle \hat H^{(2)}(\beta) \rangle
& = & 
{N-1\over 2} \left(   \varepsilon - (N-3) V \cos\beta \right) \sin\beta\  \langle \hat I \rangle
\nonumber\\
& + & 
{1\over 2} \left(    \varepsilon - 3 (N-1) V \cos\beta  \right) \sin\beta\  \langle \hat Z \rangle
\nonumber\\
& + & 
{\sqrt{N}\over 2} \left(  \varepsilon \cos \beta - (N-1) V \cos 2\beta \right)\   \langle \hat X \rangle
\ , 
\nonumber\\
\nabla_\theta \langle \hat H^{(2)}(\beta) \rangle & = & 
- {N-1\over 4} \left(  (N-3) V \sin^2\beta +2 \varepsilon \cos\beta \right) \nabla_\theta \langle \hat I \rangle 
\nonumber\\
& - & 
{1\over 4} \left(  3 (N-1) V \sin^2\beta +2 \varepsilon \cos\beta \right) \nabla_\theta \langle \hat Z \rangle 
\nonumber\\
& + & 
{\sqrt{N}\over 2} \left(  \varepsilon -(N-1) V \cos\beta\right)\sin\beta\  \nabla_\theta \langle\hat X \rangle 
\ \ , \nonumber \\
\label{eq:H1q}
\end{eqnarray}
where $\braket{ \hat{O} } \equiv \,^{(2)}\langle \Psi ({\bm \theta}) |\hat{O}| \Psi ({\bm \theta})  \rangle^{(2)}$.
Matrix elements of $\hat Z$ are found from the difference of probabilities in the computational basis, 
while matrix elements of $\hat X$ require further action by a Hadamard-gate $\hat {\rm H}$ prior to measurement.
Derivatives of the expectation values of the Pauli operators with respect to ${\bm \theta}$, which are computed using the finite difference method, require two additional circuits per operator. Therefore six different ensembles of quantum circuits are needed. 
More details are provided in App.~\ref{app:HLVQEnQ1}.
\\
\\
\indent
For the 2-qubit systems, the four-dimensional Hamiltonian in the $\Lambda=4$ effective model space has a 
Pauli decomposition of the form,
\begin{eqnarray}
H^{(4)}(\beta) & = & 
h^{(4)}_{II}(\beta) 
\ +\ 
h^{(4)}_{xx}(\beta) \   \hat X\otimes \hat X  
\nonumber\\
& + & h^{(4)}_{xz}(\beta) \   \hat X\otimes \hat Z  
 + h^{(4)}_{xI}(\beta) \   \hat X\otimes \hat I  
\nonumber\\
& + & h^{(4)}_{yy}(\beta) \   \hat Y\otimes \hat Y  
 + h^{(4)}_{zx}(\beta) \   \hat Z\otimes \hat X  
\nonumber\\
& + & h^{(4)}_{zz}(\beta) \   \hat Z\otimes \hat Z  
 + h^{(4)}_{zI}(\beta) \   \hat Z\otimes \hat I  
\nonumber\\
& + & h^{(4)}_{Ix}(\beta) \   \hat I\otimes \hat X  
 + h^{(4)}_{Iz}(\beta) \   \hat I\otimes \hat Z  
 \ \ \ ,
 \label{eq:ham2pauliALL1}
 \end{eqnarray}
with
\begin{eqnarray}
&& h^{(4)}_{II}(\beta) \ =\  \frac{1}{4} (N-3) \left((N-7) V \sin ^2(\beta )-2 \
\varepsilon  \cos (\beta )\right)
 \nonumber\\
&& h^{(4)}_{xx}(\beta) \ =\  \frac{\sqrt{N-1} \sin (\beta ) (\varepsilon +(N-3) V \
\cos (\beta ))}{2 \sqrt{2}}
 \nonumber\\
&&h^{(4)}_{xz}(\beta) \ =\ \frac{\left(\sqrt{N}-\sqrt{3} \sqrt{N-2}\right) \sqrt{\
N-1} V (\cos (2 \beta )+3)}{8 \sqrt{2}}
 \nonumber\\
&& h^{(4)}_{xI}(\beta) \ =\  \frac{\left(\sqrt{N} + \sqrt{3} \sqrt{N-2}\right) \sqrt{\
N-1} V (\cos (2 \beta )+3)}{8 \sqrt{2}}
 \nonumber\\
&&h^{(4)}_{yy}(\beta) \ =\  h^{(4)}_{xx}(\beta)
 \nonumber\\
&& h^{(4)}_{zx}(\beta) \ =\  \frac{1}{4} \sin (\beta ) 
\left(
\varepsilon  \left(\sqrt{N}-\sqrt{3} \sqrt{N-2}\right)
\right.
 \nonumber\\
&& \ \  +  \left.
\left(N^{3/2}-\sqrt{3} \sqrt{N-2} N-\sqrt{N}+5 \sqrt{3} \sqrt{N-2}\right) V \cos (\beta )
\right) 
\nonumber \\
&& h^{(4)}_{zz}(\beta) \ =\  \frac{3}{2} V \sin ^2(\beta )
 \nonumber\\
&& h^{(4)}_{zI}(\beta) \ =\  \frac{3}{2} (N-3) V \sin ^2(\beta )-\varepsilon  \cos (\beta )
 \nonumber\\
&& h^{(4)}_{Ix}(\beta) \ =\  \frac{1}{4} \sin (\beta ) 
\left(
\varepsilon  \left(\sqrt{N} + \sqrt{3} \sqrt{N-2}\right)
 \right.
 \nonumber\\
&& \ \  +  \left.
\left(N^{3/2}+\sqrt{3} \sqrt{N-2} N-\sqrt{N}-5 \sqrt{3} \sqrt{N-2}\right) V \cos (\beta )
\right)
 \nonumber\\
&& h^{(4)}_{Iz}(\beta) \ =\  \frac{1}{4} \left(3 (N-3) V \sin ^2(\beta )-2 \varepsilon  \
\cos (\beta )\right)
\ \ \ ,
\label{eq:ham2pauliALL2}
\end{eqnarray}
with derivatives with respect to $\beta$ that can be evaluated straightforwardly.
The explicit forms for the expectation values of $H^{(4)}(\beta)$ are given in Eqs.~(\ref{eq:ham2pauliEXP}) and (\ref{eq:psi2mats})
with products of Pauli operators
$\hat I$, $\hat X$, $\hat Y$ or $\hat Z$
acting on each qubit.
For qubits with a $\hat X$, the action of $\hat {\rm H}$ is further required, while for 
$\hat Y$, the action of $\hat S^\dagger \hat {\rm H}$ is required prior to measurement.
For each operator, there are six additional circuits per operator 
to provide derivatives for the three angles $\theta_0$, $\theta_1$ and $\theta_2$, 
leading to thirty-five (35) physics ensemble measurements per iteration step.
Further details can be found in the App.~\ref{app:HLVQEnQ2}.
\\
\\
\indent
With multiple circuits per iteration step, and using IBM's 7-qubit quantum computers, we explored packing multiple circuits for parallel running. For the 1-qubit case, executing single qubit jobs was found to minimize errors, but at the expense of an increased number of shots.
For 2-qubit circuits, the large number of ensemble measurements motivated executing multiple circuits in parallel.  For each operator, we executed  1,2,2,2 circuits per job. After initial tuning, runs with 8k shots per ensemble, where k denotes $\times 10^3$, we used 32k shots per circuit in production on {\tt ibm\_lagos}.  
We used 100k shots per circuit on the {\tt AER} classical simulator.
IBM's measurement error correction routine was used as part of post-processing, and because of the shallow circuit depth, and the use of the $R_{ZX}$-gate, 
we did not utilize a CNOT extrapolation~\cite{PhysRevX.7.021050,PhysRevA.98.032331} or 
randomized compiling of the CNOTs (Pauli twirling)~\cite{PhysRevA.94.052325},
and we also did not use dynamic-decoupling~\cite{PhysRevA.58.2733,DUAN1999139,ZANARDI199977,PhysRevLett.82.2417},  dynamical decoupling, 
or decoherence-renormalization~\cite{Urbanek:2021oej,ARahman:2022tkr,Farrell:2022wyt}.~\footnote{
Studies of the stability of (some of) IBM's quantum computers can be found in Ref.~\cite{Yeter-Aydeniz:2022vuy}.} 
%

\subsection{Gradient descent and workflow of the calculation}
\label{sec:QuSimProduction_graddesc}
\noindent
For a given iteration $[k]$ of the procedure,
once the numerical values of 
\begin{eqnarray}
G_{\bm\beta}^{[k]} & = & 
\, ^{(\Lambda)} \langle \Psi({\bm\theta}^{[k]})| {\bm\nabla}_{\beta}  \hat H({\beta}^{[k]})|\Psi({\bm\theta}^{[k]})\rangle^{(\Lambda)}
\ \ ,
\nonumber\\
G_{\bm\theta}^{[k]} & = & {\bm\nabla}_{\bm\theta} \, ^{(\Lambda)}\langle \Psi({\bm\theta}^{[k]})|
\hat H({\beta}^{[k]})|\Psi({\bm\theta}^{[k]})\rangle^{(\Lambda)}
\ \ ,
\label{eq:Gdef}
\end{eqnarray}
have been determined from combined contributions from the classical and quantum computers,
the values of the variational parameters at the next iterative step are determined using gradient-descent,
\begin{eqnarray}
{\beta}^{[k+1]} & = & 
{\beta}^{[k]}\ -\ \eta\ G_{\bm\beta}^{[k]} /{\cal G}^{[k]}
\ \ ,
\nonumber\\
{\bm\theta}^{[k+1]} & = & 
{\bm\theta}^{[k]}\ -\ \eta\ G_{\bm\theta}^{[k]} /{\cal G}^{[k]}
\ \ ,
\nonumber\\
{\cal G}^{[k]} & = & \sqrt{|G_{\beta}^{[k]}  |^2
\ +\ 
|G_{\bm\theta}^{[k]} |^2 }
\ \ ,
\label{eq:CDexp}
\end{eqnarray}
where $\eta$ is the learning rate.  
Tuning led to setting $\eta=0.07$ for production
using 1- and 2-qubits with Eq.~(\ref{eq:CDexp}).
What we have described above is slightly different from what can be found 
in App.~\ref{app:HLVQE}
and Eq.~(\ref{eq:witer})
in defining the update step,
but was found to lead to more rapid numerical convergence.
The iteration continues until 
the wavefunction amplitudes and ground-state energy
stabilize, or until the number of iterations reaches a given value, 
a number exceeding ${\rm N}_{\rm iter}\ge 80$.

\subsection{Results for 1-Qubit Truncation: $\Lambda=2$}
\begin{figure}[!ht]
\centering{
\includegraphics[width=\columnwidth]{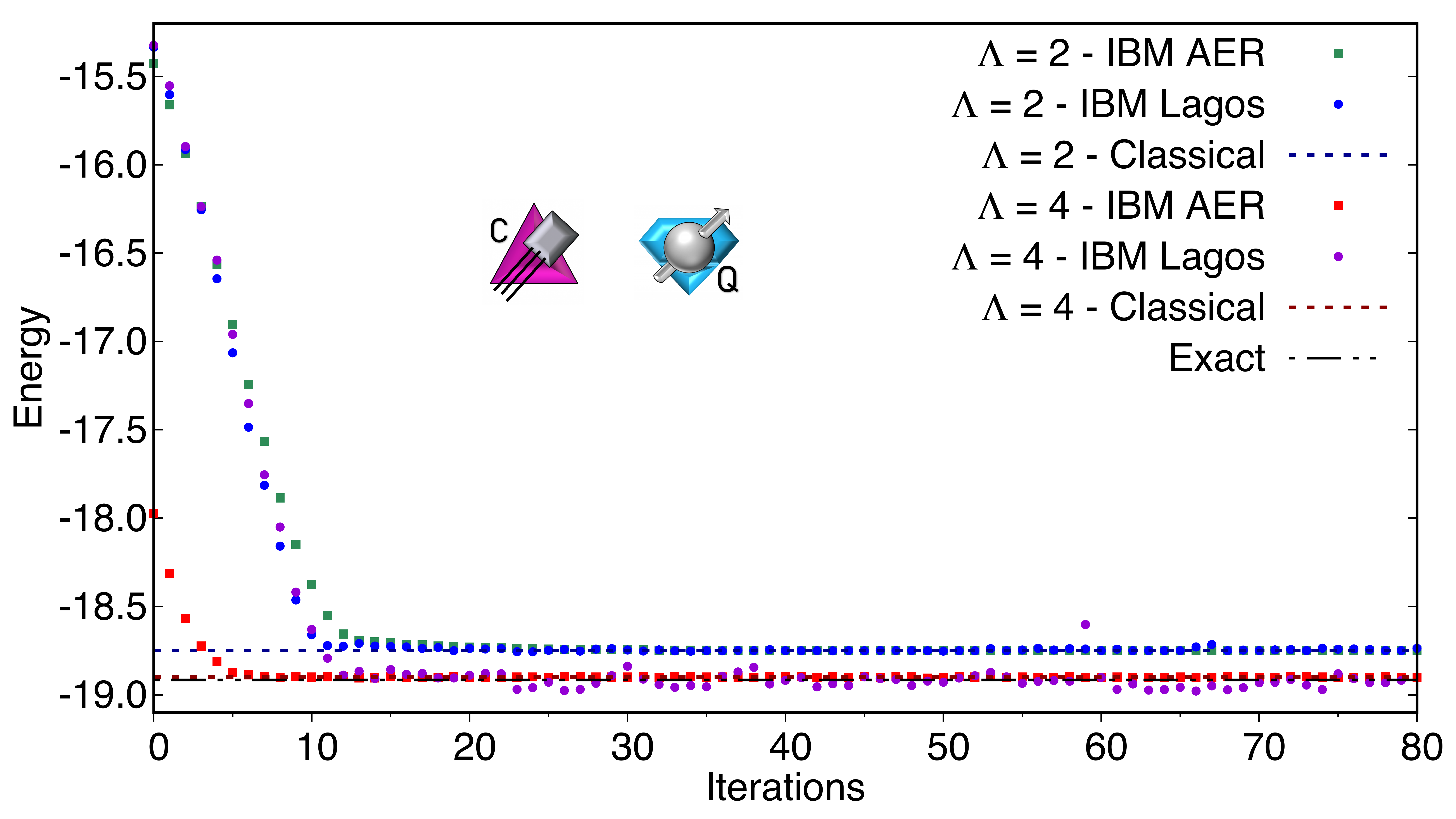}} 
\caption{
Central values of the ground-state energies (in unit of $\varepsilon$) obtained for the $\Lambda=2,4$ systems from 1- and 2-qubit simulations using {\tt AER} and {\tt ibm\_lagos} for the LMG-model parameters 
using 100k
and 32k shots, respectively.
Numerical values for these energies are given in Table~\ref{tab:AERLam1} and \ref{tab:ibmlagosLam1}.
The starting values of variational parameters of the 1-qubit simulations were:
$\beta^{[0]}= 0.2$ and $\theta_{ i}^{[0]}= 0.30$ using {\tt AER} and 
$\beta^{[0]}= 0.2$ and $\theta_{ i}^{[0]} = 0.10$ using {\tt ibm\_lagos}. 
The starting values of variational parameters of the 2-qubit simulations were:
$\beta^{[0]}= 0.8$ and $\theta_{i}^{[0]}= 0$ using {\tt AER}, and 
$\beta^{[0]}= 0.2$ and $\theta_{i}^{[0]}= 0$ using {\tt ibm\_lagos}.
The dark blue and dark red dashed lines show the results obtained from exact diagonalization in the 
 effective model spaces $\Lambda=2$ and $\Lambda=4$, respectively. The black dotted-dashed line shows the exact result, obtained from exact diagonalization in the full Hilbert space.
}
\label{fig:EiterLam24}
\end{figure}

\begin{figure}[!ht]
\centering{
\includegraphics[width=\columnwidth]{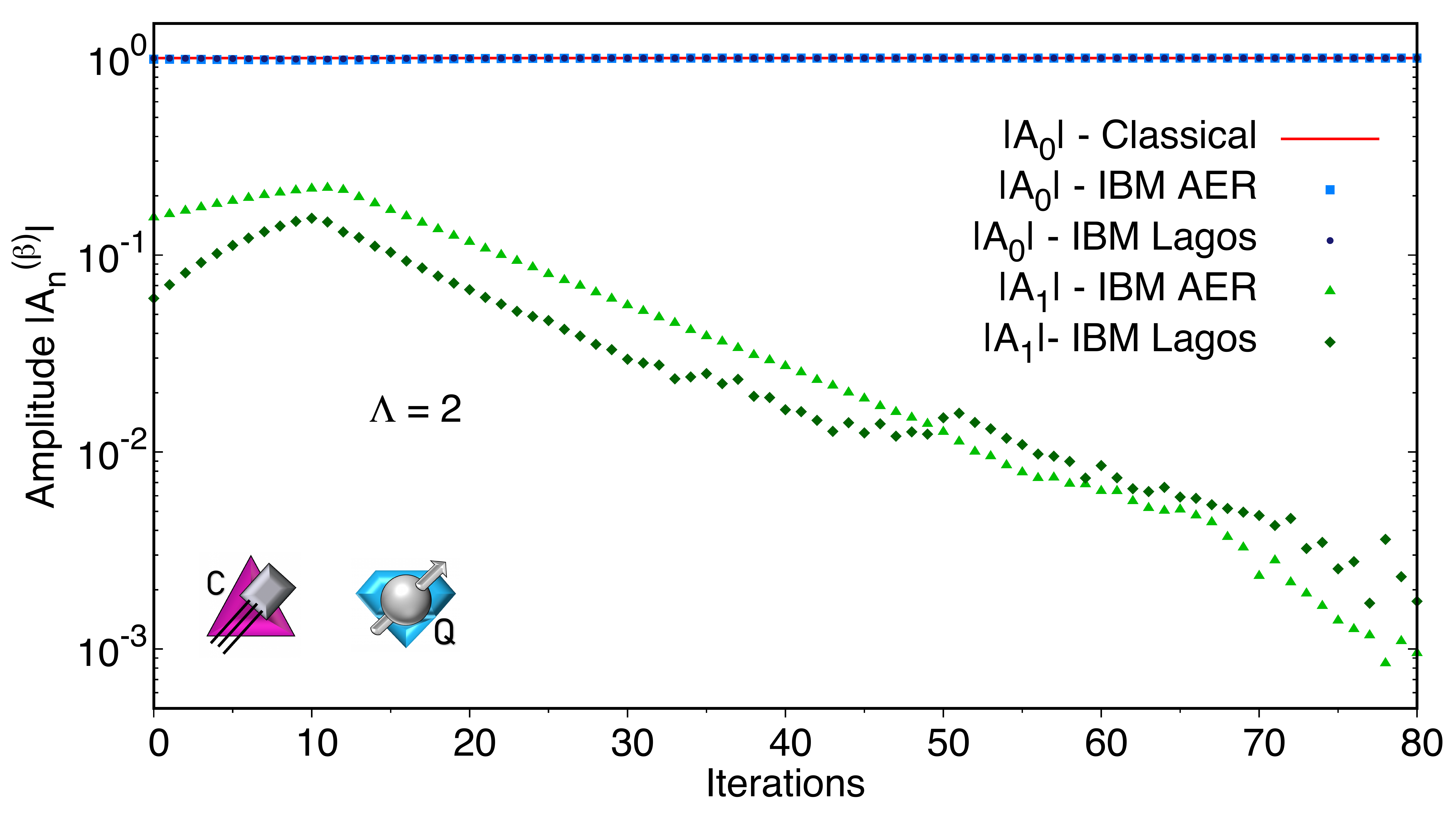}} 
\caption{
Central values of the wavefunction amplitudes obtained for the $\Lambda=2$ systems from 1-qubit simulations using {\tt AER} and {\tt ibm\_lagos} for the LMG-model parameters using 100k and 32k shots, respectively.
Numerical values for these amplitudes are given in Table~\ref{tab:AERLam1} and \ref{tab:ibmlagosLam1}.
The starting values of variational parameters of the 1-qubit  simulations were:
$\beta^{[0]}= 0.2$ and $\theta_{i}^{[0]}= 0.30$ using {\tt AER} and 
$\beta^{[0]}= 0.2$ and $\theta_{i}^{[0]} = 0.10$ using {\tt ibm\_lagos}.
Exact diagonalization in the  effective model space yields $|A_0|=1$ (red horizontal line) and $|A_1|=0$.
}
\label{fig:A0A1iterLam24}
\end{figure}

\begin{figure}[!ht]
\centering{
\includegraphics[width=\columnwidth]{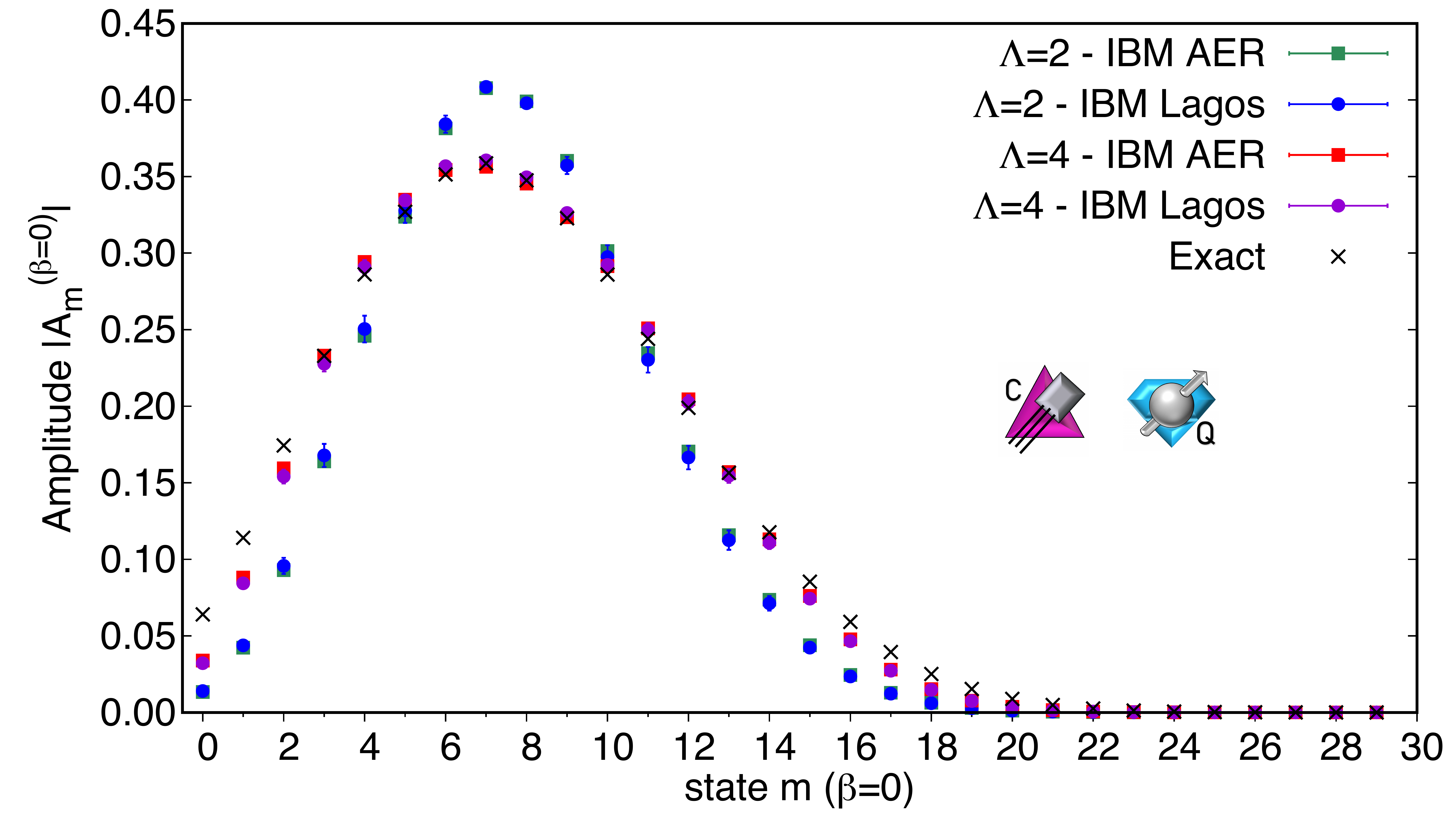}} 
\caption{
Full-space wavefunctions reconstructed from the effective model spaces with 
$\Lambda=2,4$ from 1- and 2-qubit simulations using {\tt AER} and {\tt ibm\_lagos} for the LMG-model parameters 
 using 100k  and 32k shots, respectively. 
The black crosses denote the exact full-Hilbert space wavefunctions.
Numerical values for these amplitudes are given in Table~\ref{tab:fullpsi}.
The starting values of variational parameters of the 1-qubit simulations were:
$\beta^{[0]}= 0.2$ and $\theta_{i}^{[0]}= 0.30$ using {\tt AER} and 
$\beta^{[0]}= 0.2$ and $\theta_{i}^{[0]} = 0.10$ using {\tt ibm\_lagos}.
The starting values of variational parameters of the 2-qubit simulations were:
$\beta^{[0]}= 0.8$ and $\theta_{i}^{[0]}= 0$ using {\tt AER}, and 
$\beta^{[0]}= 0.2$ and $\theta_{i}^{[0]}= 0$ using {\tt ibm\_lagos}.
The uncertainties are determined from the standard deviation derived from 
a correlated propagation of the last 20 iterations.
}
\label{fig:PsiFullIBM}
\end{figure}

\begin{figure}[!ht]
\centering{
\includegraphics[width=\columnwidth]{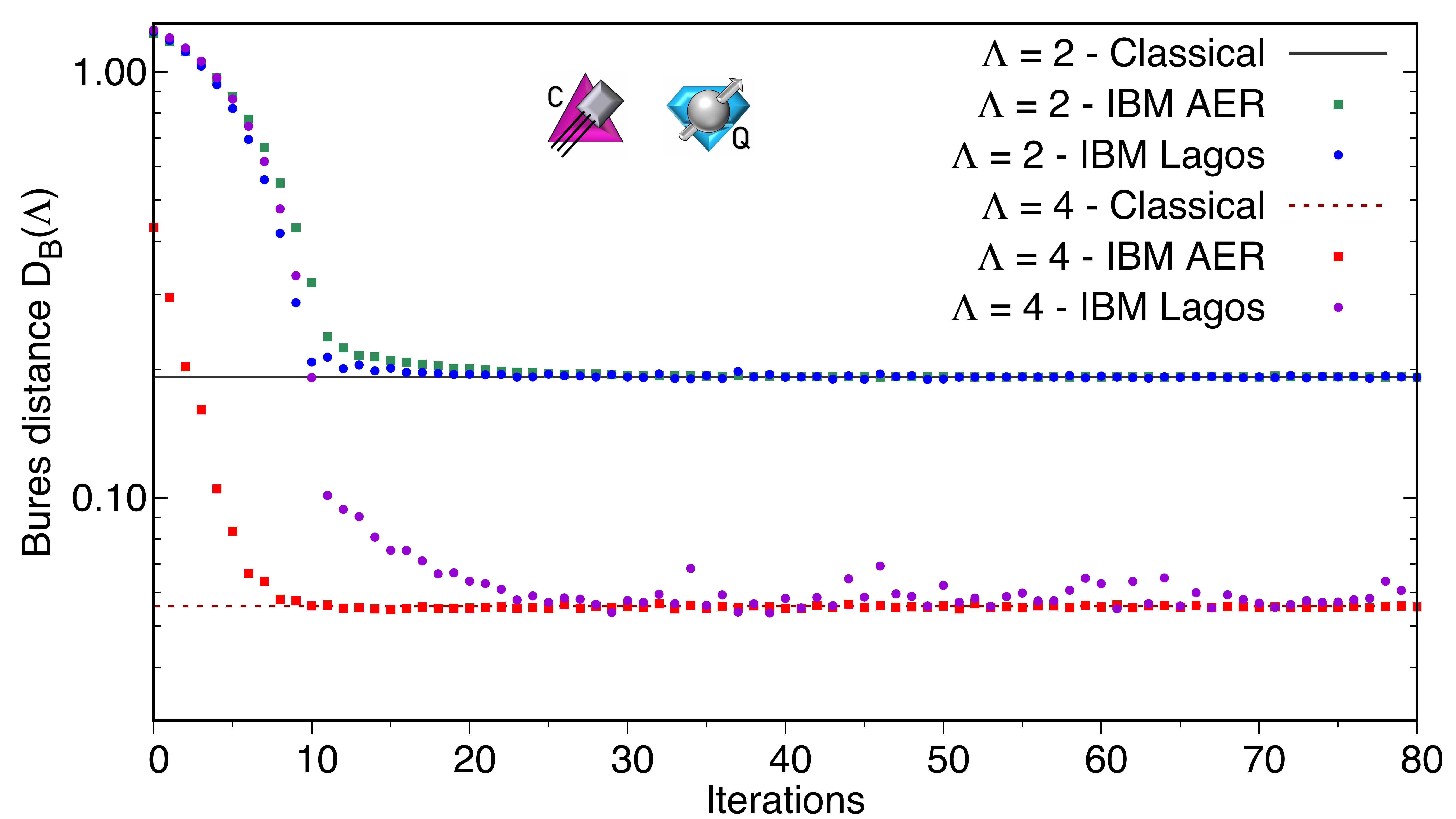}} 
\caption{
Central values of the 
Bures distance defined in Eq.~(\ref{eq:fidelityDB}),
obtained for the $\Lambda=2,4$ systems from 1- and 2-qubit simulations using {\tt AER} and {\tt ibm\_lagos} for the LMG-model parameters 
 using 100k  and 32k shots, respectively. 
Numerical values for these distances are given in Table~\ref{tab:AERLam1} and \ref{tab:ibmlagosLam1}.
The starting values of variational parameters of the 1-qubit simulations were:
$\beta^{[0]}= 0.2$ and $\theta_{i}^{[0]}= 0.30$ using {\tt AER} and 
$\beta^{[0]}= 0.2$ and $\theta_{i}^{[0]} = 0.10$ using {\tt ibm\_lagos}.
The starting values of variational parameters of the 2-qubit simulations were:
$\beta^{[0]}= 0.8$ and $\theta_{i}^{[0]}= 0$ using {\tt AER}, and 
$\beta^{[0]}= 0.2$ and $\theta_{i}^{[0]}= 0$ using {\tt ibm\_lagos}.
}
\label{fig:BurasiterLam24}
\end{figure}

\noindent
For our production using IBM's QExperience
superconducting quantum computer {\tt ibm\_lagos} and the simulator
{\tt AER}, 
one set of model parameters was used: $N=30$, $\overline{v}=V(N-1)/\varepsilon=2.0$ and $\varepsilon=1.0$.
These parameters 
will allow for a comparison with the results
obtained with explicit diagonalizations 
that were presented in Sec.~\ref{sec:LGM_eff}.
The quantum circuits, HL-VQE algorithm and workflow described above were implemented
using a learning-rate of $\eta=0.07$ and starting values 
of the parameters $\beta^{[0]}= 0.2$ and $\theta_{i}^{[0]} = 0.10$.
Typically, 80 iterations were performed, with stable results being obtained after many fewer.
\\
\\
\indent The results obtained 
from {\tt ibm\_lagos} and {\tt AER}
for the central values of the ground-state energy (in unit of $\varepsilon$) 
as a function of iteration are shown in Fig.~\ref{fig:EiterLam24},
and numerical values are given in Table~\ref{tab:AERLam1} and \ref{tab:ibmlagosLam1}.
The uncertainty associated with each iteration is not shown, 
but can be deduced, in combination with other uncertainties, 
from the fluctuations about the mean value 
(as each iteration is statistically independent).
The ground-state energy of the $\Lambda=2$ systems obtained from {\tt AER} and {\tt ibm\_lagos} are consistent with each other, and with the exact diagonalization in the (truncated) effective model space performed in Sec.~\ref{sec:LGM_eff} (see Table~\ref{tab:Lam2summary}).
As different starting values of parameters in the HL-VQE were used in production on {\tt AER} and {\tt ibm\_lagos}, a direct and meaningful comparison of the two approaches to the ground-state energy should not be made, and only the asymptotic values and uncertainties should be compared.
\\
\\
\indent The wavefunction amplitudes as a function of iteration are shown in Fig.~\ref{fig:A0A1iterLam24} and numerical values are given in Table~\ref{tab:AERLam1} and \ref{tab:ibmlagosLam1}.
For an effective model space with $\Lambda=2$ states, the effective 
Hamiltonian is defined by the lowest-lying $2\times 2$ block of $H(\beta)^{(2)}$. Thus for fixed values of $N$, $\varepsilon$ and $\overline{v}$, it is a $2\times 2$ numerical matrix with entries that are functions of $\beta$.
For the value of $\overline{v}$ we are considering, the structure of the Hamiltonian is such that the vanishing of the off-diagonal element, 
$\Xi= \varepsilon - (N-1) V \cos\beta =0$, corresponds to the minimum of the lowest-lying eigenvalue. This is most easily seen in a $N$-expansion followed by a considerations of perturbations around $\Xi=0$.   
As such the ground state of this system corresponds to 
$|\Psi(\theta)\rangle^{(2)}$ in Eq.~(\ref{eq:psi12}) 
with $\theta=0$, and therefore without mixing between 
0p-0h and 1p-1h configurations in the truncated basis,
corresponding to the HF state.
See App.~\ref{app:GSforms} for more details.
We see from Fig.~\ref{fig:A0A1iterLam24} that the amplitudes obtained from both {\tt AER} and {\tt ibm\_lagos} are consistent with these expectations, as we observe that the amplitude of the 1p-1h state is systematically reducing with iteration step. 
The precision of extracted amplitudes, at any iteration step, are limited by the precision used in the workflow, along with the learning-rate.  
The results obtained with {\tt ibm\_lagos} show signs 
of the impact of simulation errors 
at the percent level, as is to be expected,
with similar fluctuations occurring an order of magnitude smaller for {\tt AER}.  
The convergence to the known result is consistent with an
exponential over a large number of iteration steps.
\\
\\
\indent 
The full-space wavefunctions (expressed in the unrotated $\beta=0$ basis) 
reconstructed from the previous $\Lambda=2$ effective model space using Eq.~(\ref{eq:rotation_mat}) is shown in Fig.~\ref{fig:PsiFullIBM} (green and blue points), 
with numerical results given in Table~\ref{tab:fullpsi}.
The Bures distance of the reconstructed full-space wavefunction, 
given in
Eq.~(\ref{eq:fidelityDB}), is displayed in Fig.~\ref{fig:BurasiterLam24}.
While the convergence of the amplitudes and energies towards the expected (classical) values 
are consistent with single exponentials,
the convergence of the $\Lambda=2$ ground-state wavefunction  has a different functional form, with initial fidelities only slowly converging for a number of iteration steps. 
This is followed by an interval of rapid convergence to the wavefunction that is exact for the effective model space.
\\
\\
\indent
A summary of the results obtained for the $\Lambda=2$-state effective model space
from IBM's classical simulator {\tt AER} 
and from their superconducting-qubit quantum computer {\tt ibm\_lagos}, 
along with
results obtained by exact diagonalization, in effective and full model spaces, are given in Table~\ref{tab:Lam2summary}.
 \begin{table}[!ht]
 \centering
 \resizebox{\columnwidth}{!}{  
 \begin{tabular}{ccccc} 
 \hline   
 \\
  & & $\Lambda=2$  & &  
 \\
 \\
 \hline
 {\rm Quantity} 
 & {\tt AER} 
 &  {\tt ibm\_lagos}  
 & {\rm Exact Effective} 
 & {\rm Exact Full}
  \\
 \hline
 $E_{\rm g.s.}$ & -18.7500(0)  & -18.7490(37) & -18.750000 & -18.916414 \\
 $\beta$ & 1.0472(14) & 1.041(13) & 1.0471975 & 0\\
 $|A_0|$ & 1.00000(2) & 1.00000(5) & 1 & - \\
 $|A_1|$ & 0.0018(12) & 0.0035(16) & 0 & - \\
 $D_B$ & 0.1926(7) & 0.1922(15) &   0.1922023 & - \\
 \hline
 \hline
 \end{tabular}
 }
 \caption{
 Results obtained for $N=30$, $\varepsilon=1.0$ and $\overline{v}=2.0$ using IBM's
{\tt AER} 
 and 
 {\tt ibm\_lagos}, and results obtained by exact diagonalization,
 in the $\Lambda=2$ effective model space 
 and in the full Hilbert space.
 The central values and uncertainties are derived from iterations steps between 70 and 80.
 The central value is the mean in this interval, while the uncertainty is the half the difference between the maximum and minimum values.
 The values at each iteration step from which these results are obtained, are given in Table~\ref{tab:AERLam1} and \ref{tab:ibmlagosLam1}, and displayed in 
 Figs.~\ref{fig:EiterLam24}, \ref{fig:A0A1iterLam24} and \ref{fig:BurasiterLam24}.
 }
 \label{tab:Lam2summary}
 \end{table}
Good agreement is found between the results obtained from {\tt AER} and {\tt ibm\_lagos}, 
and those from exact diagonalizations within the model space.

\subsection{Results for 2-Qubit Truncation: $\Lambda=4$}
\begin{figure}[!ht]
\centering{
\includegraphics[width=\columnwidth]{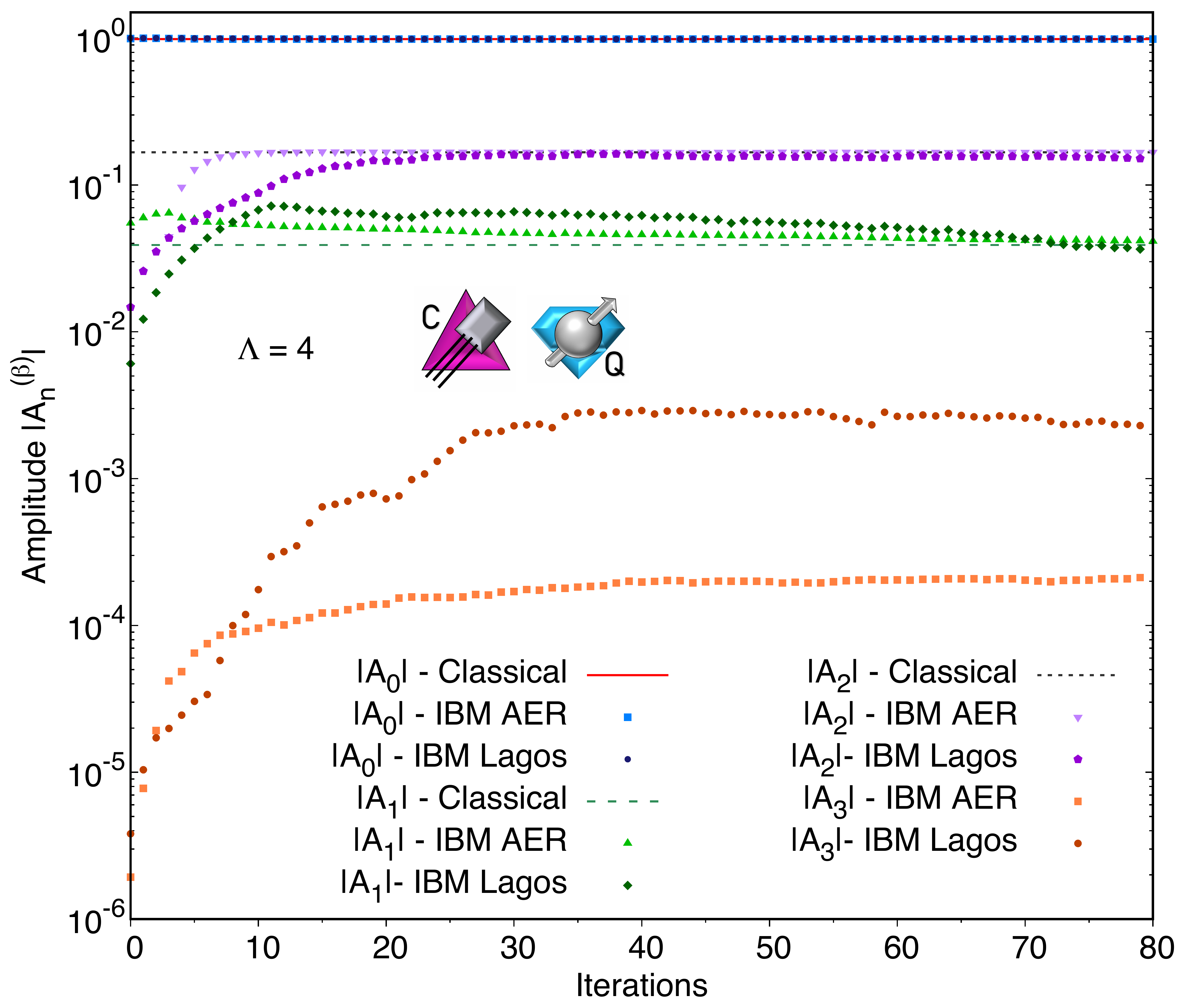}} 
\caption{
Central values of the wavefunction amplitudes obtained for the $\Lambda=4$ systems from 2-qubit simulations using {\tt AER} and {\tt ibm\_lagos} for the LMG-model parameters 
 using 100k and 32k shots, respectively.
Numerical values for these amplitudes are given in Table~\ref{tab:AERLam4} and \ref{tab:ibmlagosLam4}.
The starting values of variational parameters of the 2-qubit simulations were:
$\beta^{[0]}= 0.8$ and $\theta_{i}^{[0]}= 0$ using {\tt AER}, and 
$\beta^{[0]}= 0.2$ and $\theta_{i}^{[0]}= 0$ using {\tt ibm\_lagos}.
Exact diagonalization in the truncated effective space yields the values of $|A_0|$, $|A_1|$ 
and $|A_2|$ shown with horizontal lines, and $|A_3|=0$.
}
\label{fig:A0A1A2A3iterLam4}
\end{figure}

\noindent
The HL-VQE algorithm is applied to the $\Lambda=4$ system in the same way that is was applied to the 
$\Lambda=2$ system, with the obtained
ground-state energy as a function of iteration shown in Fig.~\ref{fig:EiterLam24} 
(red and purple points),
with a converged value provided in Table~\ref{tab:Lam4summary}.
 \begin{table}[!ht]
 \centering
 \resizebox{\columnwidth}{!}{  
 \begin{tabular}{ccccc} 
 \hline   
 \\
  & & $\Lambda=4$  & &  
 \\
 \\
 \hline
 {\rm Quantity} 
 & {\tt AER} 
 &  {\tt ibm\_lagos}  
 & {\rm Exact Effective} 
 & {\rm Exact Full}
  \\
 \hline
 $E_{\rm g.s.}$ & -18.9000(12)  & -18.929(44) & -18.900130 & -18.916414 \\
 $\beta$ & 1.01479(39)  & 1.0160(96)  & 1.0162245  & 0\\
 $|A_0|$ &  0.98500(02) & 0.98666(85)  & 0.98516 & - \\
 $|A_1|$ &  0.04193(18) & 0.0469(98)  & 0.03901 & - \\
 $|A_2|$ &  0.16739(06) & 0.1557(34)  & 0.16711 & - \\
 $|A_3|$ &  0.00020(01) & 0.00258(28)  & 0 & - \\
 $D_B$ & 0.05559(12)  & 0.0579(42)   & 0.05578    & - \\
 \hline
 \hline
 \end{tabular}
 }
 \caption{
 Results obtained for $N=30$, $\varepsilon=1.0$ and $\overline{v}=2.0$ using IBM's
 {\tt AER} 
 and 
 {\tt ibm\_lagos}, and results obtained by exact diagonalization,
 from the $\Lambda=4$ effective model space
 and in the full Hilbert space.
 The central values and uncertainties are derived from iterations steps between 70 and 80.
 The central value is the mean in this interval, while the uncertainty is the half the difference between the maximum and minimum values.
 The values at each iteration step from which these results are obtained, are given in Table~\ref{tab:AERLam4} and \ref{tab:ibmlagosLam4}, and displayed in 
 Figs.~\ref{fig:EiterLam24}, \ref{fig:A0A1iterLam24} and \ref{fig:BurasiterLam24}.
 }
 \label{tab:Lam4summary}
 \end{table}
The results obtained with both {\tt AER} and {\tt ibm\_lagos} are consistent 
with the exact value in the effective model space, but the result from 
{\tt AER} are an order of magnitude more precise than that from {\tt ibm\_lagos}.
The extracted value of $\beta$ from {\tt AER} is a few standard deviations away from the expected value, while that from {\tt ibm\_lagos} is consistent, but with a larger uncertainty, as seen in 
Table~\ref{tab:Lam4summary}.
\\
\\
\indent
Figure~\ref{fig:A0A1A2A3iterLam4} shows the amplitudes $|A_n^{(\beta)}|$ of the $\Lambda=4$ states in the effective model space determined with {\tt AER} and {\tt ibm\_lagos}.
With the effective Hamiltonian $H^{(4)}(\beta)$ (see Eq.~(\ref{eq:ham2pauliALL1})) 
truncated for $\Lambda=4$ states, and with the parameters used in the $\Lambda=4$ 
effective model space, the amplitude of the 3p-3h state is expected to vanish by the arguments given in App.~\ref{app:GSforms3p3h}.
However, this expectation was not enforced in carrying out the HL-VQE, 
and, in addition to $\beta$, the three angles $\theta_0$, $\theta_1$ and $\theta_2$, in $|\Psi({\bm \theta})\rangle^{(4)}$ (Eq.~(\ref{eq:psi12})) were varied.
The amplitude of the 3p-3h state is found to be much smaller than the other amplitudes from both simulations, consistent with its vanishing in the exact diagonalization in the effective model space, along with the statistical and systematic errors implicit in the simulations.
The results obtained for $|A_{0,1,2,3}^{(\beta)}|$ from {\tt AER} and {\tt ibm\_lagos} at large iteration number
are consistent with each other and with expectations at the few-$\sigma$ level, as can be seen from
Fig.~\ref{fig:A0A1A2A3iterLam4} and Tables~\ref{tab:Lam4summary}, \ref{tab:AERLam4} and \ref{tab:ibmlagosLam4}.
\\
\\
\indent
The full-space wavefunctions reconstructed from $|A_{0,1,2,3}^{(\beta)}|$
using Eqs.~(\ref{eq:innerD}) and (\ref{eq:rotation_mat})
are shown in Fig.~\ref{fig:PsiFullIBM} (red and purple points)
and given in Table~\ref{tab:fullpsi}.
The corresponding Bures distance to the exact full-space wavefunction is shown in Fig.~\ref{fig:BurasiterLam24} 
and given in Tables~\ref{tab:AERLam4} and \ref{tab:ibmlagosLam4}.
The Bures distance for $\Lambda=4$ is significantly improved over that for 
$\Lambda=2$, consistent with the exponential scaling for $\Lambda \ll N$ 
that is anticipated from the exact calculations in 
effective model space.

\subsection{The Scaling of Quantum Resources}

It is helpful to consider the scaling of quantum resources required 
for these effective model space simulations in order to estimate the practicality of our HL-VQE algorithm.
For the present calculations using {\tt ibm\_lagos}, the workflow that we implemented 
for $\Lambda=4$ used 1,2,2,2 circuits per ensemble run, 
with 32k shots (measurements) per ensemble, for a total of 128k shots.
For the 80 iterations that were run, this corresponds to $\sim $ 10.2M shots.
Of course, this number could be reduced using more sophisticated classical search algorithms, 
implementing more aggressive convergence criteria, 
implementing a dynamical shot estimator between iterations, and more,
but the order-of-magnitude quantum resource cost will remain.

The exponential growth in the number of angles required 
to define the variational wavefunction, scaling with the number of states 
in the Hilbert space, nominally renders VQE algorithms unsuitable for use at scale.
However, with the double exponential improvement in ground-state energy and wavefunction fidelity 
with increasing number of qubits supporting the effective model spaces for $\Lambda\ll N$,
the scaling of the VQE algorithm, and in this case the HL-VQE algorithm, 
becomes favorable for increasing the size of effective model spaces 
until threshold effects become significant.

\section{Summary, Conclusions and Outlook}
\label{sec:conclusions}
\noindent
We have performed a detailed study of the use of effective model spaces and effective Hamiltonians for 
the preparation of low-lying states of the Lipkin-Meshkov-Glick model, 
a well-known solvable nuclear physics model of multi-nucleon systems, using quantum computers.
In this model, while variational ground states for systematically truncated Hamiltonians exhibit power-law convergence to the exact results,
additionally optimizing the truncated Hamiltonian from a variational principle is shown to exponentially improve convergence throughout the regime where the optimal energy surface leads to spontaneous symmetry breaking.
We have demonstrated these gains using exact and approximate classical calculations.
These findings 
indicate that quantum simulations using small model spaces with variationally-improved Hamiltonians
can furnish precise and accurate results, and have utility in bounded-error simulations with reduced quantum resource requirements. 
Borrowing ideas from quantum chemistry, we have suggested and explored an algorithm that combined Hamiltonian learning with VQE 
to be used in classical-quantum hybrid computing environments, which we call HL-VQE. 
In this algorithm, parameters defining an effective Hamiltonian used in an effective model space (or EFT), 
and the parameters defining its ground-state wavefunction, are simultaneously determined through minimization of an associated cost function, which we took to be the ground-state energy.
In the regime where the convergence is scaling exponentially with increasing model space dimensionality
because of the localized wavefunction in the Hilbert space, 
the HL-VQE algorithm is efficient, 
with the addition of a single qubit,
and associated state preparation resource requirements,
leading to an exponential improvement in ground-state energy and wavefunction fidelity, compared to traditional VQE.
This was demonstrated by classical calculations and 
by performing quantum simulations of 1- and 2-qubit systems using IBM's simulator and quantum computer, {\tt AER} and {\tt ibm\_lagos}, which reproduced the expected classical results within (small) uncertainties.
\\
\\
\indent
There are obvious connections and differences between this work and unitary coupled-clusters 
calculations of the LMG model.
The HL-VQE algorithm involves performing a unitary transformation on the states, 
mixing particle-hole states, which leads to a unitary-equivalent Hamiltonian.
In addition to limiting the form of the transformation, the Hilbert space is truncated and the Hamiltonian optimized in the truncated space.
To add some further context to the HL-VQE method, while the previous coupled-cluster 
simulations of systems of $N$ particles required $N$ qubits, the HL-VQE method requires $\log\Lambda$,
and so a LMG model
with $N \sim 100$, which is generically impractical today for the coupled-cluster VQE algorithms, 
can be simulated with exponential precision with just two or three qubits using HL-VQE. 
In some ways, HL-VQE makes a bridge between coupled-cluster expansions and EFT techniques.
It is conceivable that a combination of ADAPT-VQE and HL-VQE algorithms could also be useful due to their complementary nature.
Exploring and better understanding these connections, similarities and differences is the subject of future works.
\\
\\
\indent
This study of the Lipkin-Meshkov-Glick model has provided insights and demonstrations of the 
utility of Hamiltonian learning in effective model spaces.  
As this model exhibits symmetries of simple and well-structured interactions,
the question remains as to the utility beyond this model for more realistic 
effective interactions.  We do not address this question in this work,  
however, we find it likely that for systems where there are clear separations of scale(s),  
some or all of the methods developed and employed in this work can also be used, and will be the subject of future work.
An important element of this framework is that the optimal value of the variational parameter in the Hamiltonian is such that, in the transformed basis, the multi-particle-hole correlations are strongly localized around the mean-field (0p-0h) solution, and rapidly converge with increasing model space.
\\
\\
\indent
In general, the non-commutivity of the global orbital transformation with the truncation to effective model spaces is the feature that provides an opportunity for variational improvements through Hamiltonian learning.
This is one  unitary transformation that rearranges matrix elements and entanglement in nuclear many-body systems.
Another such unitary transformation is the SRG, in which a flow that, in its simplest form, introduces a Gaussian suppression of high momentum modes in the system.
Consequently, the Hamiltonian matrix is evolved toward a block diagonal form.
While somewhat dissimilar in form, the SRG flow and a truncation do not generally commute, and the utility 
for SRG-flowed systems remains to be explored.
Similarly, the HL-VQE algorithm, used together with a desired ansatz for the nuclear state, could be potentially utilized to perform various versions of In-Medium SRG (IMSRG) on a quantum computer. A further possibility that comes to mind, would be to use the variational orbital transformation, which reduces the coupling between effective model space and the remaining part of the Hilbert space \cite{Robin:2020aeh}, as pre-processing step before the IMSRG procedure.
\\
\\
\indent
The HL-VQE algorithm combined with the use of effective model spaces 
is likely to have applicability beyond nuclear many-body systems.
EFTs are a well-defined and systematic ways to recover S-matrix elements 
and bound states from a (more-)complete theory through operator and Hamiltonian matching 
or through reproducing experimental results.
As such, variationally optimizing non-perturbative leading-order contributions, 
when the one-body and higher-body interactions make comparable contributions,
though the HL-VQE algorithm may have applicability to EFT calculations of 
few-body systems.
It could also be the case that these methods will be useful in simulating 
quantum field theories, {\it e.g.} Refs.~\cite{Banuls:2019bmf,Klco:2021lap},
in particular, lattice scalar field theory and gauge theories.   
For example, 
local truncations in the representations of the SU($N_c$) in each link space could be mitigated
by improving the Kogut-Susskind Hamiltonian~\cite{PhysRevD.11.395} and learning the coefficients of 
gauge-invariant operators.  
A challenge for this application is to transform the states in the local Hilbert spaces in such a way to preserve Gauss's law at each vertex, {\it e.g.} Refs.~\cite{Pardo:2022hrp}, 
or to include protocols that constrain the wavefunction to the gauge-invariant space.
In addition,  the methods developed in this work may also be useful in designing quantum sensors,
where VQE is being utilized in encoding and decoding sensors to enhance sensitivity to external fields~\cite{PhysRevX.11.041045}.
These applications are the subject of future work.
\\
\\
\indent
The part of this paper that has focused on quantum simulation 
has addressed implementations using digital devices. 
With the rapid convergence with increasing dimension of the effective model space,
it is useful to consider implementations on other devices such as analog simulators and quantum annealers.
For the $\Lambda=2,4$ systems that we have considered, and for much larger effective model spaces,
analogous calculations could be performed using, for example, D-Wave's quantum annealers, such as {\tt Advantage}~\cite{DwaveSystemDocs}.  
This too will be the subject of future work.
\\
\\
\indent
In closing, with the approximate symmetries of QCD and the emergent 
separations of scales, which in some sense define nuclear physics, 
implementing the HL-VQE algorithm in low-energy effective model spaces 
using quantum computers may have utility for calculations involving realistic nuclear forces, 
and also for improving convergence of quantum simulations of quantum field theories relevant to the Standard Model.

\section*{Acknowledgments}

We would like to thank Anthony Ciavarella, Jon Engel and Marc Illa for helpful discussions,
and for all of our other 
colleagues and collaborators that provide the platform from which this work has emerged.
Caroline Robin would like to thank the Inqubator for Quantum Simulation (IQuS) at the University of Washington for kind hospitality during this work.
This work was supported, in part,  
by Universit\"at Bielefeld and ERC-885281-KILONOVA Advanced Grant,
and, in part, by U.S. Department of Energy, Office of Science, Office of Nuclear Physics, Inqubator for Quantum Simulation (IQuS) under Award Number DOE (NP) Award DE-SC0020970.
We acknowledge the use of IBM Quantum services for this work. 
The views expressed are those of the authors, 
and do not reflect the official policy or position of IBM or the IBM Quantum team.
It was enabled, in part, by
the use of advanced computational, storage and networking infrastructure provided by the Hyak supercomputer system at the University of Washington 
\footnote{\url{https://itconnect.uw.edu/research/hpc}},
and was also supported, in part, through the Department of Physics 
\footnote{\url{https://phys.washington.edu}}
and the College of Arts and Sciences 
\footnote{\url{https://www.artsci.washington.edu}}
at the University of Washington.

\clearpage
\appendix

\section{The Lipkin-Meshkov-Glick Model: More Details} 
\label{app:Lipkin_formalism}
\noindent
In this appendix, we provide more details about computing matrix elements of the 
effective Hamiltonian in the LMG model.
The  Hamiltonian 
can be written as
\begin{eqnarray}
\hat{H} &=& \varepsilon \, \hat{J}_z - \frac{V}{2} \left( \hat{J}_+^2 + \hat{J}_-^2 \right) \; ,
\label{eq:H_app}
\end{eqnarray}
where  the quasi-spin operators $\hat{\vec{J}}$ are defined in Eq.~(\ref{eq:Js}).
A rotation by an angle $\beta$ around the $y$-axis, 
\begin{eqnarray}
\hat{\vec{J}} & \equiv & \hat{\vec{J}}(\beta=0) \mapsto \hat{\vec{J}}(\beta) \; ,
\label{eq:Jequiv}
\end{eqnarray}
in quasi-spin space can be obtained from the
general transformation properties of rank-1 spherical tensors
\begin{eqnarray}
T^{(1)}_q (\beta) & = &  \hat U^\dagger (\beta)\ \hat T^{(1)}_q\ \hat U (\beta)
\ ,
\nonumber\\
& = & 
\sum_{q^\prime = 0,\pm}  d^{1}_{q q^\prime} (\beta) \ 
\hat T^{(1)}_{q^\prime} 
\ \ ,
\end{eqnarray}
where the spherical components are related via
$\hat{J}_z =  \hat T^{(1)}_0$,
$\hat{J}_\pm = \pm \sqrt{2} \hat T^{(1)}_\pm$,
and the expression for $d^{1}_{q q^\prime} (\beta)$ is given in Eq.~(\ref{eq:rotation_mat}).
This leads to
\begin{eqnarray}
\hat{J}_z &=& \cos(\beta) J_z(\beta)  + \frac{1}{2} \sin\beta \left( J_+(\beta)  + J_-(\beta)\right) \; , \\
\hat{J}_+ &=& \frac{1}{2} \Bigl[ - 2 \sin(\beta) J_z(\beta)  \nonumber \\
&& + \bigl( \cos\beta +1 \bigr) J_+(\beta)  + \bigl( \cos\beta -1 \bigr) J_-(\beta) \Bigr] \; , \\
\hat{J}_- &=& (J_+)^\dagger =  \frac{1}{2} \Bigl[ - 2\sin\beta J_z(\beta)  \nonumber \\ 
&& + \bigl( \cos\beta +1 \bigr) J_-(\beta) + \bigl( \cos\beta -1 \bigr) J_+(\beta) \Bigr] \; ,
\label{eq:Js_relations}
\end{eqnarray}
corresponding to
\begin{eqnarray}
\hat{J}_z(\beta) &=& \frac{1}{2} \sum_{p\sigma} \sigma c(\beta) ^\dagger_{p\sigma}c(\beta) _{p\sigma} \; , \\
\hat{J}_+(\beta)  &=&  \sum_{p}  c(\beta) ^\dagger_{p+} c(\beta) _{p-} \; , \\
\hat{J}_-(\beta)  &=& (\hat{J}_+)^\dagger =  \sum_{p} c(\beta) ^\dagger_{p-} c(\beta) _{p+} \; ,
\label{eq:Js_beta}
\end{eqnarray}
where the operators $c(\beta)$, $c(\beta)^\dagger$ 
are related to the $c$, $c^\dagger$  
via Eq.~(\ref{eq:transfo}). 
\\ \\
\indent
Inserting Eq.~(\ref{eq:Js_relations}) into Eq.~(\ref{eq:H_app}), 
the effective  Hamiltonian becomes
\begin{eqnarray}
\hat{H}(\beta) =&&  \varepsilon \; \left[ \cos\beta \hat{J}_z(\beta) + \frac{1}{2} \sin\beta \bigl( \hat{J}_+(\beta) + \hat{J}_-(\beta) \bigr) \right]  \nonumber \\
&& - \frac{V}{4} \Bigl[ \sin^2\beta \bigl( 4 \hat{J}_z(\beta)^2 - \{ \hat{J}_+(\beta), \hat{J}_-(\beta) \} \bigr) \nonumber \\
&& \hspace{1cm} + (1 + \cos^2\beta) \bigl(\hat{J}_+(\beta)^2 + \hat{J}_-(\beta)^2 \bigr) \nonumber \\
&& \hspace{1cm} - 2 \sin\beta \cos\beta \bigl( \{ \hat{J}_z(\beta),\hat{J}_+(\beta)\} \nonumber \\
&& \hspace{1cm} + \{ \hat{J}_z(\beta), \hat{J}_-(\beta) \} \bigr) \Bigr] \; .
\label{eq:eff_H_app}
\end{eqnarray}
Matrix elements of the Hamiltonian between the many-body basis states $\ket{ n ,\beta}$ are determined by 
{\footnotesize
\begin{eqnarray}
&&\braket{n',\beta | \hat{J}_z(\beta) | n , \beta} = M \delta_{n',n} \; ,
\nonumber \\
&&\braket{n',\beta | \hat{J}_\pm (\beta) | n , \beta} = \sqrt{J (J+1) - M (M \pm 1)} \delta_{n',n \pm 1} \; ,
\nonumber \\
&&\braket{n',\beta | \hat{J}_z(\beta)^2 | n , \beta} = M^2 \delta_{n',n} \; ,
\nonumber \\
&&\braket{n',\beta | \hat{J}_\pm(\beta)^2 | n , \beta} = \sqrt{J (J+1) - M (M \pm 1)} \nonumber \\
&& \hspace{2.5cm} \times \sqrt{J (J+1) - (M\pm 2) (M \pm 1)} \delta_{n',n\pm 2} \; ,
\nonumber \\ 
&&\braket{n',\beta | \{ \hat{J}_z(\beta), \hat{J}_\pm (\beta) \}  | n , \beta} = (2M \pm 1) \nonumber \\
&& \hspace{3.5cm} \times \sqrt{J (J+1) - M (M\pm1)}  \delta_{n', n \pm 1} \; ,
\nonumber \\
&&\braket{n',\beta | \{ \hat{J}_+(\beta), \hat{J}_- (\beta)\}  | n , \beta} = \left[ 2J(J+1) - 2 M^2 \right] \delta_{n' , n}   
\; ,
\nonumber \\
\label{app:Jmats}
\end{eqnarray}
}
where $J = N/2$ and $M = n-J = n - N/2$. 
This leads to
{\footnotesize
\begin{eqnarray} 
 && \braket{n',\beta | \hat{H}(\beta) | n,\beta }  = \nonumber \\
&&  \left(  \varepsilon \cos(\beta) (n-\frac{N}{2}) - \frac{V}{4} \sin^2(\beta) \left[ N^2 + 6n^2  -6 n N  -N\right] \right) \delta_{n',n}\nonumber \\
&& + \frac{1}{2} \sqrt{(N-n)(n+1)} \sin(\beta) \left[  \varepsilon - V  \cos(\beta) (N -2n-1) \right] \delta_{n',n+1} \nonumber \\
&& + \frac{1}{2} \sqrt{n(N-n+1)} \sin(\beta) \left[  \varepsilon - V  \cos(\beta) (N-2n+1) \right] \delta_{n',n-1} \nonumber \\
&& - \frac{V}{4} (1+\cos^2(\beta)) \sqrt{(N-n)(n+1)} \sqrt{(N-n-1)(n+2)}  \delta_{n',n+2} \nonumber \\
&& - \frac{V}{4} (1+\cos^2(\beta)) \sqrt{(N-n+2)(n-1)} \sqrt{n(N-n+1)} \delta_{n',n-2} 
\ .
\nonumber \\
\label{eq:H_ME}
\end{eqnarray}
}
When $\beta=0$, the Hamiltonian in Eq.~(\ref{eq:H_app}), 
which does not connect configurations characterized by $n'=n \pm 1$,
is recovered.

\section{Hamiltonian Learning - Variational Quantum Eigensolver (HL-VQE): More Details}
\label{app:HLVQE}
\noindent
In this appendix, we detail the hybrid algorithm that combines VQE with Hamiltonian 
learning to optimize
the efficacy of effective model spaces in recovering
the lowest-lying states of the LMG model, as described in the main text.
The Hamiltonian is uniquely defined with $\beta=0$ in the full model space, and the  
lowest-lying states can be found by using, for example, the VQE algorithm
with a quantum computer, or variants thereof, if the system lies beyond the capabilities of classical computing.
If available quantum resources are insufficient to accommodate the full Hamiltonian, 
then truncated systems can be implemented, 
but the Hamiltonian in the effective model space will include operators with coefficients
that must be determined 
in addition to the parameters defining the ground state, or low-lying excited state wavefunctions.
For a general system, we define the cost-function for the variation to be the expectation value of the energy,
which for the ground state is 
(with one variational parameter in the effective model space Hamiltonian)
\begin{eqnarray}
E(\beta, {\bm\theta})  
& = & 
\langle\Psi ({\bm\theta}) | \hat H(\beta) | \Psi ({\bm\theta}) \rangle
\ \ \ .
\label{eq:expEa}
\end{eqnarray}
We assume
that the functional dependence of 
$\hat H(\beta)$ on $\beta$ can be determined classically,
as is the case for the LMG model, and that
the derivative ${\partial\over\partial\beta} \hat H(\beta)$ can also be determined classically.
While not required, it is further assumed that the Pauli-decomposition of both can be readily determined classically.
The usual arguments motivating the use of VQE to determine matrix elements of Pauli-operators in a quantum many-body wavefunction hold in this case, and therefore a quantum computer enables evaluation of parts of
\begin{eqnarray}
E(\beta, {\bm\theta})  \ ,\ 
{\partial\over\partial\beta}  E(\beta, {\bm\theta})  \ ,\ 
{\partial\over\partial\theta_{i}} E(\beta, {\bm\theta}) 
\ \ \ .
\label{eq:grads}
\end{eqnarray}
Using gradient descent to learn the parameters that minimize 
$E(\beta, {\bm\theta})$, 
and starting from initial values, $\beta^{[k=0]}$ and $\theta_{i}^{[k=0]}$, 
sets of evaluations are used to determine 
updated values for the next iteration
\begin{eqnarray}
\beta^{[k+1]} & = & \beta^{[k]} - \eta {\partial\over\partial\beta}  E(\beta^{[k]}, {\bm\theta}^{[k]}) 
\ \ ,
\nonumber\\
\theta_{i}^{[k+1]} & = & \theta_{i}^{[k]} - \eta {\partial\over\partial\theta_{i}}  E(\beta^{[k]}, {\bm\theta}^{[k]}) 
\ \ \ ,
\label{eq:learni}
\end{eqnarray}
using a linear learning function,
with a learning rate $\eta$ 
iteratively tuned for optimal convergence.
If the starting points for the parameters are within the domain of attraction of the optimal value, this
system will iteratively converge to the optimal values.
This simultaneous iteration and variation of parameters identifies 
the optimal variational ground-state wavefunction and learns the Hamiltonian in the effective model space
by determining $\beta$.
This method generalizes straightforwardly to multiple parameters defining 
an effective Hamiltonian.

\subsection{1-Qubit Quantum Circuits ($\Lambda=2$ States)}
\label{app:HLVQEnQ1}
\noindent
An effective model space defined by one qubit is straightforward to study.
The $\beta$-transformation
described previously leads to an effective two-state Hamiltonian represented on one qubit that has a Pauli-decomposition,
\begin{eqnarray}
\hat H^{(2)}(\beta) & = & h^{(2)}_x(\beta)\  \hat X + h^{(2)}_y(\beta)\  \hat Y 
\nonumber\\
&& + h^{(2)}_z(\beta) \ \hat Z + h^{(2)}_I(\beta) \ \hat I 
\ \ .
\label{eq:h1gen}
\end{eqnarray}
Defining the (real) wavefunction through a rotation about the y-axis,
\begin{eqnarray}
|\Psi(\theta)\rangle  
& = &  \cos {\theta\over 2} |0\rangle \ + \  \sin {\theta\over 2} |1\rangle \; ,
\end{eqnarray}
matrix elements of the Pauli operators are
\begin{eqnarray}
\langle \hat X \rangle_\theta & = & \sin\theta ,\ 
\langle \hat Y \rangle_\theta  =  0 ,\ 
\langle \hat Z \rangle_\theta  =  \cos\theta ,\ 
\langle \hat I_2 \rangle_\theta  =  1 ,\ \nonumber \\
\label{eq:psi2}
\end{eqnarray}
where $\braket{\hat O}_{\theta} \equiv \braket{\Psi(\theta) | \hat O |\Psi(\theta)}$.
The computation of derivatives is straightforward using a finite-difference relation,
\begin{eqnarray}
{\partial\over\partial\theta} \langle \hat X \rangle_\theta
& = & {1\over 2}\left( \langle \hat X \rangle_{\theta+\pi/2} - \langle \hat X \rangle_{\theta-\pi/2} \right)
\nonumber\\
{\partial\over\partial\theta} \langle \hat Z \rangle_\theta
& = & {1\over 2}\left( \langle \hat Z \rangle_{\theta+\pi/2} - \langle \hat Z \rangle_{\theta-\pi/2} \right)
\ \ \ ,
\label{eq:Drivs}
\end{eqnarray}
and is directly implementable on a quantum computer with two different 
ensemble measurements per operator.
From these relations, the expectation value of the Hamiltonian is
\begin{eqnarray}
\langle\hat H^{(2)}(\beta)\rangle_\theta 
& = &  h^{(2)}_I(\beta)  +  h^{(2)}_x(\beta) \ \langle \hat X \rangle_\theta 
\nonumber\\
&& 
\qquad + h^{(2)}_z(\beta) \ \langle \hat Z \rangle_\theta 
\ ,
\nonumber\\
{\partial\over\partial\beta} \langle\hat H^{(2)}(\beta)\rangle_\theta 
& = & {\partial\over\partial\beta} h^{(2)}_I(\beta) + {\partial\over\partial\beta} h^{(2)}_x(\beta) \ \langle \hat X \rangle_\theta 
\nonumber\\
&& 
\qquad + {\partial\over\partial\beta} h^{(2)}_z(\beta) \ \langle \hat Z \rangle_\theta 
\ ,
\\
{\partial\over\partial\theta} 
\langle\hat H^{(2)}(\beta)\rangle_\theta 
& = & h^{(2)}_x(\beta) {\partial\over\partial\theta} \ \langle \hat X \rangle_\theta 
+ h^{(2)}_z(\beta) {\partial\over\partial\theta} \ \langle \hat Z \rangle_\theta 
\ \ .
\nonumber
\label{eq:h1genMAT}
\end{eqnarray}
A gradient-descent learning of $\beta, \theta$ involves iteratively applying
\begin{eqnarray}
\beta^{[k+1]} & = & \beta^{[k]} - \eta {\partial\over\partial\beta} \langle\hat H^{(2)}(\beta^{[k]})\rangle_{\theta^{[k]}}
\nonumber\\
\theta^{[k+1]} & = & \theta^{[k]} - \eta {\partial\over\partial\theta} \langle\hat H^{(2)}(\beta^{[k]})\rangle_{\theta^{[k]}}
\ \ \ ,
\label{eq:learn1}
\end{eqnarray}
until the parameters have reached stable values.
The value of the learned parameters are found from the mean and standard deviation~\footnote{
Alternately, half of the difference between the maximum and minimum values in that interval could be quoted.}
of the last $n_{\rm sel}$ parameters steps after the parameter sets have converged, where $n_{\rm sel}$ is typically taken to be 
$n_{\rm sel}\gtrsim 10$, and determined by the quantum resources.
The learning rate, $\eta$, is selected based upon the convergence behavior of the iteration. 
As is well known, a value that is too large leads to instability, and a value that is 
too small, leads to large resource requirements.
\\ \\
\indent
Quantum circuits are used to prepare ensembles of states 
$|\Psi(\theta^{[k]})\rangle$,
$|\Psi(\theta^{[k]}+{\pi\over 2})\rangle$,
$|\Psi(\theta^{[k]}-{\pi\over 2})\rangle$,
in which expectation values of 
$\hat X$ and $\hat Z$ are measured.
The latter is in the computational basis and requires no further operations 
before measurement, 
while the former requires an application of a Hadamard-gate.
With  classical computation of the 
$h^{(2)}_a(\beta^{[k]})$ and derivatives,
this is sufficient to implement the learning in Eq.~(\ref{eq:learn1}), which will have efficacy dictated, in part, by the statistical and systematic errors associated with the quantum computations.

\subsection{2-Qubit Quantum Circuits ($\Lambda=4$ States)}
\label{app:HLVQEnQ2}
\noindent
For an effective model space contained in two qubits, 
the protocol is analogous to that of one qubit, described above.
A real wavefunction on two qubits can 
 be prepared by the quantum circuit shown in 
Fig.~\ref{fig:QCprep12}
in terms of three angles $\theta_{0,1,2}$.
The 
$R_{ZX}(\theta)$-gate is available for the IBM's quantum computers, and we find it to be convenient for our purposes,
\begin{eqnarray}
R_{ZX}(\theta) & = & e^{-i \theta \hat X\otimes \hat Z /2}
\ \ \ .
\end{eqnarray}

The most general Hamiltonian for this system has the form, 
restricting ourselves again to one unknown parameter defining the Hamiltonian, $\beta$,
\begin{eqnarray}
H^{(4)}(\beta) & = & 
\sum_{i,j}^4 
h^{(4)}_{ij}(\beta) \ 
\overline{\sigma}_i\otimes\overline{\sigma}_j
\ \ \ ,
\label{eq:ham2pauli}
\end{eqnarray}
where
$\overline{\sigma}=\{ \hat X , \hat Y , \hat Z , \hat I \}$.
There are 10  
operator structures contributing to the Hamiltonian, 
with Pauli decomposition,
\begin{eqnarray}
\langle H^{(4)}(\beta) \rangle_{\bm\theta} & = & 
h^{(4)}_{II}(\beta) \ +\ 
h^{(4)}_{xx}(\beta) \ \langle \hat X\otimes \hat X \rangle_{\bm\theta} 
\nonumber
\\
& + & h^{(4)}_{xz}(\beta) \ \langle \hat X\otimes \hat Z \rangle_{\bm\theta}
 + h^{(4)}_{xI}(\beta) \ \langle \hat X\otimes \hat I \rangle_{\bm\theta}
\nonumber\\
& + & h^{(4)}_{yy}(\beta) \ \langle \hat Y\otimes \hat Y \rangle_{\bm\theta}
 + h^{(4)}_{zx}(\beta) \ \langle \hat Z\otimes \hat X \rangle_{\bm\theta}
\nonumber\\
& + & h^{(4)}_{zz}(\beta) \ \langle \hat Z\otimes \hat Z \rangle_{\bm\theta}
 + h^{(4)}_{zI}(\beta) \ \langle \hat Z\otimes \hat I \rangle_{\bm\theta}
\nonumber\\
& + & h^{(4)}_{Ix}(\beta) \ \langle \hat I\otimes \hat X \rangle_{\bm\theta}
 + h^{(4)}_{Iz}(\beta) \ \langle \hat I\otimes \hat Z \rangle_{\bm\theta}
\ \ \ , \nonumber \\
\label{eq:ham2pauliEXP}
\end{eqnarray}
with explicit expressions for the $h^{(2)}_{ab}(\beta)$ 
given in Eq.~(\ref{eq:ham2pauliALL2}), 
and with 
expectation values
in the 
wavefunction
in Eq.~(\ref{eq:psi12}),
\begin{eqnarray}
\langle \hat X\otimes \hat X \rangle_{\bm\theta} & = & 
\sin\theta_0 \ \sin \theta_2
\nonumber\\
\langle \hat X\otimes \hat Z \rangle_{\bm\theta} & = & 
\cos\theta_0 \cos\theta_1 \sin\theta_2 - 
\sin\theta_1 \cos\theta_2
\nonumber\\
\langle \hat X\otimes \hat I \rangle_{\bm\theta} & = & 
\cos\theta_1\sin\theta_2
- \cos\theta_0 \sin\theta_1 \cos\theta_2
\nonumber\\
\langle \hat Y\otimes \hat Y \rangle_{\bm\theta} & = & 
-\sin\theta_0 \sin\theta_1
\nonumber\\
\langle \hat Z\otimes \hat X \rangle_{\bm\theta} & = & 
\sin\theta_0 \cos\theta_2
\nonumber\\
\langle \hat Z\otimes \hat Z \rangle_{\bm\theta} & = & 
\cos\theta_0\cos\theta_1\cos\theta_2
+ \sin\theta_1\sin\theta_2
\nonumber\\
\langle \hat Z\otimes \hat I \rangle_{\bm\theta} & = & 
\cos\theta_0\sin\theta_1\sin\theta_2
+ \cos\theta_1\cos\theta_2
\nonumber\\
\langle \hat I\otimes \hat X \rangle_{\bm\theta} & = & 
\sin\theta_0 \cos\theta_1
\nonumber\\
\langle \hat I\otimes \hat Z \rangle_{\bm\theta} & = & 
\cos\theta_0 
\nonumber\\
\langle \hat I\otimes \hat I \rangle_{\bm\theta} & = & 
1
\ \ \ .
\label{eq:psi2mats}
\end{eqnarray}
\\ \\
\indent
Derivatives of $\langle H^{(4)}(\beta) \rangle_{\bm\theta}$ with respect to the $\theta_i$ 
($i=0,1,2$)
can be found from finite-differences, analogous to the single qubit case, {\it e.g.},
{\small
\begin{eqnarray}
&&{\partial\over\partial\theta_0} \langle \hat X\otimes \hat X \rangle_{\bm\theta} = \nonumber \\
&& \hspace{1cm} {1\over 2} 
\left(
\langle \hat X\otimes \hat X \rangle_{(\theta_0+{\pi\over 2},\theta_1,\theta_2)}
- 
\langle \hat X\otimes \hat X \rangle_{(\theta_0-{\pi\over 2},\theta_1,\theta_2)}
\right)
\ \ \ .
\nonumber\\
\label{eq:2Qder}
\end{eqnarray}
}
This relation can be used for all of the operators for $\theta_{0,1,2}$,
made possible by the structure of the quantum circuit in Fig.~\ref{fig:QCprep12},
where each angle only appears in one gate in the operator.
\\ \\
\indent
In our quantum simulations,
for a given set of $\theta_i$, 
the circuit in Fig.~\ref{fig:QCprep12} 
prepares the wavefunction on the 2-qubits.
This is performed for 7 quantum
circuits, with parameters
$\left( \theta_0, \theta_1, \theta_2 \right)$,
$\left( \theta_0\pm\pi/2, \theta_1, \theta_2\right)$,
$\left( \theta_0, \theta_1\pm \pi/2, \theta_2\right)$ and 
$\left( \theta_0, \theta_1, \theta_2\pm\pi/2\right)$.
Matrix elements of the operators 
$\hat Z\otimes \hat Z$, 
$\hat I\otimes \hat Z$ and 
$\hat Z\otimes \hat I$ 
can be constructed from 
combinations of measurements of the circuit in the computational basis. 
Matrix elements of operators 
$\hat X\otimes \hat X$, 
$\hat I\otimes \hat X$, 
$\hat X\otimes \hat I$, 
$\hat Z\otimes \hat X$ and 
$\hat X\otimes \hat Z$
can be found by adding a Hadamard gate, $\hat {\rm H}$, 
to the qubits with $\hat X$ operators, and then performing measurements in the computational basis.
For $\hat Y\otimes \hat Y$, adding $\hat S^\dagger$ followed by $\hat {\rm H}$ to both qubits
rotates the system into the computational basis.
If $\{ p_{00} , p_{11} , p_{01} , p_{10} \}$ are the probabilities of outcomes in the computational basis, 
then contracting with the vectors 
$\{1,1,-1,-1\}$ gives the matrix element of 
$\hat X\otimes \hat X$ 
or $\hat Y\otimes \hat Y$ 
or $\hat Z\otimes \hat Z$ 
or $\hat X\otimes \hat Z$ 
or $\hat Z\otimes \hat X$,
$\{1,-1,-1,1\}$ gives the matrix element of 
$\hat Z\otimes \hat I$ 
or $\hat X\otimes \hat I$,
and 
$\{1,-1,1,-1\}$ gives the matrix element of 
$\hat I\otimes \hat Z$ 
or $\hat I\otimes \hat X$.
\\ \\
\indent
The gradient-descent takes an analogous form to that for one qubit,
iterating on $\beta$ and $\theta_{0,1,2}$,
\begin{eqnarray}
\beta^{[k+1]} & = & \beta^{[k]} - \eta {\partial\over\partial\beta} \langle\hat H^{(4)}(\beta^{[k]} )\rangle_{\bm \theta^{[k]} }
\nonumber\\
\theta_{0}^{[k+1]} & = & \theta_{0}^{[k]}  - \eta {\partial\over\partial\theta_0} \langle\hat H^{(4)}(\beta^{[k]})\rangle_{\bm \theta^{[k]} }
\nonumber\\
\theta_{1}^{[k+1]} & = & \theta_{1}^{[k]}  - \eta {\partial\over\partial\theta_1} \langle\hat H^{(4)}(\beta^{[k]})\rangle_{\bm \theta^{[k]} }
\nonumber\\
\theta_{2}^{[k+1]} & = & \theta_{2}^{[k]}  - \eta {\partial\over\partial\theta_2} \langle\hat H^{(4)}(\beta^{[k]})\rangle_{\bm \theta^{[k]} }
\ \ \ .
\label{eq:learn2}
\end{eqnarray}
%

\section{Ground States in the Effective Model Spaces}
\label{app:GSforms}
\noindent
The ground states of the LMG model in the effective model spaces have a particular form for the parameters that we 
have chosen to work with.
In particular, the highest-lying odd p-h configuration has a vanishing amplitude independent of the size of the space.

\subsection{Vanishing of the 1p-1h Amplitude for $\Lambda=2$}
\label{app:GSforms1p1h}
\noindent
For the effective model space defined on a single qubit, and
for the parameters we are working with, 
the Hamiltonian in Eq.~(\ref{eq:H1q}) can be written as, for 
$\overline{v} 
= (N-1)V/\varepsilon >0$,
\begin{eqnarray}
    H^{(2)}(\beta) & = & 
    \left(
    \begin{array}{cc}
    h^{(2)}_{11} & h^{(2)}_{12} \\
h^{(2)}_{12} & h^{(2)}_{22} 
\end{array}
    \right)
  \; ,
  \nonumber\\
    h^{(2)}_{11} & = & -\frac{1}{4} N \varepsilon  \left(2 \cos (\beta )+
    |\overline{v}|
     \sin ^2(\beta  )\right)
  \; ,
        \nonumber\\
    h^{(2)}_{12} & = &-\frac{1}{2} \sqrt{N} \varepsilon  \sin (\beta  ) (|\overline{v}| \cos (\beta )-1)
  \; ,
     \nonumber\\
    h^{(2)}_{22} & = &h^{(1)}_{11}  
    + \varepsilon \left( \cos (\beta  )+\frac{3}{2} |\overline{v}| \sin ^2(\beta ) \right)
    \; ,
    \label{eq:1stab}
\end{eqnarray}
from which we see that
$h^{(2)}_{11},h^{(2)}_{22} \sim N $ and $h^{(2)}_{12} \sim\sqrt{N}$ and 
$h^{(2)}_{11}-h^{(2)}_{22} \sim 1 $.
The minimum eigenvalue of this system is obtained at a value of $\beta$ such that $h^{(2)}_{12}=0$. 
This is not so obvious from the matrix itself, but can 
be
found from a stability argument around this condition.
Writing $\beta = \cos^{-1}\left({1\over|\overline{v}| }\right) + \delta_\beta$, the eigenvalues are found to be 
\begin{eqnarray}
E_{\rm g.s.} & = & -\frac{N \left(|\overline{v}|^2+1\right) \varepsilon }{4 |\overline{v}| }
\ +\ 
\delta_\beta^2
\frac{N |\overline{v}| \left(|\overline{v}|^2+1\right) \varepsilon }{4(3 |\overline{v}|^2-1)}
\; ,
\nonumber\\
E_1 & = & E_{\rm g.s.} + \frac{\left(3 |\overline{v}|^2-1\right) \varepsilon }{2 |\overline{v}|}
\ -\ 2 \varepsilon \delta_\beta
    \ \ \ .
    \label{eq:1stabevals}
\end{eqnarray}
From these expressions, it is clear that for 
$\delta_\beta\sim 1/\sqrt{N}$, 
the eigenvalues do not invert and the extremum occurs at 
$\delta_\beta =0$.
Consequently, the eigenvector at the minimum is entirely 0p-0h configuration, corresponding to the HF wavefunction.
Numerically, this is found to be true for the explored parameter space even for small $N$.

\subsection{Vanishing of the 3p-3h Amplitude for $\Lambda=4$}
\label{app:GSforms3p3h}
\noindent
The structure of the Hamiltonian, particularly the degree of symmetry that renders the LMG-model solvable,  leads to ground states with some properties that are naively unexpected.
One of those properties is the vanishing of the 3p-3h amplitude in the 4-state effective model space, and generally the vanishing of the $(\Lambda-1)p-(\Lambda-1)h$ amplitude in the $\Lambda$ model space.
The origin of this can be seen by considering the $N$-scaling of the elements of the Hamiltonian, 
\begin{eqnarray}
H^{(4)}(\beta) &\sim &
\left(
\begin{array}{cccc}
N & \sqrt{N} & 1 & 0 \\
\sqrt{N} & N & \sqrt{N} & 1 \\
1 & \sqrt{N} & N & \sqrt{N}  \\
0 & 1 & \sqrt{N} & N 
\end{array}
\right)
\ \ \ ,
\label{eq:H4Nscale}
\end{eqnarray}
where the entries denote the maximum power of $N$.
The scaling of the diagonal entries remains the same after subtracting the identity contribution.
Choosing to diagonalize the upper-left $3\times 3$ block of this matrix 
with a transformation, $\mathcal{S}$,
\begin{eqnarray}
\mathcal{S} & = & 
  \left(\begin{array}{@{}c|c@{}}
    {\bf U} & {\bf 0}   \\  \hline
    {\bf 0} & 1 \\
   \end{array}\right)
   \; , \nonumber\\
\overline{H}^{(4)}(\beta) & = &  \mathcal{S}.H^{(4)}(\beta).\mathcal{S}^\dagger 
\ =\ 
  \left(\begin{array}{@{}c|c@{}}
    H^{(4)}_{3,{\rm diag}}(\beta) & \Gamma  \\  \hline
    \Gamma^\dagger & \delta \\
   \end{array}\right)
\  ,
\end{eqnarray}
where $ H^{(4)}_{3,{\rm diag}}(\beta) $ is a diagonal $3\times 3$ matrix, and $\delta~\sim N$.
Because of the hierarchy explicit in Eq.~(\ref{eq:H4Nscale}), 
the transformation ${\bf U}$ is of the form
${\bf U}\sim I_3 + \xi^a T^a$, where $\xi^a\sim 1/\sqrt{N}$ and $T^a$ are the 
generators of SU(3).
Thus, the elements of $\Gamma$ scale at most as $\Gamma\sim \sqrt{N}$, 
and the element of $\Gamma$ corresponding to the lowest eigenvalue, $E_{{\rm g.s.}, 3}$
scales attached  most as 
$\Gamma_1 \sim 1$ (compared to the diagonal element that scales as $\sim N$).
Returning to diagonalizing 
$\overline{H}^{(4)}(\beta)$, 
by considering perturbations in $1/N$, 
the lowest eigenvalue 
corresponding to an eigenvector of the form
${\bf d}^{(2)} = (1,0,0,\alpha)^T/\sqrt{1+|\alpha|^2}$,
is of the form,
\begin{eqnarray}
E^{(4)}_{\rm g.s.} & = & 
{1\over 1+|\alpha|^2} \left[\ E_{{\rm g.s.}, 3} + |\alpha|^2 \Gamma_1\right]
\ \ \ .
\end{eqnarray}
For $\Gamma_1 > E_{{\rm g.s.}, 3} $, which is manifest from the $N$ scaling,
the minimum energy is obtained for 
$\alpha=0$, corresponding to  
$E^{(4)}_{\rm g.s.}=E_{{\rm g.s.}, 3}$.
\\ \\
\indent
If we had assumed from the start that the amplitude of the 3p-3h configuration vanished, 
only
two parameters would have been required to parameterize the wavefunction. 
From the previous subsection, enforcing the vanishing of the 3p-3h configuration requires $\theta_1+\theta_2=0$.
As such, a more compact circuit could have been used with just two independent angles defining the wavefunction, 
or the circuit in Fig.~\ref{fig:QCprep12} can be used, with a modified gradient-descent.

\section{Comments on the Methods used in the Quantum Simulations using IBM's Quantum Computers}
\label{app:IBMresults}
\noindent
Standard techniques and {\tt qiskit}~\cite{matthew_treinish_2022_6924865} 
packages were used to produce results from 
IBM's QExperience~\cite{IBMQ}
{\tt AER} and {\tt ibm\_lagos}.
{\tt Python}~\cite{python3} and
{\tt jupyter} notebooks~\cite{PER-GRA:2007} were used as an interface
after 
developing
codes with {\tt Mathematica}~\cite{Mathematica}.
The Hamiltonian coefficients and derivatives were included as 
function calls,
as 
were
the state-preparation quantum circuits.
Measurement errors were mitigated using the standard packages in {\tt qiskit}~\cite{matthew_treinish_2022_6924865}, 
{\tt TensoredMeasFitter} and {\tt meas\_fitter},
but the errors associated with entangling operations and decoherence 
were not  mitigated as the implemented circuits are sufficiently shallow.
A higher precision simulation would require such mitigation.   
The quantum circuits were transpiled onto the qubits with the highest fidelity (as determined from the most recent calibration) consistent with not using swap gates,
for example, 
{\tt\small qc2t = transpile(qc2a,backend=backend,optimization\_level=3)}.
When multiple circuits were executed, the measurements related to one of the circuits were determined by tracing over the other qubits.
\\ \\
\indent
The workflow was implemented with an initialization of the IBMQ environment, 
functions and subroutines.
Gradient descent was accomplished with a {\tt for}-loop that generated the quantum circuits, 
submitted them to the device queue, 
post-processed the results to determine the energy, amplitudes and gradients, 
appended them to appropriate arrays,
determined updated values of the variational parameters, 
and then repeated until the iteration limit was exceeded.
\\ \\
\indent
The following code-snippets were used to determine the matrix elements of,
as an example,
$\hat X \hat X$
from the variational wavefunction in the 2-qubit simulations,
\scriptsize\begin{verbatim}
     # XX        
    qc0 = QuantumCircuit(nQ, nQ)
    qc0 = iniwaveRZX(qc0,theta0,theta1,theta2,cq0,cq1)
    qc0.h(cq0)
    qc0.h(cq1)
    qc0.measure([0,1,2,3],[0,1,2,3])
    for ntnt in range(nTwirl):
        qc0a = qc0
        qc0t = transpile(qc0a,backend=backend,optimization_level=3)    
        circ_list.append(qc0t)

    .
    .
    .
    .
    
    qc3 = QuantumCircuit(nQ, nQ)
    qc3 = iniwaveRZX(qc3,theta0,theta1,theta2+npihalf,cq0,cq1)
    qc3 = iniwaveRZX(qc3,theta0,theta1,theta2-npihalf,cq2,cq3)
    qc3.h(cq0)
    qc3.h(cq1)
    qc3.h(cq2)
    qc3.h(cq3)
    qc3.measure([0,1,2,3],[0,1,2,3])
    for ntnt in range(nTwirl):
        qc3a = qc3
        qc3t = transpile(qc3a,backend=backend,optimization_level=3)    
        circ_list.append(qc3t)
   
    \end{verbatim}
\normalsize using the function,
\scriptsize\begin{verbatim}
def iniwaveRZX(qc,ttt0,ttt1,ttt2,s0,s1):
    qc.ry(ttt0,s1)
    qc.s(s0)
    qc.rzx(ttt1,s1,s0)
    qc.sdg(s0)
    qc.ry(ttt2,s0)
    return qc
    \end{verbatim}
\normalsize corresponding to Fig.~\ref{fig:QCprep12}.

\section{Excited States}
\label{app:excited States}
\noindent
Finding the excited states in the effective model space is straightforward 
using chemical potentials,
as demonstrated in 
Ref.~\cite{Illa:2022jqb}.
A chemical potential can be given to the ground-state wavefunction
of sufficient size to move 
the state
high into the spectrum,
above the first excited state, and the ground state of this new system is the first excited state.
Adding a chemical potential, $\mu_0$
to Eq.~(\ref{eq:HPauli})
for the ground state $|\Psi\rangle^{(\Lambda)}$,
\begin{eqnarray}
\hat H^\prime({\beta}_0) & = & 
\hat H({\beta}_0) 
\ + \ \mu_0\  |\Psi\rangle^{(\Lambda)} {}^{(\Lambda)}\langle \Psi |
\ \ \ ,
\label{eq:HchemGS}
\end{eqnarray}
where $\beta = \beta_0$ is the value optimized for the ground state,
furnishes a Hamiltonian which can be used with the VQE algorithm (or others) to variationally find the first excited state.
Tracing against products of Pauli matrices,
$\overline{\sigma}_{i_1}\otimes\ ... \otimes \overline{\sigma}_{i_M}$,
allows this modified Hamiltonian to be written as 
\begin{eqnarray}
&& \hat H^\prime({\beta}_0) \ =\ 
\sum_{i_1,.., i_M =1}^4\ h^\prime_{i_i,.., i_M}({ \beta}_0) 
\ \overline{\sigma}_{i_1}\otimes\ ... \otimes \overline{\sigma}_{i_M}
\nonumber\\
&& h^\prime  \ =\ h 
\ +\ {\mu_0 \over 2^M}\ 
{}^{(\Lambda)}\langle \Psi | \overline{\sigma}_{i_1}\otimes\ ... \otimes \overline{\sigma}_{i_M} 
|\Psi\rangle^{(\Lambda)}
\ ,
\label{eq:HPauliPrime}
\end{eqnarray}
which can be evaluated straightforwardly as the matrix elements of the Pauli operators 
in the ground state have already been determined.
The ground state of $\hat H^\prime({\beta}_0)$ can be determined using VQE, thereby providing the 
energy and wavefunction of the first excited state in the effective model space.
Higher excited states can be determined by repeated application of this procedure.
As shown in Refs.~\cite{Illa:2022jqb} and \cite{Farrell:2022wyt}, the repeated application can lead to an accumulation of errors in the effective Hamiltonian for highly excited states.
\\ \\
\indent
We note that the procedure described above means that the same effective Hamiltonian determines the ground and excited states, which are orthonormal.
In principle, the Hamiltonian parameter could be optimized for each state, which would lead to non-orthogonal 
states, but would provide a better approximation to the exact states. 
From an EFT perspective, this corresponds to resumming contributions from an infinite number of 
higher-order operators differently for each state (with the difference vanishing for increasing size of model space).
It then becomes interesting to understand better the form of perturbation theory that could be used to systematically include residual (higher-order) terms in the Hamiltonian.
One can potentially force the orthogonality of the ground and excited states via the addition of constraints~\cite{doi:10.1080/00268978200100722}.
Another approach, which appears to be widely used in quantum chemistry, is to determine the orbitals via state-averaged variational calculations~\cite{https://doi.org/10.1002/qua.560200809,doi:10.1063/1.1667468}, so that the resulting single-particle basis contains information about both ground and excited states correlations.
The latter technique provides some sort of comprise as the orthonormality of the states is automatically fulfilled, at the price of potentially degrading the quality of each individual state.

\clearpage

\section{Tables of the Results Shown in the Figures}
\label{app:resultsTabs}
\noindent


%
\begin{longtable}{ccccc} 
\hline   
\\
& $\Lambda=2$  & & {\tt AER} & 
\\
\\
\hline
\\
Iteration & Energy & $|A_0|$ & $|A_1|$ & Bures Distance\\
 \\
\hline
 1 & -15.43 & 0.9879 & 0.1550 & 1.229 \\
 2 & -15.66 & 0.9869 & 0.1613 & 1.179 \\
 3 & -15.94 & 0.9858 & 0.1678 & 1.119 \\
 4 & -16.24 & 0.9846 & 0.1746 & 1.048 \\
 5 & -16.56 & 0.9834 & 0.1814 & 0.9676 \\
 6 & -16.91 & 0.9821 & 0.1882 & 0.8762 \\
 7 & -17.24 & 0.9808 & 0.1949 & 0.7749 \\
 8 & -17.57 & 0.9795 & 0.2014 & 0.6649 \\
 9 & -17.89 & 0.9783 & 0.2074 & 0.5487 \\
 10 & -18.15 & 0.9771 & 0.2128 & 0.4307 \\
 11 & -18.37 & 0.9761 & 0.2172 & 0.3200 \\
 12 & -18.55 & 0.9757 & 0.2191 & 0.2390 \\
 13 & -18.66 & 0.9769 & 0.2138 & 0.2249 \\
 14 & -18.69 & 0.9805 & 0.1965 & 0.2162 \\
 15 & -18.70 & 0.9831 & 0.1829 & 0.2143 \\
 16 & -18.71 & 0.9856 & 0.1691 & 0.2104 \\
 17 & -18.71 & 0.9876 & 0.1573 & 0.2084 \\
 18 & -18.72 & 0.9893 & 0.1458 & 0.2059 \\
 19 & -18.72 & 0.9908 & 0.1353 & 0.2041 \\
 20 & -18.73 & 0.9921 & 0.1252 & 0.2017 \\
 21 & -18.73 & 0.9932 & 0.1165 & 0.2010 \\
 22 & -18.73 & 0.9942 & 0.1078 & 0.1997 \\
 23 & -18.73 & 0.9950 & 0.09995 & 0.1984 \\
 24 & -18.74 & 0.9956 & 0.09333 & 0.1975 \\
 25 & -18.74 & 0.9962 & 0.08662 & 0.1973 \\
 26 & -18.74 & 0.9968 & 0.07996 & 0.1958 \\
 27 & -18.74 & 0.9972 & 0.07458 & 0.1953 \\
 28 & -18.74 & 0.9976 & 0.06978 & 0.1954 \\
 29 & -18.74 & 0.9979 & 0.06461 & 0.1954 \\
 30 & -18.74 & 0.9982 & 0.05996 & 0.1946 \\
 31 & -18.75 & 0.9985 & 0.05552 & 0.1939 \\
 32 & -18.75 & 0.9986 & 0.05197 & 0.1938 \\
 33 & -18.75 & 0.9988 & 0.04821 & 0.1933 \\
 34 & -18.75 & 0.9990 & 0.04518 & 0.1940 \\
 35 & -18.75 & 0.9991 & 0.04158 & 0.1934 \\
 36 & -18.75 & 0.9993 & 0.03870 & 0.1932 \\
 37 & -18.75 & 0.9993 & 0.03634 & 0.1931 \\
 38 & -18.75 & 0.9994 & 0.03370 & 0.1939 \\
 39 & -18.75 & 0.9995 & 0.03104 & 0.1927 \\
 40 & -18.75 & 0.9996 & 0.02922 & 0.1931 \\
 41 & -18.75 & 0.9996 & 0.02728 & 0.1928 \\
 42 & -18.75 & 0.9997 & 0.02535 & 0.1926 \\
 43 & -18.75 & 0.9997 & 0.02321 & 0.1927 \\
 44 & -18.75 & 0.9998 & 0.02172 & 0.1925 \\
 45 & -18.75 & 0.9998 & 0.02003 & 0.1925 \\
 46 & -18.75 & 0.9998 & 0.01867 & 0.1930 \\
 47 & -18.75 & 0.9999 & 0.01708 & 0.1921 \\
 48 & -18.75 & 0.9999 & 0.01595 & 0.1926 \\
 49 & -18.75 & 0.9999 & 0.01493 & 0.1925 \\
 50 & -18.75 & 0.9999 & 0.01389 & 0.1925 \\
 51 & -18.75 & 0.9999 & 0.01264 & 0.1928 \\
 52 & -18.75 & 0.9999 & 0.01129 & 0.1927 \\
 53 & -18.75 & 0.9999 & 0.01003 & 0.1922 \\
 54 & -18.75 & 1.000 & 0.009508 & 0.1926 \\
 55 & -18.75 & 1.000 & 0.008560 & 0.1921 \\
 56 & -18.75 & 1.000 & 0.007880 & 0.1928 \\
 57 & -18.75 & 1.000 & 0.007372 & 0.1921 \\
 58 & -18.75 & 1.000 & 0.007416 & 0.1925 \\
 59 & -18.75 & 1.000 & 0.006891 & 0.1921 \\
 60 & -18.75 & 1.000 & 0.006850 & 0.1930 \\
 61 & -18.75 & 1.000 & 0.006336 & 0.1921 \\
 62 & -18.75 & 1.000 & 0.006331 & 0.1927 \\
 63 & -18.75 & 1.000 & 0.005625 & 0.1927 \\
 64 & -18.75 & 1.000 & 0.005190 & 0.1927 \\
 65 & -18.75 & 1.000 & 0.005027 & 0.1919 \\
 66 & -18.75 & 1.000 & 0.005102 & 0.1926 \\
 67 & -18.75 & 1.000 & 0.004761 & 0.1928 \\
 68 & -18.75 & 1.000 & 0.004395 & 0.1928 \\
 69 & -18.75 & 1.000 & 0.003703 & 0.1922 \\
 70 & -18.75 & 1.000 & 0.003271 & 0.1927 \\
 71 & -18.75 & 1.000 & 0.002348 & 0.1918 \\
 72 & -18.75 & 1.000 & 0.002814 & 0.1932 \\
 73 & -18.75 & 1.000 & 0.002179 & 0.1924 \\
 74 & -18.75 & 1.000 & 0.001911 & 0.1928 \\
 75 & -18.75 & 1.000 & 0.001652 & 0.1926 \\
 76 & -18.75 & 1.000 & 0.001393 & 0.1926 \\
 77 & -18.75 & 1.000 & 0.001261 & 0.1925 \\
 78 & -18.75 & 1.000 & 0.001172 & 0.1925 \\
 79 & -18.75 & 1.000 & 0.0008452 & 0.1923 \\
 80 & -18.75 & 1.000 & 0.001096 & 0.1929 \\
\hline
\hline
\\
\caption{
Central values of  1-qubit quantum simulations obtained using {\tt AER} simulator for the 
ground-state energy (in unit of $\varepsilon$), wavefunction amplitudes and Bures Distance (Eq.~(\ref{eq:fidelityDB}))
for the $\Lambda=2$ effective model space using 100k shots per ensemble.
The LMG model parameters are:
$N=30$, $\varepsilon=1.0$, $\overline{v}=2.0$.
The starting values of variational parameters were:
$\beta^{[0]}= 0.2$ and $\theta_i^{[0]} = 0.30$.
}
\label{tab:AERLam1}
\end{longtable}
%


%
\begin{longtable}{ccccc}
\hline   
\\
& $\Lambda=2$  & & {\tt ibm\_lagos} & 
\\
\\
\hline
\\
Iteration & Energy & $|A_0|$ & $|A_1|$ & Bures Distance\\
 \\
\hline
 1 & -15.33 & 0.9982 & 0.06027 & 1.243 \\
 2 & -15.60 & 0.9975 & 0.07069 & 1.186 \\
 3 & -15.91 & 0.9967 & 0.08118 & 1.116 \\
 4 & -16.25 & 0.9958 & 0.09165 & 1.032 \\
 5 & -16.65 & 0.9948 & 0.1018 & 0.9335 \\
 6 & -17.06 & 0.9937 & 0.1120 & 0.8204 \\
 7 & -17.49 & 0.9925 & 0.1220 & 0.6944 \\
 8 & -17.81 & 0.9913 & 0.1314 & 0.5583 \\
 9 & -18.16 & 0.9901 & 0.1403 & 0.4179 \\
 10 & -18.46 & 0.9890 & 0.1482 & 0.2873 \\
 11 & -18.66 & 0.9881 & 0.1537 & 0.2083 \\
 12 & -18.72 & 0.9891 & 0.1471 & 0.2140 \\
 13 & -18.72 & 0.9914 & 0.1310 & 0.2011 \\
 14 & -18.71 & 0.9924 & 0.1228 & 0.2052 \\
 15 & -18.72 & 0.9938 & 0.1110 & 0.1986 \\
 16 & -18.73 & 0.9946 & 0.1034 & 0.2017 \\
 17 & -18.73 & 0.9956 & 0.09329 & 0.1971 \\
 18 & -18.74 & 0.9963 & 0.08597 & 0.1970 \\
 19 & -18.73 & 0.9969 & 0.07820 & 0.1961 \\
 20 & -18.75 & 0.9974 & 0.07203 & 0.1949 \\
 21 & -18.74 & 0.9978 & 0.06672 & 0.1951 \\
 22 & -18.74 & 0.9981 & 0.06094 & 0.1945 \\
 23 & -18.74 & 0.9984 & 0.05637 & 0.1949 \\
 24 & -18.76 & 0.9987 & 0.05183 & 0.1922 \\
 25 & -18.76 & 0.9988 & 0.04881 & 0.1923 \\
 26 & -18.75 & 0.9989 & 0.04652 & 0.1952 \\
 27 & -18.74 & 0.9991 & 0.04197 & 0.1935 \\
 28 & -18.75 & 0.9992 & 0.03876 & 0.1933 \\
 29 & -18.74 & 0.9994 & 0.03522 & 0.1922 \\
 30 & -18.74 & 0.9995 & 0.03311 & 0.1946 \\
 31 & -18.75 & 0.9996 & 0.02958 & 0.1922 \\
 32 & -18.75 & 0.9996 & 0.02833 & 0.1916 \\
 33 & -18.75 & 0.9996 & 0.02764 & 0.1955 \\
 34 & -18.75 & 0.9997 & 0.02354 & 0.1906 \\
 35 & -18.75 & 0.9997 & 0.02406 & 0.1903 \\
 36 & -18.75 & 0.9997 & 0.02501 & 0.1938 \\
 37 & -18.75 & 0.9998 & 0.02224 & 0.1907 \\
 38 & -18.75 & 0.9997 & 0.02339 & 0.1981 \\
 39 & -18.75 & 0.9998 & 0.01918 & 0.1923 \\
 40 & -18.74 & 0.9998 & 0.01891 & 0.1948 \\
 41 & -18.75 & 0.9999 & 0.01641 & 0.1922 \\
 42 & -18.75 & 0.9999 & 0.01604 & 0.1923 \\
 43 & -18.75 & 0.9999 & 0.01448 & 0.1926 \\
 44 & -18.75 & 0.9999 & 0.01274 & 0.1901 \\
 45 & -18.75 & 0.9999 & 0.01409 & 0.1934 \\
 46 & -18.75 & 0.9999 & 0.01251 & 0.1900 \\
 47 & -18.75 & 0.9999 & 0.01391 & 0.1954 \\
 48 & -18.75 & 0.9999 & 0.01204 & 0.1926 \\
 49 & -18.75 & 0.9999 & 0.01266 & 0.1935 \\
 50 & -18.75 & 0.9999 & 0.01233 & 0.1898 \\
 51 & -18.75 & 0.9999 & 0.01492 & 0.1902 \\
 52 & -18.75 & 0.9999 & 0.01574 & 0.1921 \\
 53 & -18.75 & 0.9999 & 0.01413 & 0.1922 \\
 54 & -18.74 & 0.9999 & 0.01312 & 0.1924 \\
 55 & -18.75 & 0.9999 & 0.01175 & 0.1924 \\
 56 & -18.75 & 0.9999 & 0.01091 & 0.1923 \\
 57 & -18.74 & 1.000 & 0.009768 & 0.1924 \\
 58 & -18.75 & 1.000 & 0.009526 & 0.1921 \\
 59 & -18.74 & 1.000 & 0.008957 & 0.1938 \\
 60 & -18.74 & 1.000 & 0.007376 & 0.1913 \\
 61 & -18.75 & 1.000 & 0.008537 & 0.1931 \\
 62 & -18.74 & 1.000 & 0.007399 & 0.1922 \\
 63 & -18.75 & 1.000 & 0.006512 & 0.1915 \\
 64 & -18.75 & 1.000 & 0.006302 & 0.1910 \\
 65 & -18.75 & 1.000 & 0.006617 & 0.1925 \\
 66 & -18.75 & 1.000 & 0.005911 & 0.1919 \\
 67 & -18.73 & 1.000 & 0.005819 & 0.1925 \\
 68 & -18.72 & 1.000 & 0.005404 & 0.1929 \\
 69 & -18.75 & 1.000 & 0.005169 & 0.1920 \\
 70 & -18.75 & 1.000 & 0.004952 & 0.1914 \\
 71 & -18.75 & 1.000 & 0.004762 & 0.1923 \\
 72 & -18.75 & 1.000 & 0.004237 & 0.1916 \\
 73 & -18.74 & 1.000 & 0.004602 & 0.1937 \\
 74 & -18.75 & 1.000 & 0.003239 & 0.1912 \\
 75 & -18.74 & 1.000 & 0.003477 & 0.1928 \\
 76 & -18.74 & 1.000 & 0.002555 & 0.1923 \\
 77 & -18.74 & 1.000 & 0.002776 & 0.1931 \\
 78 & -18.74 & 1.000 & 0.001708 & 0.1908 \\
 79 & -18.75 & 1.000 & 0.003603 & 0.1934 \\
 80 & -18.75 & 1.000 & 0.002328 & 0.1923 \\
\hline
\hline
\\
\caption{
Central values of  1-qubit quantum simulations obtained using {\tt ibm\_lagos} for the 
ground-state energy (in unit of $\varepsilon$), wavefunction amplitudes and 
Bures Distance (Eq.~(\ref{eq:fidelityDB}))
for the $\Lambda=2$ effective model space using 32k shots per ensemble.
The LMG model parameters are:
$N=30$, $\varepsilon=1.0$, $\overline{v}=2.0$.
The starting values of variational parameters were:
$\beta^{[0]}= 0.2$ and $\theta_i^{[0]} = 0.10$
}
\label{tab:ibmlagosLam1}
\end{longtable}
%


%
\begin{longtable*}{ccccccc} 
\hline   
\\
& & & $\Lambda=4$  & & {\tt AER} & 
\\
\\
\hline
\\
Iteration & Energy & $|A_0|$ & $|A_1|$& $|A_2|$ & $|A_3|$ & Bures Distance\\
 \\
\hline
1 & -17.98 & 0.9957 & 0.05494 & -0.07397 & 1.933$\times 10^{-6}$ & 0.4320 \\
 2 & -18.32 & 0.9972 & 0.05962 & -0.04422 & 7.771$\times 10^{-6}$ & 0.2953 \\
 3 & -18.57 & 0.9980 & 0.06343 & -0.006676 & 0.00001920 & 0.2031 \\
 4 & -18.72 & 0.9969 & 0.06415 & 0.04478 & 0.00004183 & 0.1606 \\
 5 & -18.81 & 0.9935 & 0.05956 & 0.09709 & 0.00004846 & 0.1048 \\
 6 & -18.87 & 0.9900 & 0.05837 & 0.1283 & 0.00006483 & 0.08426 \\
 7 & -18.89 & 0.9879 & 0.05568 & 0.1449 & 0.00007499 & 0.06657 \\
 8 & -18.90 & 0.9862 & 0.05564 & 0.1557 & 0.00008579 & 0.06405 \\
 9 & -18.90 & 0.9855 & 0.05388 & 0.1607 & 0.00008767 & 0.05761 \\
 10 & -18.90 & 0.9849 & 0.05401 & 0.1645 & 0.00009111 & 0.05843 \\
 11 & -18.90 & 0.9847 & 0.05283 & 0.1658 & 0.00009575 & 0.05548 \\
 12 & -18.90 & 0.9845 & 0.05294 & 0.1672 & 0.0001050 & 0.05667 \\
 13 & -18.90 & 0.9845 & 0.05202 & 0.1675 & 0.0001009 & 0.05489 \\
 14 & -18.90 & 0.9844 & 0.05202 & 0.1680 & 0.0001081 & 0.05598 \\
 15 & -18.90 & 0.9845 & 0.05128 & 0.1678 & 0.0001130 & 0.05491 \\
 16 & -18.90 & 0.9844 & 0.05118 & 0.1682 & 0.0001217 & 0.05545 \\
 17 & -18.90 & 0.9845 & 0.05062 & 0.1681 & 0.0001216 & 0.05497 \\
 18 & -18.90 & 0.9844 & 0.05053 & 0.1683 & 0.0001279 & 0.05529 \\
 19 & -18.90 & 0.9845 & 0.05015 & 0.1681 & 0.0001343 & 0.05515 \\
 20 & -18.90 & 0.9845 & 0.04993 & 0.1682 & 0.0001389 & 0.05536 \\
 21 & -18.90 & 0.9845 & 0.04951 & 0.1681 & 0.0001400 & 0.05504 \\
 22 & -18.90 & 0.9845 & 0.04935 & 0.1682 & 0.0001538 & 0.05530 \\
 23 & -18.90 & 0.9846 & 0.04893 & 0.1680 & 0.0001561 & 0.05512 \\
 24 & -18.90 & 0.9846 & 0.04878 & 0.1681 & 0.0001551 & 0.05535 \\
 25 & -18.90 & 0.9846 & 0.04847 & 0.1680 & 0.0001555 & 0.05501 \\
 26 & -18.90 & 0.9846 & 0.04836 & 0.1681 & 0.0001546 & 0.05535 \\
 27 & -18.90 & 0.9846 & 0.04810 & 0.1681 & 0.0001561 & 0.05524 \\
 28 & -18.90 & 0.9846 & 0.04785 & 0.1681 & 0.0001623 & 0.05517 \\
 29 & -18.90 & 0.9846 & 0.04764 & 0.1680 & 0.0001607 & 0.05515 \\
 30 & -18.90 & 0.9846 & 0.04751 & 0.1681 & 0.0001686 & 0.05526 \\
 31 & -18.90 & 0.9847 & 0.04735 & 0.1679 & 0.0001703 & 0.05531 \\
 32 & -18.90 & 0.9847 & 0.04711 & 0.1680 & 0.0001756 & 0.05538 \\
 33 & -18.90 & 0.9847 & 0.04687 & 0.1679 & 0.0001738 & 0.05518 \\
 34 & -18.90 & 0.9847 & 0.04668 & 0.1679 & 0.0001804 & 0.05553 \\
 35 & -18.90 & 0.9847 & 0.04637 & 0.1678 & 0.0001792 & 0.05518 \\
 36 & -18.90 & 0.9847 & 0.04635 & 0.1679 & 0.0001825 & 0.05551 \\
 37 & -18.90 & 0.9848 & 0.04608 & 0.1678 & 0.0001847 & 0.05525 \\
 38 & -18.90 & 0.9847 & 0.04592 & 0.1678 & 0.0001865 & 0.05548 \\
 39 & -18.90 & 0.9848 & 0.04575 & 0.1677 & 0.0001945 & 0.05528 \\
 40 & -18.90 & 0.9848 & 0.04561 & 0.1677 & 0.0002003 & 0.05574 \\
 41 & -18.90 & 0.9848 & 0.04527 & 0.1676 & 0.0001977 & 0.05524 \\
 42 & -18.90 & 0.9848 & 0.04521 & 0.1678 & 0.0001995 & 0.05558 \\
 43 & -18.90 & 0.9848 & 0.04501 & 0.1676 & 0.0002030 & 0.05538 \\
 44 & -18.90 & 0.9848 & 0.04480 & 0.1676 & 0.0002015 & 0.05558 \\
 45 & -18.90 & 0.9849 & 0.04461 & 0.1675 & 0.0001946 & 0.05550 \\
 46 & -18.90 & 0.9848 & 0.04447 & 0.1677 & 0.0001984 & 0.05537 \\
 47 & -18.90 & 0.9848 & 0.04437 & 0.1677 & 0.0002005 & 0.05539 \\
 48 & -18.90 & 0.9849 & 0.04425 & 0.1677 & 0.0001999 & 0.05553 \\
 49 & -18.90 & 0.9849 & 0.04412 & 0.1677 & 0.0001998 & 0.05540 \\
 50 & -18.90 & 0.9849 & 0.04406 & 0.1676 & 0.0001998 & 0.05561 \\
 51 & -18.90 & 0.9849 & 0.04388 & 0.1674 & 0.0001987 & 0.05559 \\
 52 & -18.90 & 0.9849 & 0.04372 & 0.1675 & 0.0001946 & 0.05546 \\
 53 & -18.90 & 0.9849 & 0.04361 & 0.1675 & 0.0001972 & 0.05569 \\
 54 & -18.90 & 0.9849 & 0.04345 & 0.1675 & 0.0001945 & 0.05544 \\
 55 & -18.90 & 0.9849 & 0.04345 & 0.1676 & 0.0001951 & 0.05558 \\
 56 & -18.90 & 0.9849 & 0.04338 & 0.1676 & 0.0001980 & 0.05552 \\
 57 & -18.90 & 0.9849 & 0.04328 & 0.1676 & 0.0002018 & 0.05555 \\
 58 & -18.90 & 0.9849 & 0.04319 & 0.1675 & 0.0002033 & 0.05551 \\
 59 & -18.90 & 0.9849 & 0.04311 & 0.1674 & 0.0002049 & 0.05579 \\
 60 & -18.90 & 0.9850 & 0.04290 & 0.1673 & 0.0002034 & 0.05547 \\
 61 & -18.90 & 0.9849 & 0.04286 & 0.1675 & 0.0002040 & 0.05571 \\
 62 & -18.90 & 0.9850 & 0.04267 & 0.1674 & 0.0002031 & 0.05540 \\
 63 & -18.90 & 0.9850 & 0.04274 & 0.1674 & 0.0002060 & 0.05598 \\
 64 & -18.90 & 0.9850 & 0.04253 & 0.1673 & 0.0002066 & 0.05547 \\
 65 & -18.90 & 0.9850 & 0.04253 & 0.1675 & 0.0002075 & 0.05564 \\
 66 & -18.90 & 0.9850 & 0.04244 & 0.1674 & 0.0002078 & 0.05553 \\
 67 & -18.90 & 0.9850 & 0.04241 & 0.1675 & 0.0002079 & 0.05578 \\
 68 & -18.90 & 0.9850 & 0.04223 & 0.1675 & 0.0002052 & 0.05553 \\
 69 & -18.90 & 0.9850 & 0.04225 & 0.1675 & 0.0002076 & 0.05558 \\
 70 & -18.90 & 0.9850 & 0.04221 & 0.1675 & 0.0002079 & 0.05548 \\
 71 & -18.90 & 0.9850 & 0.04218 & 0.1675 & 0.0002032 & 0.05561 \\
 72 & -18.90 & 0.9850 & 0.04211 & 0.1675 & 0.0002007 & 0.05561 \\
 73 & -18.90 & 0.9850 & 0.04205 & 0.1673 & 0.0001984 & 0.05560 \\
 74 & -18.90 & 0.9850 & 0.04202 & 0.1673 & 0.0002026 & 0.05565 \\
 75 & -18.90 & 0.9850 & 0.04196 & 0.1674 & 0.0002035 & 0.05553 \\
 76 & -18.90 & 0.9850 & 0.04191 & 0.1674 & 0.0002033 & 0.05559 \\
 77 & -18.90 & 0.9850 & 0.04180 & 0.1674 & 0.0002079 & 0.05553 \\
 78 & -18.90 & 0.9850 & 0.04183 & 0.1673 & 0.0002082 & 0.05565 \\
 79 & -18.90 & 0.9850 & 0.04173 & 0.1674 & 0.0002081 & 0.05555 \\
 80 & -18.90 & 0.9850 & 0.04171 & 0.1674 & 0.0002119 & 0.05571 \\
\hline
\hline
\\
\caption{
Central values of 2-qubit quantum simulations obtained using {\tt AER} simulator for the 
ground-state energy (in unit of $\varepsilon$), wavefunction amplitudes and Bures Distance (Eq.~(\ref{eq:fidelityDB}))
for the $\Lambda=4$ effective model space using 100k shots per ensemble.
The LMG model parameters are:
$N=30$, $\varepsilon=1.0$, $\overline{v}=2.0$.
The starting values of variational parameters were:
$\beta^{[0]}= 0.2$ and $\theta_i^{[0]} = 0.0$
}
\label{tab:AERLam4}
\end{longtable*}
%


%
\begin{longtable*}{ccccccc} 
\hline   
\\
& & & $\Lambda=4$  & & {\tt ibm\_lagos} & 
\\
\\
\hline
\\
Iteration & Energy & $|A_0|$ & $|A_1|$& $|A_2|$ & $|A_3|$ & Bures Distance\\
 \\
\hline
 1 & -15.32 & 0.9999 & 0.006064 & 0.01473 & 3.807$\times 10^{-6}$ & 1.256 \\
 2 & -15.55 & 0.9996 & 0.01218 & 0.02589 & 0.00001039 & 1.203 \\
 3 & -15.90 & 0.9992 & 0.01849 & 0.03509 & 0.00001715 & 1.138 \\
 4 & -16.24 & 0.9987 & 0.02476 & 0.04357 & 0.00001987 & 1.061 \\
 5 & -16.54 & 0.9982 & 0.03083 & 0.05049 & 0.00002447 & 0.9690 \\
 6 & -16.96 & 0.9977 & 0.03698 & 0.05680 & 0.00003040 & 0.8638 \\
 7 & -17.35 & 0.9971 & 0.04354 & 0.06307 & 0.00003385 & 0.7457 \\
 8 & -17.76 & 0.9963 & 0.05009 & 0.06948 & 0.00005767 & 0.6161 \\
 9 & -18.05 & 0.9956 & 0.05601 & 0.07560 & 0.00009977 & 0.4770 \\
 10 & -18.42 & 0.9947 & 0.06213 & 0.08194 & 0.0001186 & 0.3326 \\
 11 & -18.63 & 0.9938 & 0.06759 & 0.08814 & 0.0001753 & 0.1916 \\
 12 & -18.79 & 0.9925 & 0.07196 & 0.09833 & 0.0002948 & 0.1013 \\
 13 & -18.89 & 0.9914 & 0.07163 & 0.1097 & 0.0003181 & 0.09405 \\
 14 & -18.87 & 0.9907 & 0.07054 & 0.1160 & 0.0003484 & 0.09040 \\
 15 & -18.91 & 0.9902 & 0.06760 & 0.1219 & 0.0004992 & 0.08092 \\
 16 & -18.86 & 0.9894 & 0.06633 & 0.1290 & 0.0006422 & 0.07533 \\
 17 & -18.88 & 0.9888 & 0.06595 & 0.1342 & 0.0006689 & 0.07526 \\
 18 & -18.88 & 0.9887 & 0.06438 & 0.1353 & 0.0007022 & 0.07114 \\
 19 & -18.90 & 0.9878 & 0.06406 & 0.1418 & 0.0007734 & 0.06634 \\
 20 & -18.90 & 0.9871 & 0.06346 & 0.1471 & 0.0007934 & 0.06667 \\
 21 & -18.89 & 0.9875 & 0.06104 & 0.1456 & 0.0007279 & 0.06377 \\
 22 & -18.88 & 0.9874 & 0.06016 & 0.1461 & 0.0007626 & 0.06292 \\
 23 & -18.88 & 0.9871 & 0.06019 & 0.1485 & 0.0009840 & 0.06102 \\
 24 & -18.97 & 0.9860 & 0.06244 & 0.1543 & 0.001076 & 0.05763 \\
 25 & -18.96 & 0.9856 & 0.06469 & 0.1563 & 0.001311 & 0.05889 \\
 26 & -18.93 & 0.9854 & 0.06447 & 0.1573 & 0.001550 & 0.05686 \\
 27 & -18.98 & 0.9855 & 0.06487 & 0.1568 & 0.001828 & 0.05821 \\
 28 & -18.97 & 0.9852 & 0.06480 & 0.1587 & 0.002056 & 0.05782 \\
 29 & -18.94 & 0.9852 & 0.06438 & 0.1588 & 0.002052 & 0.05624 \\
 30 & -18.89 & 0.9848 & 0.06371 & 0.1614 & 0.002102 & 0.05383 \\
 31 & -18.84 & 0.9849 & 0.06574 & 0.1604 & 0.002289 & 0.05740 \\
 32 & -18.91 & 0.9853 & 0.06504 & 0.1582 & 0.002322 & 0.05684 \\
 33 & -18.94 & 0.9855 & 0.06414 & 0.1568 & 0.002349 & 0.05940 \\
 34 & -18.96 & 0.9856 & 0.06216 & 0.1569 & 0.002225 & 0.05648 \\
 35 & -18.95 & 0.9851 & 0.06419 & 0.1597 & 0.002649 & 0.06830 \\
 36 & -18.95 & 0.9849 & 0.06237 & 0.1614 & 0.002799 & 0.05596 \\
 37 & -18.89 & 0.9845 & 0.06350 & 0.1633 & 0.002843 & 0.05919 \\
 38 & -18.87 & 0.9849 & 0.06108 & 0.1620 & 0.002706 & 0.05396 \\
 39 & -18.84 & 0.9847 & 0.06264 & 0.1627 & 0.002848 & 0.05645 \\
 40 & -18.94 & 0.9851 & 0.06052 & 0.1612 & 0.002817 & 0.05367 \\
 41 & -18.92 & 0.9851 & 0.06204 & 0.1607 & 0.002910 & 0.05809 \\
 42 & -18.90 & 0.9856 & 0.05983 & 0.1583 & 0.002756 & 0.05516 \\
 43 & -18.96 & 0.9855 & 0.06083 & 0.1581 & 0.002885 & 0.05836 \\
 44 & -18.94 & 0.9854 & 0.06041 & 0.1590 & 0.002891 & 0.05584 \\
 45 & -18.95 & 0.9858 & 0.06025 & 0.1568 & 0.002908 & 0.06455 \\
 46 & -18.90 & 0.9861 & 0.05771 & 0.1560 & 0.002773 & 0.05849 \\
 47 & -18.91 & 0.9861 & 0.05807 & 0.1556 & 0.002822 & 0.06921 \\
 48 & -18.91 & 0.9865 & 0.05494 & 0.1541 & 0.002727 & 0.05952 \\
 49 & -18.95 & 0.9858 & 0.05695 & 0.1578 & 0.002868 & 0.05873 \\
 50 & -18.92 & 0.9860 & 0.05616 & 0.1569 & 0.002755 & 0.05573 \\
 51 & -18.93 & 0.9861 & 0.05624 & 0.1565 & 0.002733 & 0.06232 \\
 52 & -18.91 & 0.9861 & 0.05514 & 0.1566 & 0.002694 & 0.05688 \\
 53 & -18.89 & 0.9864 & 0.05450 & 0.1552 & 0.002717 & 0.05817 \\
 54 & -18.87 & 0.9861 & 0.05481 & 0.1570 & 0.002853 & 0.05565 \\
 55 & -18.90 & 0.9861 & 0.05484 & 0.1568 & 0.002843 & 0.05866 \\
 56 & -18.93 & 0.9866 & 0.05316 & 0.1541 & 0.002646 & 0.05978 \\
 57 & -18.92 & 0.9866 & 0.05261 & 0.1543 & 0.002559 & 0.05728 \\
 58 & -18.92 & 0.9866 & 0.05189 & 0.1545 & 0.002451 & 0.05736 \\
 59 & -18.92 & 0.9867 & 0.05059 & 0.1546 & 0.002325 & 0.06070 \\
 60 & -18.60 & 0.9868 & 0.05234 & 0.1531 & 0.002831 & 0.06481 \\
 61 & -18.90 & 0.9864 & 0.05136 & 0.1562 & 0.002656 & 0.06291 \\
 62 & -18.97 & 0.9861 & 0.04991 & 0.1583 & 0.002647 & 0.05492 \\
 63 & -18.94 & 0.9861 & 0.04994 & 0.1582 & 0.002713 & 0.06368 \\
 64 & -18.97 & 0.9867 & 0.04787 & 0.1553 & 0.002674 & 0.05649 \\
 65 & -18.97 & 0.9862 & 0.04959 & 0.1578 & 0.002785 & 0.06491 \\
 66 & -18.96 & 0.9865 & 0.04724 & 0.1567 & 0.002695 & 0.05571 \\
 67 & -18.98 & 0.9868 & 0.04627 & 0.1554 & 0.002636 & 0.05992 \\
 68 & -18.95 & 0.9864 & 0.04527 & 0.1577 & 0.002587 & 0.05517 \\
 69 & -18.97 & 0.9864 & 0.04607 & 0.1577 & 0.002678 & 0.05925 \\
 70 & -18.96 & 0.9867 & 0.04401 & 0.1567 & 0.002654 & 0.05776 \\
 71 & -18.93 & 0.9870 & 0.04284 & 0.1550 & 0.002585 & 0.05663 \\
 72 & -18.93 & 0.9865 & 0.04322 & 0.1579 & 0.002611 & 0.05534 \\
 73 & -18.91 & 0.9869 & 0.04026 & 0.1563 & 0.002455 & 0.05618 \\
 74 & -18.94 & 0.9870 & 0.03932 & 0.1559 & 0.002338 & 0.05738 \\
 75 & -18.97 & 0.9870 & 0.03816 & 0.1559 & 0.002348 & 0.05688 \\
 76 & -18.88 & 0.9872 & 0.03828 & 0.1547 & 0.002434 & 0.05698 \\
 77 & -18.91 & 0.9872 & 0.03856 & 0.1547 & 0.002462 & 0.05768 \\
 78 & -18.93 & 0.9874 & 0.03741 & 0.1539 & 0.002335 & 0.05807 \\
 79 & -18.93 & 0.9876 & 0.03743 & 0.1527 & 0.002344 & 0.06376 \\
 80 & -18.92 & 0.9878 & 0.03655 & 0.1515 & 0.002298 & 0.06062 \\
\hline
\hline
\\
\caption{
Central values of 2-qubit quantum simulations obtained using {\tt ibm\_lagos} for the 
ground-state energy (in unit of $\varepsilon$), wavefunction amplitudes and Bures Distance (Eq.~(\ref{eq:fidelityDB}))
for the $\Lambda=4$ effective model space using 32k shots per ensemble.
The LMG model parameters are:
$N=30$, $\varepsilon=1.0$, $\overline{v}=2.0$.
The starting values of variational parameters were:
$\beta^{[0]}= 0.2$ and $\theta_i^{[0]} = 0.0$
}
\label{tab:ibmlagosLam4}
\end{longtable*}
%


%
\begin{longtable*}{cccccc} 
\hline   
\\
 &  &  & Full-Space Wavefunctions (Unprojected) & &  
\\
\\
\hline
\\
State, $n$  
& Exact 
& {\tt AER} $\Lambda=2$ 
& {\tt ibm\_lagos} $\Lambda=2$ 
& {\tt AER} $\Lambda=4$
& {\tt ibm\_lagos} $\Lambda=4$ 
\\
 \\
\hline
1 & 0.064046 & \text{0.01337(15)} & \text{0.0139(16)} & \text{0.03390(62)} & \text{0.0321(33)} \\
 2 & 0.11405 & \text{0.04229(42)} & \text{0.0437(44)} & \text{0.0881(13)} & \text{0.0843(70)} \\
 3 & 0.17438 & \text{0.09298(78)} & \text{0.0957(82)} & \text{0.1594(19)} & \text{0.1541(99)} \\
 4 & 0.23289 & \text{0.1639(11)} & \text{0.167(11)} & \text{0.2330(20)} & \text{0.227(11)} \\
 5 & 0.28608 & \text{0.2459(13)} & \text{0.250(13)} & \text{0.2942(18)} & \text{0.290(10)} \\
 6 & 0.32693 & \text{0.3237(12)} & \text{0.327(12)} & \text{0.3348(12)} & \text{0.3344(76)} \\
 7 & 0.35138 & \text{0.38146(87)} & \text{0.3842(87)} & \text{0.35422(66)} & \text{0.3567(46)} \\
 8 & 0.35860 & \text{0.40774(31)} & \text{0.4085(28)} & \text{0.35635(27)} & \text{0.3605(18)} \\
 9 & 0.34752 & \text{0.39910(30)} & \text{0.3978(35)} & \text{0.34535(43)} & \text{0.3495(17)} \\
 10 & 0.32276 & \text{0.36020(82)} & \text{0.3572(87)} & \text{0.32343(84)} & \text{0.3261(46)} \\
 11 & 0.28598 & \text{0.3013(11)} & \text{0.297(11)} & \text{0.2914(12)} & \text{0.2922(75)} \\
 12 & 0.24395 & \text{0.2345(12)} & \text{0.230(12)} & \text{0.2508(15)} & \text{0.2498(98)} \\
 13 & 0.19892 & \text{0.1703(11)} & \text{0.166(11)} & \text{0.2045(17)} & \text{0.202(11)} \\
 14 & 0.15647 & \text{0.11572(97)} & \text{0.1125(98)} & \text{0.1570(17)} & \text{0.154(10)} \\
 15 & 0.11767 & \text{0.07361(73)} & \text{0.0712(73)} & \text{0.1130(15)} & \text{0.1107(97)} \\
 16 & 0.085371 & \text{0.04389(50)} & \text{0.0422(50)} & \text{0.0760(12)} & \text{0.0743(78)} \\
 17 & 0.059140 & \text{0.02453(31)} & \text{0.0235(31)} & \text{0.04778(93)} & \text{0.0465(57)} \\
 18 & 0.039465 & \text{0.01285(18)} & \text{0.0122(18)} & \text{0.02796(63)} & \text{0.0272(38)} \\
 19 & 0.025074 & \text{0.00630(10)} & \text{0.00599(99)} & \text{0.01522(38)} & \text{0.0148(23)} \\
 20 & 0.015295 & \text{0.002892(51)} & \text{0.00273(49)} & \text{0.00770(22)} & \text{0.0074(13)} \\
 21 & 0.0088408 & \text{0.001238(23)} & \text{0.00116(23)} & \text{0.00361(11)} & \text{0.00351(67)} \\
 22 & 0.0048782 & \text{0.000493(10)} & \text{0.000463(98)} & \text{0.001569(54)} & \text{0.00152(31)} \\
 23 & 0.0025309 & \text{0.0001821(40)} & \text{0.000170(38)} & \text{0.000627(23)} & \text{0.00061(13)} \\
 24 & 0.0012416 & \text{$<$ 0.0001} & \text{$<$ 0.0001} & \text{0.0002302(93)} & \text{0.000224(54)} \\
 25 & 0.00056553 & \text{$<$ 0.0001} & \text{$<$ 0.0001} & \text{$<$ 0.0001} & \text{$<$ 0.0001} \\
 26 & 0.00023956 & \text{$<$ 0.0001} & \text{$<$ 0.0001} & \text{$<$ 0.0001} & \text{$<$ 0.0001} \\
 27 & \text{$<$ 0.0001} & \text{$<$ 0.0001} & \text{$<$ 0.0001} & \text{$<$ 0.0001} & \text{$<$ 0.0001} \\
 28 & \text{$<$ 0.0001} & \text{$<$ 0.0001} & \text{$<$ 0.0001} & \text{$<$ 0.0001} & \text{$<$ 0.0001} \\
 29 & \text{$<$ 0.0001} & \text{$<$ 0.0001} & \text{$<$ 0.0001} & \text{$<$ 0.0001} & \text{$<$ 0.0001} \\
 30 & \text{$<$ 0.0001} & \text{$<$ 0.0001} & \text{$<$ 0.0001} & \text{$<$ 0.0001} & \text{$<$ 0.0001} \\
\hline
\hline
\\
\caption{
Unprojected wavefunctions expressed in the full $\beta=0$ basis, reconstructed from classical and quantum simulations in 
$\Lambda=2,4$ effective model spaces, along with the exact wavefunction.
The LMG model parameters are:
$N=30$, $\varepsilon=1.0$, $\overline{v}=2.0$.
The projected ground-state wavefunctions can be obtained by setting the odd-$n$ components to zero, and rescaling, as described in Sec.~\ref{sec:LGM_eff}.
}
\label{tab:fullpsi}
\end{longtable*}
%


%
\begin{table}[!ht]
\centering
\begin{tabular}{cccc} 
\hline   
\\
 &  & $N=32$ &  
\\
\\
\hline
\\
 $\Lambda$ & {\rm Naive} & {\rm Effective} & {\rm Effective Projected   }
\\
 \\
\hline
 2 & 4.1650 & 1.6497 $\times 10^{-1}$ & 1.6497$\times 10^{-1}$ \\
 4 & 3.4144 & 1.5623$\times 10^{-2}$ & 1.5619$\times 10^{-2}$ \\
 6 & 2.4775 & 2.0314$\times 10^{-3}$ & 1.9222$\times 10^{-3}$ \\
 8 & 1.6218 & 6.9980$\times 10^{-4}$ & 2.3089$\times 10^{-4}$ \\
 10 & 9.5918$\times 10^{-1}$ & 5.5488$\times 10^{-4}$ & 2.4225$\times 10^{-5}$ \\
 12 & 4.9318$\times 10^{-1}$ & 5.3602$\times 10^{-4}$ & 4.0713$\times 10^{-6}$ \\
 14 & 2.0974$\times 10^{-1}$ & 5.3282$\times 10^{-4}$ & 9.5794$\times 10^{-7}$ \\
 16 & 6.9504$\times 10^{-2}$ & 5.3203$\times 10^{-4}$ & 2.3434$\times 10^{-7}$ \\
 18 & 1.7265$\times 10^{-2}$ & 5.3168$\times 10^{-4}$ & 1.2486$\times 10^{-7}$ \\
 20 & 3.6812$\times 10^{-3}$ & 5.3131$\times 10^{-4}$ & 3.5171$\times 10^{-7}$ \\
 22 & 8.9597$\times 10^{-4}$ & 5.2763$\times 10^{-4}$ & 1.0558$\times 10^{-5}$ \\
 24 & 7.4972$\times 10^{-5}$ & 7.4972$\times 10^{-5}$ & 7.4972$\times 10^{-5}$ \\
 26 & 3.7240$\times 10^{-6}$ & 3.7240$\times 10^{-6}$ & 3.7240$\times 10^{-6}$ \\
 28 & 9.6145$\times 10^{-8}$ & 9.6145$\times 10^{-8}$ & 9.6145$\times 10^{-8}$ \\
 30 & 9.7540$\times 10^{-10}$ & 9.7541$\times 10^{-10}$ & 9.7541$\times 10^{-10}$ \\
 32 & 0 & 0 & 0 \\
\hline
\hline
\end{tabular}
\caption{
The difference between the exact ground state energy and that obtained using HL-VQE 
with different truncations obtained with classical simulations (in unit of $\varepsilon$).
"Naive" indicates truncation of the Hamiltonian without modification, 
"Effective" indicates results from the effective model space,
and "Effective Projected" indicates results from the projected effective wavefunction.
The LMG model parameters are:
$N=32$, $\varepsilon=1.0$, $\overline{v}=2.0$.
}
\label{tab:N32Ediff}
\end{table}
%


%
\begin{table}[!ht]
\centering
\begin{tabular}{cccc} 
\hline   
\\
 &  & $N=64$ &  
\\
\\
\hline
\\
 $\Lambda$ & {\rm Naive} & {\rm Effective} & {\rm Effective Projected  }
\\
 \\
\hline
\hline
2 & 8.1569 & 1.5689$\times 10^{-1}$ & 1.5689$\times 10^{-1}$ \\
 4 & 7.4157 & 1.3157$\times 10^{-2}$ & 1.3157$\times 10^{-2}$ \\
 6 & 6.3471 & 1.0902$\times 10^{-3}$ & 1.0902$\times 10^{-3}$ \\
 8 & 5.2706 & 9.0117$\times 10^{-5}$ & 9.0117$\times 10^{-5}$ \\
 10 & 4.2615 & 7.7929$\times 10^{-6}$ & 7.7886$\times 10^{-6}$ \\
 12 & 3.3392 & 8.2237$\times 10^{-7}$ & 7.8705$\times 10^{-7}$ \\
 14 & 2.5243 & 2.0110$\times 10^{-7}$ & 9.6436$\times 10^{-8}$ \\
 16 & 1.8297 & 1.4181$\times 10^{-7}$ & 6.4595$\times 10^{-9}$ \\
 18 & 1.2606 & 1.3566$\times 10^{-7}$ & 9.1655$\times 10^{-10}$ \\
 20 & 8.1584$\times 10^{-1}$ & 1.3495$\times 10^{-7}$ & 1.0981$\times 10^{-10}$ \\
 22 & 4.8843$\times 10^{-1}$ & 1.3486$\times 10^{-7}$ & 1.6069$\times 10^{-11}$ \\
 24 & 2.6531$\times 10^{-1}$ & 1.3485$\times 10^{-7}$ & 2.9514$\times 10^{-12}$ \\
 26 & 1.2783$\times 10^{-1}$ & 1.3484$\times 10^{-7}$ & 6.7541$\times 10^{-13}$ \\
 28 & 5.3358$\times 10^{-2}$ & 1.3484$\times 10^{-7}$ & 1.8833$\times 10^{-13}$ \\
 30 & 1.8905$\times 10^{-2}$ & 1.3484$\times 10^{-7}$ & 6.1047$\times 10^{-14}$ \\
 32 & 5.6094$\times 10^{-3}$ & 1.3484$\times 10^{-7}$ & 2.5971$\times 10^{-14}$ \\
 34 & 1.3850$\times 10^{-3}$ & 1.3484$\times 10^{-7}$ & 8.9075$\times 10^{-15}$ \\
 36 & 2.8397$\times 10^{-4}$ & 1.3484$\times 10^{-7}$ & 9.2551$\times 10^{-15}$ \\
 38 & 4.8379$\times 10^{-5}$ & 1.3484$\times 10^{-7}$ & 7.0723$\times 10^{-15}$ \\
 40 & 6.9896$\times 10^{-6}$ & 1.3484$\times 10^{-7}$ & 4.9407$\times 10^{-15}$ \\
 42 & 1.0368$\times 10^{-6}$ & 1.3484$\times 10^{-7}$ & 4.7592$\times 10^{-14}$ \\
 44 & 2.3949$\times 10^{-7}$ & 1.3477$\times 10^{-7}$ & 2.7661$\times 10^{-10}$ \\
 46 & 1.9755$\times 10^{-8}$ & 1.9755$\times 10^{-8}$ & 1.9755$\times 10^{-8}$ \\
 48 & 1.2852$\times 10^{-9}$ & 1.2852$\times 10^{-9}$ & 1.2852$\times 10^{-9}$ \\
 50 & 6.445$\times 10^{-11}$ & 6.445$\times 10^{-11}$ & 6.445$\times 10^{-11}$ \\
 52 & 2.417$\times 10^{-12}$ & 2.417$\times 10^{-12}$ & 2.417$\times 10^{-12}$ \\
54 & 6.5$\times 10^{-14}$ & 6.5$\times 10^{-14}$ & 6.5$\times 10^{-14}$ \\
 56 & 1.1$\times 10^{-15}$ & 1.1$\times 10^{-15} $ & 1.1$\times 10^{-15}$ \\
 \hline
\end{tabular}
\caption{
The difference between the exact ground state energy and that obtained using HL-VQE 
with different truncations obtained with classical simulations (in unit of $\varepsilon$).
"Naive" indicates truncation of the Hamiltonian without modification, 
"Effective" indicates results from the effective model space,
and "Effective Projected" indicates results from the projected effective wavefunction.
The LMG model parameters are:
$N=64$, $\varepsilon=1.0$, $\overline{v}=2.0$.
}
\label{tab:N64Ediff}
\end{table}
%


%
\begin{table}[!ht]
\centering
\begin{tabular}{cccc} 
\hline   
\\
 &  & $N=96$ &  
\\
\\
\hline
\\
 $\Lambda$ & {\rm Naive} & {\rm Effective} & {\rm Effective Projected   }
\\
 \\
\hline
\hline
2 & 1.2155$\times 10^1$ & 1.5459$\times 10^{-1}$ & 1.5459$\times 10^{-1}$ \\
 4 & 1.1416$\times 10^1$ & 1.2629$\times 10^{-2}$ & 1.2629$\times 10^{-2}$ \\
 6 & 1.0304$\times 10^1$ & 9.9641$\times 10^{-4}$ & 9.9641$\times 10^{-4}$ \\
 8 & 9.1508 & 7.6372$\times 10^{-5}$ & 7.6372$\times 10^{-5}$ \\
 10 & 8.0154 & 5.8604$\times 10^{-6}$ & 5.8604$\times 10^{-6}$ \\
 12 & 6.9164 & 4.5826$\times 10^{-7}$ & 4.5826$\times 10^{-7}$ \\
 14 & 5.8757 & 3.6940$\times 10^{-8}$ & 3.6940$\times 10^{-8}$ \\
 16 & 4.9088 & 3.1168$\times 10^{-9}$ & 3.1151$\times 10^{-9}$ \\
 18 & 4.0261 & 2.9945$\times 10^{-10}$ & 2.9098$\times 10^{-10}$ \\
 20 & 3.2338 & 5.4361$\times 10^{-11}$ & 3.3224$\times 10^{-11}$ \\
 22 & 2.5358 & 3.2170$\times 10^{-11}$ & 3.6422$\times 10^{-12}$ \\
 24 & 1.9334 & 2.9776$\times 10^{-11}$ & 2.1585$\times 10^{-13}$ \\
 26 & 1.4260 & 2.9563$\times 10^{-11}$ & 3.9432$\times 10^{-14}$ \\
 28 & 1.0109 & 2.9534$\times 10^{-11}$ & 9.3732$\times 10^{-15}$ \\
 30 & 6.8338$\times 10^{-1}$ & 2.9527$\times 10^{-11}$ & 2.5873$\times 10^{-15}$ \\
 32 & 4.3622$\times 10^{-1}$ & 2.9541$\times 10^{-11}$ & - \\
 34 & 2.5994$\times 10^{-1}$ & 2.9527$\times 10^{-11}$ & - \\
 36 & 1.4276$\times 10^{-1}$ & 2.9527$\times 10^{-11}$ & - \\
 38 & 7.1321$\times 10^{-2}$ & 2.9520$\times 10^{-11}$ & - \\
 40 & 3.2037$\times 10^{-2}$ & 2.9520$\times 10^{-11}$ & - \\
 42 & 1.2829$\times 10^{-2}$ & 2.9520$\times 10^{-11}$ & - \\
 44 & 4.5583$\times 10^{-3}$ & 2.9520$\times 10^{-11}$ & - \\
 46 & 1.4350$\times 10^{-3}$ & 2.9534$\times 10^{-11}$ & - \\
 48 & 4.0038$\times 10^{-4}$ & 2.9520$\times 10^{-11}$ & - \\
 50 & 9.9070$\times 10^{-5}$ & 2.9527$\times 10^{-11}$ & - \\
 52 & 2.1741$\times 10^{-5}$ & 2.9527$\times 10^{-11}$ & - \\
 54 & 4.2285$\times 10^{-6}$ & 2.9527$\times 10^{-11}$ & - \\
 56 & 7.2773$\times 10^{-7}$ & 2.9520$\times 10^{-11}$ & - \\
 58 & 1.1060$\times 10^{-7}$ & 2.9520$\times 10^{-11}$ & - \\
 60 & 1.4833$\times 10^{-8}$ & 2.9534$\times 10^{-11}$ & - \\
 62 & 1.7895$\times 10^{-9}$ & 2.9520$\times 10^{-11}$ & - \\
 64 & 2.3580$\times 10^{-10}$ & 2.9527$\times 10^{-11}$ & - \\
 66 & 5.3561$\times 10^{-11}$ & 2.9527$\times 10^{-11}$ & - \\
 68 & 4.4042$\times 10^{-12}$ & 4.4096$\times 10^{-12}$ & 4.4096$\times 10^{-12}$ \\
 70 & 3.1050$\times 10^{-13}$ & 5.0871$\times 10^{-13}$ & 5.0871$\times 10^{-13}$ \\
 72 & 1.8596$\times 10^{-14}$ & 1.8431$\times 10^{-14}$ & 1.8431$\times 10^{-14}$ \\
 \hline
\end{tabular}
\caption{
The difference between the exact ground state energy and that obtained using HL-VQE 
with different truncations obtained with classical simulations (in unit of $\varepsilon$).
"Naive" indicates truncation of the Hamiltonian without modification, 
"Effective" indicates results from the effective model space,
and "Effective Projected" indicates results from the projected effective wavefunction.
A dash indicates a value below $10^{-15}$, and all values for $74<\Lambda<96$ are below this value.
The LMG model parameters are:
$N=96$, $\varepsilon=1.0$, $\overline{v}=2.0$.
}
\label{tab:N96Ediff}
\end{table}

\clearpage
\bibliography{biblio,biblioKRS}

\end{document}